\documentclass[12pt]{article}

\pdfoutput=1
\usepackage[latin9]{inputenc}
\usepackage[top=80pt,bottom=85pt,left=65pt,right=65pt]{geometry}
\usepackage{amssymb,amsmath}
\usepackage{xypic,subfigure}
\usepackage{graphicx,color}
\usepackage{float}
\usepackage{cite}
\usepackage[debug,pageanchor=false]{hyperref}
\definecolor{link}{rgb}{.8,.15,.1}
\hypersetup{colorlinks=true,linkcolor=link,citecolor=link,urlcolor=link,linktocpage}

\makeatletter
\@addtoreset{equation}{section}
\makeatother

\renewcommand{\theequation}{\thesection.\arabic{equation}}

\newcommand{\beq}{\begin{equation}}
\newcommand{\eeq}{\end{equation}}
\newcommand{\bea}{\begin{eqnarray}}
\newcommand{\eea}{\end{eqnarray}}
\newcommand{\nn}{\nonumber}

\newcommand{\eq}{\begin{equation}}
\newcommand{\feq}{\end{equation}}
\newcommand{\eqn}{\begin{eqnarray}}
\newcommand{\feqn}{\end{eqnarray}}

\newcommand{\ma}[1]{\mbox{$\mathcal{#1}$}}

\newcommand{\mrm}[1]{\mbox{$\mathrm{#1}$}}

\begin{document}
\begin{titlepage}

\begin{center}

\vskip .5in 
\noindent

{\Large \bf{New AdS$_3$/CFT$_2$ pairs in massive IIA with $(0,4)$ and $(4,4)$ supersymmetries}}

\bigskip\medskip

Yolanda Lozano$^{a,b}$\footnote{ylozano@uniovi.es}, Niall T. Macpherson$^{c,d}$\footnote{ntmacpher@gmail.com},  Nicol\`o Petri$^e$\footnote{petri@post.bgu.ac.il}, Cristian Risco$^{a,b}$\footnote{cristianrg96@gmail.com}  \\

\bigskip\medskip
{\small 

a: Department of Physics, University of Oviedo,
Avda. Federico Garcia Lorca s/n, 33007 Oviedo
\\
and
\\
b: Instituto Universitario de Ciencias y Tecnolog\'ias Espaciales de Asturias (ICTEA), Calle de la Independencia 13, 33004 Oviedo, Spain}
\bigskip\medskip
{\small 

c: Departamento de F\'{i}sica de Part\'{i}culas, Universidade de Santiago de Compostela
\\
and
\\

d: Instituto Galego de F\'{i}sica de Altas Enerx\'{i}as (IGFAE),\\
R\'{u}a de Xoaqu\'{i}n D\'{i}az de R\'{a}bago s/n
E-15782 Santiago de Compostela, Spain
}
\bigskip\medskip
{\small 

e: Department of Physics, Ben-Gurion University of the Negev, Beer-Sheva 84105, Israel.}

\vskip 1cm 

     	{\bf Abstract }
     	\end{center}
     	\noindent
	
We construct a new class of $\text{AdS}_3\times $S$^3\times $M$_4$ solutions of massive Type IIA supergravity with $(0,4)$ supersymmetries and SU(3) structure. We study in detail two subclasses of these solutions. The first subclass is when M$_4=$S$^2\times \Sigma_2$, with $\Sigma_2$ a 2d Riemann surface, and the geometry is foliated over the $\Sigma_2$. We interpret these solutions as duals to surface defect CFTs within the 6d $(1,0)$ CFTs dual to the $\text{AdS}_7\times$S$^2\times I$ solutions of massive IIA supergravity. The second subclass is when M$_4=\mathbb{T}^3\times I$ and the geometry is foliated over the interval.  In this case supersymmetry is enhanced to $(4,4)$ in the massless limit, and the solutions are holographically dual to $(4,4)$ CFTs living in two dimensional D2-NS5-D4 Hanany-Witten brane set-ups. In turn, in the massive case the solutions find an interpretation as D2-D4 branes embedded in Type I' string theory. We construct explicit quiver gauge theories from the different brane set-ups that flow in the IR to the 2d dual CFTs dual to the solutions. We check the validity of our proposals with the matching between the field theory and holographic central charges.

\noindent

\vfill
\eject

\end{titlepage}

\setcounter{footnote}{0}

\tableofcontents

\setcounter{footnote}{0}
\renewcommand{\theequation}{{\rm\thesection.\arabic{equation}}}

\section{Introduction}

The study of the AdS/CFT correspondence in low dimensions has seen renewed interest in the last few years \cite{Tong:2014yna}-\cite{Couzens:2022agr}. On the AdS  side of the correspondence, a plethora of new AdS$_3$ and AdS$_2$ solutions of Type II and eleven dimensional supergravities with different amounts of supersymmetries have been constructed. In turn, on the CFT side it has been possible to identify the 2d and 1d CFTs dual to some of these solutions as IR fixed points of explicit quiver field theories, from where it has been possible to explore some of their properties, in particular to compute their central charge. These AdS/CFT pairs thus represent perfect scenarios where the Bekenstein-Hawking entropy of black strings and black holes can be computed microscopically. This is particularly promising for the large classes of black strings and black holes with $\mathcal{N}=(0,4)$ and $\mathcal{N}=4$ supersymmetries constructed in \cite{Couzens:2017way,Couzens:2017nnr,Lozano:2019jza,Lozano:2019zvg,Lozano:2019ywa,Lozano:2020bxo,Faedo:2020nol,Lozano:2020txg,Lozano:2020sae,Faedo:2020lyw,Lozano:2021rmk,Ramirez:2021tkd,Lozano:2021fkk,Couzens:2021veb,Passias:2019rga,Passias:2020ubv}, which enable extensions of the seminal studies in \cite{Strominger:1996sh}-\cite{Minasian:1999qn}.

Another interesting interpretation of low dimensional AdS spaces is as holographic duals of CFTs describing defects within higher dimensional CFTs \cite{Karch:2000gx,DeWolfe:2001pq,Aharony:2003qf,DHoker:2006vfr,Lunin:2007ab}. Notable examples of such realisations for AdS$_3$ and AdS$_2$ spaces have been reported in 
\cite{DHoker:2007mci,Chiodaroli:2009yw,Chiodaroli:2009xh,Dibitetto:2017tve,Dibitetto:2017klx,Dibitetto:2018gtk,Dibitetto:2018iar,Chen:2020mtv,Faedo:2020nol,Lozano:2020sae,Faedo:2020lyw,Dibitetto:2020bsh,Ramirez:2021tkd,Lozano:2021fkk}. A hint that this interpretation may be possible is when the low dimensional AdS space flows asymptotically (locally) in the UV to a higher dimensional AdS geometry, which contains extra fluxes. These fluxes partially break the isometries (and typically also the supersymmetries) of the higher dimensional AdS space, and can be associated to extra {\it defect} branes embedded in the geometry. We will see that some of the AdS$_3$ solutions constructed in this paper allow for an interpretation as surface defects within  
6d $(1,0)$ CFTs dual to AdS$_7$ geometries.

AdS$_2$/CFT$_1$ holography features particular challenges not shared by higher dimensional AdS/CFT. These have to do mainly with the non-connectedness of the boundary of AdS$_2$ and  with the interpretation of the central charge of the dual super-conformal quantum mechanics (SCQM), which does not allow for finite energy excitations \cite{Maldacena:1998uz,Strominger:1998yg,Balasubramanian:2003kq,Hartman:2008dq,Balasubramanian:2009bg}. Therefore directly applying AdS$_2$/CFT$_1$ holography to the microscopic description of extremal black holes is not straightforward, and interesting alternative ways to make this possible have been proposed in the literature (see for example \cite{Almheiri:2014cka,Maldacena:2016hyu,Maldacena:2016upp,Harlow:2018tqv}).
 Recently, it has been shown  \cite{Lozano:2020txg} that for AdS$_2$ spaces related to AdS$_3$ through compactification or T-duality an understanding of the SCQM as a chiral half of a 2d CFT (following the ideas in  \cite{Balasubramanian:2003kq,Balasubramanian:2009bg}) allows one to sidestep these difficulties, providing explicit set-ups where the microscopic description program can been carried out in detail. It is likely that the solutions that we construct in this paper will allow for similar applications.
 
 In this paper we construct new AdS$_3$ solutions with small $(0,4)$ supersymmetry in massive Type IIA supergravity. The small ${\cal N}=4$ superconformal algebra is characterised by an SU(2) R-symmetry with generators transforming in the $\textbf{2}\oplus \overline{\textbf{2}}$, as such backgrounds realising this algebra should respect this isometry which requires an S$^2$ factor (either round or with U(1)s fibred over it). Small ${\cal N}=(0,4)$ backgrounds of Type II supergravity of the warped product form AdS$_3\times$S$^2\times $M$_5$ were recently classified across \cite{Lozano:2019emq,Macpherson:2022sbs} under the assumptions that M$_5$ contains no necessary isometries and S$^2$ does not experience an enhancement to S$^3$. Our focus here will be on solutions that lie outside these assumptions\footnote{Though they are related to classes in \cite{Macpherson:2022sbs} via T-duality.}, namely solutions containing a warped AdS$_3\times$S$^3$ factor. These have the benefit of being compatible with an enhancement to small ${\cal N}=(4,4)$, a maximal case for AdS$_3$ with relatively few known examples. We are aware of only the U-duality orbits of the D1-D5 near horizon \cite{Maldacena:1997re}, the $d=11$ solution of \cite{Dibitetto:2020bsh} and the type IIB class of \cite{Lin:2004nb}, albeit with no explicit examples. This enhancement is of course not guaranteed by the presence of an S$^3$ factor and indeed the class that we construct generically supports just $(0,4)$ supersymmetry. However an enhancement to ${\cal N}=(4,4)$ is possible when the class is suitably restricted, which allows us to find explicit examples with both $(0,4)$ and $(4,4)$ supersymmetry that we shall study in some detail.

The paper is organised as follows. In section \ref{eq:the massiveclass} we construct the general class of AdS$_3\times$S$^3\times$M$_4$ solutions of massive type IIA supergravity with $\mathcal{N}=(0,4)$ supersymmetries that are the focus of the paper. We do this by generalising the 
Minkowski$_6$ solutions constructed in \cite{Legramandi:2019ulq} to also include D2 and D4-branes. We check the supersymmetries and provide the explicit brane intersection, consisting on D2-D4 branes ending on D6-NS5-D8 bound states \cite{Imamura:2001cr}, from which the AdS$_3$ solutions arise in the near horizon limit. We further show that any solution to minimal $\mathcal{N}=2$ supergravity in 6d gives rise to a solution of massive IIA supergravity sharing the same warping and internal space as  our class.  This may be highly relevant towards the construction of superstrata solutions, as in \cite{Bena:2015bea}\footnote{We thank the referee for stressing this point.}.
In section  \ref{defectsin6d} we show that when M$_4=~$S$^2\times \Sigma_2$, with $\Sigma_2$ a 2d Riemann surface, and the geometry is foliated over the $\Sigma_2$, the AdS$_3$ solutions flow asymptotically in the UV to the AdS$_7\times $S$^2\times I$ solutions of massive IIA supergravity constructed in \cite{Apruzzi:2013yva}, dual to 6d $(1,0)$ CFTs living in D6-NS5-D8 intersections \cite{Gaiotto:2014lca,Cremonesi:2015bld}. This allows us to interpret this subclass of solutions as holographic duals of 2d $(0,4)$ CFTs describing D2-D4 defects inside the 6d CFTs. We construct the 2d $(0,4)$ quiver gauge theories that flow in the IR to the duals of our solutions, and show that they can be embedded within the 6d quivers constructed in \cite{Cremonesi:2015bld}. This extends (and corrects, in the precise sense discussed in the paper) the constructions in \cite{Faedo:2020nol} for the massless case. In section \ref{D2D4NS5} we focus on the subclass of solutions for which M$_4=\mathbb{T}^3\times I$ and the geometry is foliated over the interval, first in massless IIA. We show that these solutions arise in the near horizon limit of D2-D4-NS5 brane intersections, and enjoy an enhancement  to $\mathcal{N}=(4,4)$ supersymmetry. Our constructions represent a key step forward in the identification of the holographic duals of $(4,4)$ 2d CFTs living in D2-D4-NS5 Hanany-Witten brane set-ups, studied long ago in \cite{Brodie:1997wn,Alishahiha:1997cm}. As a consistency check of our proposal we show that the holographic and field theory central charges are in exact agreement. In section \ref{TypeI'} we complete the analysis of the AdS$_3\times$S$^3\times \mathbb{T}^4\times I$ solutions in the presence of Romans mass. We show that these backgrounds are associated to D2-D4-D8 intersections preserving $(0,4)$ supersymmetries, that can be globally embedded in Type I' string theory. We perform this explicit construction and check the matching between the field theory and holographic central charges. Section \ref{conclusions} contains our conclusions, where we summarise our results and discuss future lines of investigation, in particular the possibility of constructing new AdS$_2$ solutions with $\mathcal{N}=4$ by acting with Abelian and non-Abelian T-dualities on our new classes of solutions \cite{us}. Finally in Appendix \ref{appendix} we complement our analysis in section \ref{defectsin6d} with the construction of a domain wall solution to 7d minimal supergravity that flows to the AdS$_7$ vacuum asymptotically.

\section{A new class of $\ma N=(0,4)$ AdS$_3$ solutions in massive IIA}\label{eq:the massiveclass}

A class of solutions in massive IIA that has born much fruit over the years is the D8-D6-NS5 flat-space brane intersection \cite{Imamura:2001cr}. This is a class of $\frac{1}{4}$ BPS warped Minkowski$_6$ solutions which support an SU(2) R-symmetry realised by a round 2-sphere in the internal space. All AdS$_7$ solutions in Type II supergravity are contained in this class  as well as examples of compact Mink$_4\times \mathbb{T}^2$ vacua \cite{Macpherson:2016xwk,Bobev:2016phc}.  A  generalisation of this class without the 2-sphere was found in \cite{Legramandi:2019ulq}, where solutions with O planes back-reacted on a torus were found. The metric and fluxes of solutions in this generalised class take the local form
\begin{align}\label{eq:massiveclassmetricoriginal}
ds^2&=\frac{1}{\sqrt{h}}ds^2(\mathbb{R}^{1,5})+ g \bigg[\frac{1}{\sqrt{h}} d\rho^2+\sqrt{h}ds^2(\mathbb{R}^3)\bigg],~~~~e^{-\Phi}=\frac{h^{\frac{3}{4}}}{\sqrt{g}},\\[2mm]
F_0&= \frac{\partial_{\rho}h}{g},~~~~F_2= -\star_3 d_3 h ,~~~~H_3=\partial_{\rho}(h g)\text{vol}(\mathbb{R}^3)-(\star_3 d_3 g)\wedge d\rho,\nn
\end{align}
where $h,g$ have support on $(\rho,\mathbb{R}^3)$ and $(d_3,\star_3)$ are the exterior derivative and Hodge dual on $\mathbb{R}^3$. Away from the loci of possible sources the Bianchi identies of the 2 and 3-form impose that
\begin{align}\label{eq:Bianchiidentitiesorg}
&\partial_{\rho}(\frac{\partial_{\rho}h}{g})=0,~~~~\nabla_3^2 g +\partial_{\rho}^2 (gh)=0,~~~~\nabla_3^2 h +F_0\partial_{\rho} (gh)=0,
\end{align}
with any solution to this system giving rise to a solution of massive IIA supergravity, provided any localised source terms are also calibrated.\\
~~\\
In this section we will present a generalised version of this class for which 
\beq
\mathbb{R}^{1,5} \to  \text{AdS}_3\times \text{S}^3,~~~~ (F_0,F_2,H_3)\to(F_0,F_2,F_4,H_3),
\eeq
giving rise to AdS$_3$ vacua of massive IIA preserving small ${\cal N}=(0,4)$ supersymmetry, as explained in section \ref{sec:SUSY}. We shall construct a system of
intersecting branes in flat space giving rise to these AdS$_3$ vacua in a near horizon limit in section \ref{branepicture}, and finally establish that in fact any solution of minimal ${\cal N}=2$ supergravity in $d=6$ can be embedded into massive IIA with a similar ansatz for the metric and fluxes in section \ref{sec:uplift}.

\subsection{A small ${\cal N}=(0,4)$ AdS$_3$ class with source D8-D6-NS5 branes}\label{sec:SUSY}
In this section we present a new class of AdS$_3$ solutions preserving small ${\cal N}=(0,4)$ supersymmetry with possible D8-D6-NS5 sources.\\
~~\\
The general form of the metric and dilaton of solutions in this class is nothing more than \eqref{eq:massiveclassmetricoriginal} with $\mathbb{R}^{1,5} \to  \text{AdS}_3\times \text{S}^3$,
\begin{align}\label{eq:massiveclassmetric}
ds^2=\frac{q}{\sqrt{h}}\bigg[ds^2(\text{AdS}_3)+ds^2(\text{S}^3)\bigg]+ g \bigg[\frac{1}{\sqrt{h}} d\rho^2+\sqrt{h}\bigg(dz_1^2+dz_2^2+dz_3^2\bigg)\bigg],~~~~e^{-\Phi}=\frac{h^{\frac{3}{4}}}{\sqrt{g}},
\end{align}
where AdS$_3$ and S$^3$ both have  unit radius and $q$ is a redundant constant we keep to make contact with later sections more smooth. We have introduced $(z_1,z_2,z_3)$ coordinates spanning the $\mathbb{R}^3$ factor for later convenience. The fluxes this solution supports are
\begin{subequations}
\begin{align}
F_0&= \frac{\partial_{\rho}h}{g},~~~~F_4=2\,q\bigg(\text{vol}(\text{AdS}_3)+\text{vol}(\text{S}^3)\bigg)\wedge d\rho,\label{eq:flux1}\\[2mm]
F_2&= -(\partial_{z_1}h dz_2\wedge dz_3+\partial_{z_2}h dz_3\wedge dz_1+\partial_{z_3}h dz_1\wedge dz_2),\label{eq:flux2}\\[2mm]
H_3&=\partial_{\rho}(h g)dz_1\wedge dz_2\wedge dz_3-(\partial_{z_1}g dz_2\wedge dz_3+\partial_{z_2}g dz_3\wedge dz_1+\partial_{z_3}g dz_1\wedge dz_2)\wedge d\rho,\label{eq:flux3}
\end{align}
\end{subequations}
where the  additional 4-form with respect to \eqref{eq:massiveclassmetricoriginal} is to be expected given that the external space has been replaced with a curved product space. The Bianchi identities of the fluxes, in regular regions of the internal space, require that $F_0$ is constant and 
\begin{align}
&(\partial_{z_1}^2+\partial_{z_2}^2+\partial_{z_3}^2)g +\partial_{\rho}^2 (gh)=0,\nn\\[2mm]
&(\partial_{z_1}^2+\partial_{z_2}^2+\partial_{z_3}^2)h +F_0\partial_{\rho} (gh)=0,\label{eq:Bianchiidentities}
\end{align}
which exactly reproduce \eqref{eq:Bianchiidentitiesorg} and define solutions in this class. Actually these constraints give rise to two local classes depending on whether $F_0=0$ or not. As $F_0=0$ demands $\partial_{\rho}h=0$, the governing PDEs reduce to those of a flat space D6-NS5 brane intersection. On the other hand when $F_0\neq 0$, one is free to divide by it and take
\beq\label{imamuraeq1}
g= \frac{\partial_{\rho}h}{F_0}.
\eeq
Given this one can then show that \eqref{eq:Bianchiidentities} reduce to a single PDE
\beq\label{imamuraeq2}
(\partial_{z_1}^2+\partial_{z_2}^2+\partial_{z_3}^2)h+\frac{1}{2}\partial_{\rho}^2( h^2)=0,
\eeq
reproducing the novel behaviour of \cite{Imamura:2001cr} when we impose SO(3) invariance in $(z_1,z_2,z_3)$. Let us now move on to address the amount of supersymmetry solutions in this class preserve.

\subsubsection{Supersymmetry}\label{supersymmetry}
The preservation of supersymmetry for AdS$_3$ vacua in massive IIA can be phrased in terms of differential bi-spinor relations first introduced for ${\cal N}=(0,1)$ in \cite{Dibitetto:2018ftj}. In the conventions of \cite{Macpherson:2021lbr} for a solution decomposing as
\beq
ds^2= e^{2A}ds^2(\text{AdS}_3)+ ds^2(\text{M}_7),~~~~F=  f_++ e^{3A}\text{vol}(\text{AdS}_3)\wedge \star_7 \lambda f,
\eeq
with purely magnetic NS flux, dilaton $\Phi$ and where $\lambda f_n=  (-1)^{[\frac{n}{2}]}f_n$, these are\footnote{We are also assuming unit radius AdS$_3$ and have fixed an arbitrary constant below. The truly general conditions are given in \cite{Macpherson:2021lbr}. Note that we have inverted what is referred to as ${\cal N}=(1,0)$ and ${\cal N}=(0,1)$ with respect to that reference.}
\begin{align}
&d_{H_3}(e^{A-\Phi}\Psi_-)=0,~~~~d_{H_3}(e^{2A-\Phi}\Psi_+)- 2 e^{A-\Phi}\Psi_{-}=\frac{1}{8}e^{3A}\star_7\lambda(f_{+}),\nn\\[2mm]
&(\Psi_{-}\wedge \lambda f_{+})\bigg\lvert_7= - \frac{ 1}{2} e^{-\Phi}\text{vol}(\text{M}_7),\label{eq:BPSgen}
\end{align}
where $\Psi_{\pm}$ can be defined in term of spinors supported by M$_7$. However one does not need to make specific reference to these, it is sufficient that $\Psi_{\pm}$ realises a  G$_2\times $G$_2$-structure. For our purposes it will be sufficient to consider a restricted case where the intersection of these two G$_2$'s is a strict SU(3)-structure for which one may parameterise
\beq
\Psi_+=-\text{Im}\left(e^{-i J}\right)+V\wedge \text{Re}\Omega,~~~~\Psi_-= -\text{Im}\Omega- V\wedge \text{Re}\left(e^{-i J}\right),
\eeq
where $V$ is a real 1-form defining a vielbein direction in M$_7$, while $(J,\Omega)$ can be written in terms of a further 3 complex vielbein directions $E_1,E_2,E_3$ as
\beq
J= E_1\wedge\overline{E}_1+E_2\wedge\overline{E}_2+E_3\wedge\overline{E}_3,~~~~\Omega= E_1\wedge E_2\wedge E_3.
\eeq
The class of solutions of the previous section preserves ${\cal N}=(0,4)$ supersymmetry if it preserves 4 independent SU(3)-structures which each obey \eqref{eq:BPSgen}. As the class contains an S$^3$ factor one can define 1-forms $(L_a,R_a)$ for $a=1,2,3$ such that 
\beq
dL_a=\frac{1}{2}\epsilon_{abc}L_b\wedge L_c,~~~~dR_a=-\frac{1}{2}\epsilon_{abc}R_b\wedge R_c,~~~~ ds^2(\text{S}^3)=\frac{1}{4}(L_a)^2=\frac{1}{4}(R_a)^2,
\eeq
with $L_a$ a singlet/triplet under the SO(3)$_{L/R}$ subgroup of SO(4) $=$ SO(3)$_L\times $SO(3)$_R$, with the charge of $R_a$ the opposite. It is possible to show that  the SU(3)-structure defined through the vielbein
\beq\label{eq:specificvielnbein}
E_a=  -\sqrt{g}h^{\frac{1}{4}}dx_a+ i \frac{1}{2\mu h^{\frac{1}{4}}} L_a,~~~V= \frac{\sqrt{g}}{h^{\frac{1}{4}}}d\rho
\eeq
solves \eqref{eq:BPSgen}, realising ${\cal N}=(0,1)$ explicitly. This gets enhanced to ${\cal N}=(0,4)$ because $\Psi_{\pm}$ depend on the 3-sphere through $L_a,dL_a$, which are SO(3)$_R$ triplets, and  $\text{vol}(\text{S}^3)$, an SO(4) invariant, with only the latter entering the physical fields. As such, if \eqref{eq:specificvielnbein} solves \eqref{eq:BPSgen} so too does the SU(3)-structure that results after performing a generic constant SO(3) rotation
 of $L_a$ in \eqref{eq:specificvielnbein}, which one can exploit to generate another 3 independent SU(3)-structures necessarily solving \eqref{eq:BPSgen} for the same physical fields, for 4 SU(3)-structures in total. That it is specifically small ${\cal N}=(0,4)$ that is realised for this class rather than some other superconformal group is obvious once one notes that any other choice would necessitate additional isometries not present in the class generically. Additionally, through Hopf fiber T-duality, it is possible to map the class of solutions to that of section 3.3 of \cite{Macpherson:2022sbs}, specialised to the case where the local coordinate $x$ there is an isometry, which proves this more rigorously. 

Given the round 3-sphere in this class one might wonder whether, or under which conditions, there is an enhancement to ${\cal N}=(4,4)$. This would require a further 4 ${\cal N}=(1,0)$ SU(3)-structures to be supported by the background, which should solve a cousin of \eqref{eq:BPSgen} with $\Psi_-\to -\Psi_-$\footnote{Beware this map does not hold in full generality, only for the restricted case we consider. See \cite{Macpherson:2021lbr} for full details.}. These need to span the 3-sphere in terms of $R_a$ as each ${\cal N}=4$ sub-sector must be a singlet with respect to the R-symmetry of the other. One can show that the vielbein
\beq
E_a=  \sqrt{g}h^{\frac{1}{4}}dx_a+ i \frac{1}{2\mu h^{\frac{1}{4}}} R_a,~~~V= -\frac{\sqrt{g}}{h^{\frac{1}{4}}}d\rho
\eeq
does give rise to an SU(3)-structure which solves the ${\cal N}=(1,0)$ conditions, with a further 3 implied by this as before. However the RR 2-form now changes sign with respect to \eqref{eq:flux2}. The only way to have the same physical fields compatible with both left and right ${\cal N}=4$ sub-sectors is to fix  $dh=0$, ie
\beq
h=\text{constant}~~~~\Rightarrow~~~~{\cal  N}=(4,4),
\eeq
which makes $F_2,F_0$ trivial. Generically however just ${\cal N}=(0,4)$ is preserved. Finally we should comment that  when $h\neq$ constant one is free to replace S$^3$ by the lens space S$^3/\mathbb{Z}_k$ without breaking any further supersymmetry. Instead, when $h =$ constant the Lens space breaks ${\cal N}=(4,4)$  to ${\cal N}=(0,4)$.

\subsection{The brane picture}\label{branepicture}

In this section we show that the class of solutions \eqref{eq:massiveclassmetric} can be obtained as the near-horizon limit of a brane intersection defined by D2-D4 branes ending on D6-NS5-D8 bound states, as depicted in Table \ref{D2D4D6D8NS5_1}.
 \begin{table}[http!]
	\begin{center}
		\begin{tabular}{| l | c | c | c| c | c | c | c | c| c | c |}
			\hline		    
			& $x^0$ & $x^1$  & $r$ & $\varphi^1$ & $\varphi^2$ & $\rho$ & $\zeta$ & $\theta^1$ & $\theta^2$ & $\theta^3$ \\ \hline
			D2 & x & x &  & & & x & & & & \\ \hline
			D4 & x & x & x & x & x & & & & & \\ \hline
			NS5 & x & x  &  &  &   & &x  & x  & x & x  \\ \hline
			D6 & x & x &   &  &  &x& x  &x   &x   &x   \\ \hline
			D8 & x & x  &x  & x &  x & &x  &x  & x & x  \\ \hline
		\end{tabular} 
	\end{center}
	\caption{$\frac18$-BPS brane intersection underlying the $\ma N=(0,4)$ AdS$_3$ solutions \eqref{eq:massiveclassmetric}. $(x^0,x^1)$ are the directions where the 2d dual CFT lives, $(r,\varphi^i)$ are spherical coordinates spanning the 3d space previously parameterised by $(z_1,z_2,z_3)$, $\zeta$ is the radial coordinate of AdS$_3$  and $\theta^i$ parameterise the S$^3$.} 
	\label{D2D4D6D8NS5_1}	
\end{table}

Imamura's D6-NS5-D8 flat-space intersection \cite{Imamura:2001cr} is described by the supergravity solution \eqref{eq:massiveclassmetricoriginal}. Adding D2-D4 branes the 10d metric becomes
\begin{equation}
\label{brane_metric_D2D4NS5D6D8}
\begin{split}
d s^2&=\,h^{-1/2}\,\left[H_{\mathrm{D}4}^{-1/2}\,H_{\mathrm{D}2}^{-1/2}\,ds^2({\mathbb{R}^{1,1}})+H_{\mathrm{D}4}^{1/2}\,H_{\mathrm{D}2}^{1/2} \,(d\zeta^2+\zeta^2ds^2(\text{S}^3))\right] \\
&+h^{-1/2}\,g\,H_{\mathrm{D}4}^{1/2}\,H_{\mathrm{D}2}^{-1/2}\,d\rho^2+h^{1/2}\,g\,H_{\mathrm{D}4}^{-1/2}\,H_{\mathrm{D}2}^{1/2}(dr^2+r^2 ds^2(\text{S}^2)) \, ,
\end{split}
\end{equation}
where we have parameterised the 2d Minkowski spacetime $\mathbb{R}^{1,1}$ with $(x^0, x^1)$, the 4d space transverse to the D2-D4 branes with coordinates $(\zeta, \theta^i)$ and the 3d space parameterised by $(z_1, z_2, z_3)$ in the previous subsections with  spherical coordinates $(r,\varphi^i)$.
 We assume isotropy in the 3-sphere directions, smear the D4's over $\rho$ and the D2's over their relative codimension with respect to the D4's, i.e. $H_{\mathrm{D}4}=H_{\mathrm{D}4}(\zeta)$ and $H_{\mathrm{D}2}=H_{\mathrm{D}2}(\zeta)$, with the functions $h(\rho,r)$ and $g(\rho,r)$ describing the D6-NS5-D8 bound state as in \eqref{eq:massiveclassmetricoriginal}\footnote{In \cite{Imamura:2001cr} the functions $g$ and $h$ are respectively called $S$ and $K$.}. We introduce the following gauge potentials and dilaton,
\begin{equation}
\begin{split}\label{brane_potentials_D2D4NS5D6D8}
&C_{3}=H_{\mathrm{D}2}^{-1}\,\text{vol}(\mathbb{R}^{1,1})\wedge d\rho\,,\\
&C_{5}=H_{\mathrm{D}4}^{-1}\,h\,g\,r^2\,\text{vol}(\mathbb{R}^{1,1})\wedge dr \wedge \text{vol}(\text{S}^2)\,,\\
&C_{7}=H_{\mathrm{D}4}\,h^{-1}\,\zeta^3\,\text{vol}(\mathbb{R}^{1,1})\wedge d\zeta \wedge \text{vol}(\text{S}^3) \wedge d\rho \,,\\
&B_{6}=H_{\mathrm{D}4}\,g^{-1}\,\zeta^3\,\text{vol}(\mathbb{R}^{1,1})\wedge  d \zeta \wedge\text{vol}(\text{S}^3)\,,\\ \vspace{0.4cm}
&e^{\Phi}=h^{-3/4}\,g^{1/2}\,H_{\mathrm{D}2}^{1/4}\,H_{\mathrm{D}4}^{-1/4}\,,
\end{split}
\end{equation}
from which the fluxes read
\begin{equation}
\begin{split}\label{brane_fluxes_D2D4NS5D6D8}
& F_{2} = -\partial_r h \,r^2\,\text{vol}(\text{S}^2)\,, \\
& H_{3} =- \partial_r g \, r^2\,d\rho\wedge\text{vol}(\text{S}^2)+H_{\mathrm{D}2}\,H_{\mathrm{D}4}^{-1}\,\partial_\rho \left(h\,g\right)\,r^2\,dr\wedge\text{vol}(\text{S}^2)\,, \\
 &F_{4}=\partial_\zeta H_{\mathrm{D}2}^{-1}\,\text{vol}(\mathbb{R}^{1,1})\wedge d\zeta\wedge d\rho-\partial_\zeta H_{\mathrm{D}4}\,\zeta^3\, \text{vol}(\text{S}^3)\wedge d\rho\,,
 \end{split}
\end{equation}
plus a Romans' mass $F_{0}$.
The equations of motion and Bianchi identities for the D2-D4 branes and the D6-NS5-D8 branes can then be solved independently, such that
\begin{equation}\label{10d-defectEOM}
H_{\mathrm{D}2}=H_{\mathrm{D}4}\qquad\text{and}\qquad \nabla^2_{\zeta}\,H_{\mathrm{D}4}=0\,,
\end{equation}
for the D2-D4 subsystem, and 
\begin{equation}\label{10d-motherbranesEOM_nh}
\partial_\rho h=F_0\,g\qquad \text{and}\qquad \nabla^2_{r}\,h+\frac{1}{2}\partial_\rho^2\,h^2=0\,,
\end{equation}
for the D6-NS5-D8 branes. Here
$\nabla_r^2$ and $\nabla_\zeta^2$ are, respectively, the Laplacians in spherical coordinates on the 3d flat space transverse to the D6-NS5-D8 branes and the 4d space transverse to the D2-D4 branes.
The equations in \eqref{10d-motherbranesEOM_nh} coincide with \eqref{eq:Bianchiidentitiesorg} and then \eqref{imamuraeq1}, \eqref{imamuraeq2}. In turn,
the equations in \eqref{10d-defectEOM} can be easily solved for
\begin{equation}
  H_{\mathrm{D}4}(\zeta)=H_{\mathrm{D}2}(\zeta)=1+\frac{q}{\zeta^2}\,,
 \end{equation}
where $q$ is an integration constant.

Taking the limit $\zeta \rightarrow 0$ the $\zeta$ coordinate becomes the radial coordinate of an AdS$_3$ factor, and the metric \eqref{brane_metric_D2D4NS5D6D8} and the fluxes \eqref{brane_fluxes_D2D4NS5D6D8} take the form\footnote{We redefined the Minkowski coordinates as $(t,x^1)\rightarrow q\,(t,x^1)$.}
\begin{equation}
\label{brane_metric_D2D4NS5D6D8_nh}
\begin{split}
ds_{10}^2 &= q\, h^{-1/2} \left[ds^2(\text{AdS}_3) + ds^2(\text{S}^3) \right] + h^{-1/2} g \, d\rho^2 + h^{1/2} g\left(dr^2 + r^2 ds^2(\text{S}^2)\right) \,, \\
F_{2} &= - \partial_r h \,r^2\,\text{vol}(\text{S}^2) \,,  \qquad \qquad  e^{\Phi} = h^{-3/4} g^{1/2} \,,\\
H_{3} &= -\partial_r g \, r^2\,d\rho\wedge\text{vol}(\text{S}^2)+\partial_\rho \left(h\,g\right)\,r^2\,dr\wedge\text{vol}( \text{S}^2)\,, \\
 F_{4} &= 2q \, \text{vol}(\text{AdS}_3)\wedge d\rho + 2q \, \text{vol}(\text{S}^3) \wedge d\rho \,,\\
\end{split}
\end{equation}
with $(h,g)$ satisfying \eqref{10d-motherbranesEOM_nh}.
Therefore, we recover the AdS$_3\times $S$^3$ backgrounds \eqref{eq:massiveclassmetric} with the 3d transverse space that was parametrised by $(z_1, z_2, z_3)$  now written in spherical coordinates $(r, \varphi^i)$. Our new class of solutions can thus be interpreted as the low-energy regime of D6-NS5-D8 bound states \cite{Imamura:2001cr} wrapping an $\mrm{AdS}_3\times $S$^3$ geometry, with the geometry associated to the bound state uniquely fixed by the functions $h$ and $g$, and the D2-D4  intersection completely resolved into the AdS$_3\times $S$^3$ geometry.

\subsection{An uplift of 6d minimal ${\cal N}=2$ ungauged supergravity}\label{sec:uplift}

The fact that the system of governing PDEs \eqref{eq:Bianchiidentities} support solutions with both a warped Mink$_6$ and an AdS$_3\times$ S$^3$ factor is highly suggestive that it should actually  work for any solution to 6d ${\cal N}=2$ ungauged supergravity with SU(2) R-symmetry (see for instance \cite{Hristov:2014eba}). In this subsection we show that this is indeed the case.  This may be relevant for the construction of more general superstrata solutions, following \cite{Bena:2015bea}.

The pseudo action of the aforementioned 6d theory is
\beq
S_6= \int d^6x\sqrt{-g_6}\bigg( R- \frac{1}{3}H^{(6)}_{abc}H^{(6)abc}\bigg),
\eeq
where $H^{(6)}$ is a closed self dual 3-form, the latter constraint needing to be imposed after varying the action. It is possible to show that this theory can be embedded into massive IIA as 
\begin{align}\label{eq:massiveclassmetric1}
ds^2= \frac{1}{\sqrt{h}}\bigg[c^{-2}ds^2_6+ g d\rho^2\bigg]+ g \sqrt{h}\bigg(dz_1^2+dz_2^2+dz_3^2\bigg),~~~~e^{-\Phi}=\frac{h^{\frac{3}{4}}}{\sqrt{g}},
\end{align}
for fluxes
\begin{align}
F_0&= \frac{\partial_{\rho}h}{g},~~~~F_4=2 c^2 H^{(6)}\wedge d\rho,\\[2mm]
F_2&= -(\partial_{z_1}h dz_2\wedge dz_3+\partial_{z_2}h dz_3\wedge dz_1+\partial_{z_3}h dz_1\wedge dz_2),\\[2mm]
H_3&=\partial_{\rho}(h g)dz_1\wedge dz_2\wedge dz_3-(\partial_{z_1}g dz_2\wedge dz_3+\partial_{z_2}g dz_3\wedge dz_1+\partial_{z_3}g dz_1\wedge dz_2)\wedge d\rho,
\end{align}
where $c$ is an arbitrary constant. We have confirmed that the 10d equations of motion are implied by those following from the 6d action together with \eqref{eq:Bianchiidentities}. Therefore, any solution to the 6d theory gives rise to a solution in massive IIA once \eqref{eq:Bianchiidentities} are imposed. All such supersymmetric solutions were classified some time ago in \cite{Gutowski:2003rg}.


\section{Defects within $\ma N=(1,0)$ 6d CFTs}\label{defectsin6d}

In this section we focus on the particular subclass of solutions featured by a (locally) AdS$_7$ asymptotics, and discuss their dual interpretation as surface defects within the 6d $\ma N=(1,0)$ CFTs dual to the AdS$_7$ solutions of massive Type IIA supergravity constructed in \cite{Apruzzi:2013yva}. 

Our first aim will be to derive the particular set of coordinates for which the AdS$_7$ asymptotics is manifest. This can be done by direct calculation in 10d or by making use of the consistent truncation of massive IIA supergravity to minimal 7d $\ma N=1$ gauged supergravity \cite{Passias:2015gya}. From the latter perspective the 10d solutions take the form of a domain wall with AdS$_3\times$S$^3$ worldvolume with a locally AdS$_7$ vacuum at infinity, that arises upon consistent truncation from the AdS$_7\times $S$^2\times I$ solutions of \cite{Apruzzi:2013yva} (see Appendix \ref{appendix}). 
In 10d one can see from the brane picture studied in subsection \ref{branepicture} that D2-D4 branes break the isometries of the $\mathbb{R}^{1,5}$ worldvolume common to the D6-NS5-D8 intersection, as
$$
\mathbb{R}^{1,5} \quad \longrightarrow \quad \text{AdS}_3 \times \text{S}^3,
$$
leaving intact the conformal symmetries of $\text{AdS}_3$. In the UV the AdS$_7$ vacuum emerges as a foliation of the $\text{AdS}_3 \times \text{S}^3$ subspace over an interval. 

With the insight coming from the supergravity analysis, we will construct 2d $(0,4)$ quiver gauge theories that flow in the IR to the CFTs dual to the AdS$_3$ solutions and show that they can be embedded within the 6d quivers constructed in 
\cite{Gaiotto:2014lca,Cremonesi:2015bld}, dual to the AdS$_7$ solutions in \cite{Apruzzi:2013yva}.

\subsection{The AdS$_7$ vacua of massive IIA and their dual 6d CFTs}


We start by briefly reviewing the main properties of the AdS$_7$ solutions of massive IIA supergravity and of their 6d dual CFTs.

 The solutions in \cite{Apruzzi:2013yva} are described by AdS$_7\times$S$^2$ foliations over an interval preserving 16 supercharges. They arise in the near horizon limit of a D6-NS5-D8 intersection, constructed in \cite{Bobev:2016phc}.
In the parametrisation of \cite{Cremonesi:2015bld} they take the form
\begin{align}
\label{AdS7vacua}
ds_{10}^2 &= \pi\sqrt{2} \bigg[ 8 \Bigl(-\frac{\alpha}{{\alpha}''}\Bigr)^{1/2} ds^2(\text{AdS}_7) + \Bigl(-\frac{{\alpha}''}{\alpha}\Bigr)^{1/2} dy^2 + \Bigl(-\frac{\alpha}{{\alpha}''}\Bigr)^{1/2} \frac{(-\alpha{\alpha}'')}{{\alpha}'^2 - 2\alpha{\alpha}''} ds^2(\text{S}^2) \bigg] \,, \\
\label{dilatonAdS7}
e^{2\Phi} &= 3^8 2^{5/2} \pi^5 \frac{(-\alpha/{\alpha}'')^{3/2}}{{\alpha}'^2 - 2\alpha{\alpha}''}\,, \\
\label{B2AdS7}
B_{2} &= \pi \Bigl(-y + \frac{\alpha{\alpha}'}{{\alpha}'^2 - 2\alpha{\alpha}''}\Bigr) \, \text{vol}(\text{S}^2) \,, \\
\label{F2AdS7}
F_{2}&=\Bigl(\frac{{\alpha}''}{162\pi^2}+\frac{\pi F_0\alpha{\alpha}'}{{\alpha}'^2-2\alpha{\alpha}''}\Bigr)\text{vol}(\text{S}^2).
\end{align}
The solutions are specified by the function $\alpha(y)$, which satisfies the differential equation
\begin{equation}\label{alpha'''1}
\alpha'''=-162\pi^3 F_0.
\end{equation}

Let us now recall the main ingredients of the 6d quivers dual to these solutions. We will follow \cite{Cremonesi:2015bld} and 
\cite{Nunez:2018ags}. Equation \eqref{B2AdS7} (see below) implies that there are (color) NS5-branes located at given positions in the $y$-direction, that can be labelled by an integer number $k$. Piecewise $\alpha(y)$ functions defined in intervals $[k,k+1]$ between NS5-branes can then be constructed, with continuous first and second derivatives, and third derivative satisfying 
\begin{equation}\label{alpha'''2}
\alpha_k'''=-81\pi^2\beta_k.
\end{equation}
We thus have at a given $[k,k+1]$ interval\footnote{We have chosen units where $\alpha^\prime=g_s=1$.}
 \begin{equation}
Q_{NS5}^{(k)}=\frac{1}{4\pi^2}\int H_3=\frac{1}{4\pi^2}\int_{\text{S}^2}\Bigl(B_2(y=k+1)-B_2(y=k)\Bigr)=1\, .
\end{equation}
Moreover, given that $Q_{D8}=2\pi F_0$ equation \eqref{alpha'''2} implies that
\begin{equation}\label{QD8charge}
Q_{D8}^{(k)}=\beta_k.
\end{equation}
$\beta_k$ are therefore integer numbers, and $(\beta_{k-1}-\beta_k)$ are the numbers of D8-branes that are introduced at each $y=k$ position.
Integrating \eqref{alpha'''2} one finds
\begin{equation}
\alpha_k(y)= -\frac{27}{2}\pi^2 \beta_k (y-k)^3+\frac12 \gamma_k (y-k)^2+\delta_k (y-k)+\mu_k,\qquad \text{for} \quad y\in [k,k+1],
\end{equation}
where $(\gamma_k, \delta_k, \mu_k)$ are constants that are determined by imposing continuity of $\alpha, \alpha', \alpha''$. The condition that $\alpha_k''=\alpha_{k-1}''$ at $y=k$ imposes that
\begin{equation}\label{gamma}
\gamma_k=-81\pi^2 \beta_{k-1}+\gamma_{k-1}=-81\pi^2 (\beta_0+\beta_1+\dots +\beta_{k-1}).
\end{equation}
This implies that the D6-brane charge at each interval, given by 
\begin{equation} \label{QD6charge}
Q_{D6}^{(k)}=\frac{1}{2\pi}\int_{\text{S}^2}\hat{F}_2,=-\frac{\gamma_k}{81\pi^2},
\end{equation}
where $\hat{F}_2=F_2-F_0\wedge B_2$ is the Page flux, defining a charge that should be integer.
In turn, $\alpha_k'=\alpha_{k-1}'$ and $\alpha_k=\alpha_{k-1}$ at $y=k$ determine, respectively,
\begin{equation}
\delta_k=-\frac{81}{2}\pi^2\beta_{k-1}+\gamma_{k-1}+\delta_{k-1}, \qquad 
\mu_k=-\frac{27}{2}\pi^2\beta_{k-1}+\frac12 \gamma_{k-1}+\delta_{k-1}+\mu_{k-1}.
\end{equation}
The continuity conditions need to be supplemented by conditions at the boundaries of the $y$-interval. For this to be geometrically well-defined the asymptotic form of the metric needs to approach one of 4 physical behaviours compatible with the metric factors, namely a regular zero or singular D6, O6 or D8/O8 behaviour. Two of these arise generically: One can choose the integration constants such that $\alpha=0$ at a boundary of the space, in which case the behaviour corresponds to fully localised D6-branes, 
or one can impose that $\alpha''=0$, in which case one finds fully localised O6-planes. The other behaviours are possible with specific tunings of $\alpha$ when $F_0\neq 0$: One can tune $\alpha$ such that in the boundary interval $\alpha=- q_2(y) \alpha''$, for $q_n=q_n(y)$ an order $n$ polynomial, then as long as $q_2$ has non degenerate zeros - the zero of $\alpha''$ is regular. Like-wise one can simultaneously impose $\alpha''=0$ and $(\alpha')^2-2 \alpha \alpha''=  q_3 \alpha''$, then the behaviour at the zero of $\alpha''=0$ is that of a localised O8, which may be coincident to additional D8s.

\begin{figure}
\centering
\includegraphics[scale=0.65]{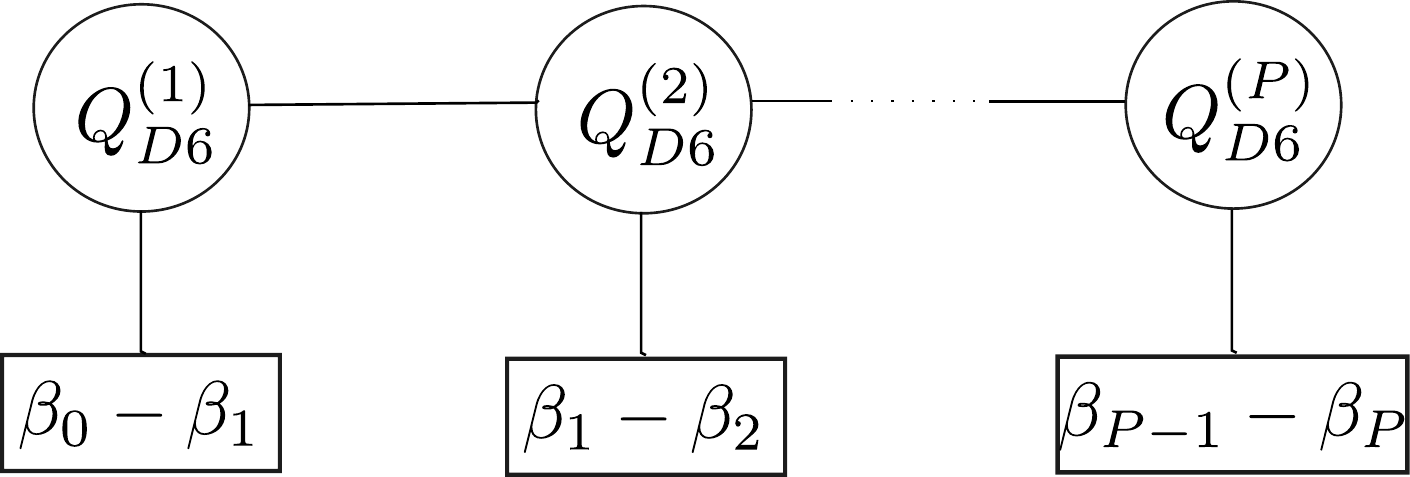}
\caption{Quiver describing the field theory living in D6-NS5-D8 intersections. The circles denote $(1,0)$ vector multiplets and the lines $(1,0)$ bifundamental matter fields. The quiver has been terminated with $(\beta_{P-1}-\beta_P)$ D8-branes at the end of the space, with $\beta_P=\frac{\gamma_P}{81\pi^2}$ and $\gamma_P=-81\pi^2\sum_{l=1}^{P-1}\beta_l$.}
\label{6dquiver}
\end{figure}

The D6-NS5-D8 brane set-up associated to the solutions is the one depicted in Table \ref{D6D8NS5}. 
 \begin{table}[ht]
	\begin{center}
		\begin{tabular}{| l | c | c | c | c| c | c| c | c| c | c |}
			\hline		    
			& $x^0$ & $x^1$  & $x^2$ & $x^3$ & $x^4$ & $x^5$ & $x^6$ & $x^7$ & $x^8$ & $x^9$\\ \hline
			D6 & x & x & x &  &  &  &x  &x   &x   &x   \\ \hline
			D8 & x & x &  &x  & x &  x &x  &x  & x & x  \\ \hline
			NS5 & x & x &  &  &  &   & x  & x  & x & x  \\ \hline
		\end{tabular} 
	\end{center}
	\caption{$\frac14$-BPS brane intersection underlying the 6d $(1,0)$ CFTs living in D6-NS5-D8 brane intersections. The directions $(x^0,x^1,x^6,x^7,x^8,x^9)$ are the directions where the 6d CFT lives. $x^2$ is the field theory direction, along which the D6-branes are stretched.  $(x^3, x^4, x^5)$ are the directions realising the SO(3) R-symmetry.}   
	\label{D6D8NS5}	
\end{table} 
Here the D6-branes play the role of colour branes while the D8-branes play the role of flavour branes \cite{Brunner:1997gk,Hanany:1997sa}. In 6d language the quantised charges give rise to the quiver depicted in Figure \ref{6dquiver}, 
as discussed in \cite{Cremonesi:2015bld,Nunez:2018ags}. One can check that 6d anomaly cancellation is fulfilled given that at each gauge node of the quiver
\begin{equation}\label{anomalycan}
2N_k=2Q_{D6}^{(k)}=N_f^k=Q_{D6}^{(k-1)}+Q_{D6}^{(k+1)}+\Delta Q_{D8}^{(k)},
\end{equation}
with $\Delta Q_{D8}^{(k)}=\beta_{k-1}-\beta_k$.

\subsection{The surface defect ansatz}\label{defectAnsatz}

In this subsection we search for a solution within the class  constructed in section \ref{eq:the massiveclass}  that is asymptotically AdS$_7$. The first step is to decide on the form of the external 7d and internal 3d spaces. We shall assume that the metric takes the form
\begin{align}
\frac{1}{\sqrt{2}\pi}ds^2&= L^2\sqrt{-\frac{\alpha}{\alpha''}}ds^2(\text{M}_{1,6})+ \Delta_1 dy^2+ \Delta_2 ds^2(\text{S}^2),\nn\\[2mm]
ds^2(\text{M}_{1,6})&= P^2\bigg[ds^2(\text{AdS}_3)+\frac{1}{m^2}ds^2(\text{S}^3)\bigg]+Q^2 dx^2,\label{eq:defectansatz}
\end{align}
where $P,Q$ are functions of $x$ only and $\Delta_{1,2}$ are functions of $x$ and $y$.  We find it convenient to fix $q=1$
in this section, which we are free to do without loss of generality.
The first step is to impose SO(3) symmetry in  \eqref{eq:massiveclassmetric}, so that $(z_1,z_2,z_3) ~\to~ (r,\text{S}^2)$.  Then we need to arrange for a change of coordinates $(r,\rho)~ \to~(x,y)$ such that \eqref{eq:defectansatz} emerges. Our experience in the previous sections suggests we take
\beq
r= q_1(x) \alpha,~~~~ \rho= -q_2(x)  \alpha'.
\eeq
By comparing \eqref{eq:massiveclassmetric} to \eqref{eq:defectansatz} we then see we must fix
\beq
h= \frac{1}{2P^4 L^4 \pi^2} \left(-\frac{\alpha''}{\alpha}\right),~~~~g=\frac{4 L^8 \pi^4 P^6 q_2^2 Q^2}{(\dot{q}_1)^2(q_1^2 (\alpha')^2-2 L^4\pi^2 P^4 q_2^2 \alpha \alpha'')}
\eeq
and solve
\beq
q_1\dot{q}_1=2 L^4\pi^2 P^4 q_2 \dot{q}_2.
\eeq
Turning our attention to the Bianchi identities, we find that $F_0=$ constant, under the assumption that $\alpha'''=-162 \pi^3F_0$, imposes that
\beq
4 q_1 \dot{P}=P \dot{q}_1,~~~~ (\dot{q}_1)^2= \frac{2\pi L^8}{3^4} P^6 Q^2 q_2,
\eeq
and implies the remaining Bianchi identities.  Modulo diffeomorphisms the 3 ODEs we have can be solved without loss of generality as
\beq
P= 2^{3/2}x,~~~~ Q=-\frac{2^{3/2}}{(c+x^4)^{\frac{1}{4}}},~~~~ q_1=\frac{64L^6}{3^4}x^4,~~~~ q_2=\frac{8L^{4}}{3^4 \pi} \sqrt{c+ x^4},~~~ dc=0.
\eeq
The NS sector of the solution then takes the form
\begin{align}\label{AdS3sliced}
\frac{ds^2}{8\sqrt{2}\pi L^2}&=\bigg[\sqrt{-\frac{\alpha}{\alpha''}}\bigg(x^2 \big(ds^2(\text{AdS}_3)+ ds^2(\text{S}^3)\big)+ \frac{dx^2}{\sqrt{c+ x^4}}\bigg)+\frac{\sqrt{c+x^4}}{x^2}\sqrt{\frac{-\alpha''}{\alpha}}\bigg(dy^2+\frac{\alpha^2 x^4}{\Delta}ds^2(\text{S}^2)\bigg)\bigg],\nn\\[2mm]
e^{-\Phi}&=\frac{L \sqrt{\Delta}}{3^4 2^{\frac{5}{4}}\pi^{\frac{5}{2}}  x (c+x^4)^{\frac{1}{4}}}\left(-\frac{\alpha''}{\alpha}\right)^{\frac{3}{4}},~~~~B_2=-L^2\pi\left(-y+ \frac{x^4 \alpha \alpha'}{\Delta}\right)\text{vol}(\text{S}^2),
\end{align}
where we have defined
\beq
\Delta=x^4\left((\alpha')^2-2 \alpha \alpha''\right)-2 c \alpha \alpha'',
\eeq
while the RR fluxes are
\begin{align}\label{AdS3slicedRR}
F_0&=-\frac{1}{162\pi^3}\alpha''',~~~~F_2= F_0 B_2- \frac{L^2}{162\pi^2}(162F_0\pi^3y+\alpha'') \text{vol}(\text{S}^2),\nn\\[2mm]
F_4&= -\frac{2^4L^4}{3^4 \pi }d(\sqrt{c+x^4}\alpha')\wedge \bigg( \text{vol}(\text{AdS}_3)+\text{vol}(\text{S}^3)\bigg),\nn\\[2mm]
F_6&=F_4\wedge B_2-\frac{2^4 L^6}{3^4}d(\sqrt{c+x^4}(\alpha-y \alpha'))\wedge \bigg( \text{vol}(\text{AdS}_3)+\text{vol}(\text{S}^3)\bigg)\wedge \text{vol}(\text{S}^2)
\end{align}
Notice that as $x \to \infty$,  $x^{-4}\Delta \to 1$ and the entire NS sector tends to that of the AdS$_7$ solutions in massive IIA reviewed in the previous subsection, where the AdS$_7$ radius is 1. The same is true for the RR 0 and 2 form fluxes, however the 4 form does not tend to zero in this limit, which reflects the presence of a D2-D4 defect. That the directions $(\text{AdS}_3,\text{S}^3,x)$, tend to AdS$_7$ can be confirmed by computing the Riemann curvature tensor. The solution is bounded from below in a way that depends on the tuning of $c$: When $c\geq 0$ $x$ is bounded to the interval $[0,\infty)$, when $c=0$  there is a curvature singularity at the lower bound that we do not recognise as physical but for  $c \neq 0$, defining $x=z^{\frac{1}{4}}$, the metric at this  locus tends to 
\beq
\frac{ds^2}{8\sqrt{2}\pi L^2}= \sqrt{-\frac{\alpha}{\alpha''}}\bigg[\sqrt{z}\bigg(ds^2(\text{AdS}_3)+ds^2(\text{S}^3)\bigg)+ \frac{1}{16  \sqrt{c}z^{\frac{3}{2}}}(dz^2+z^2 ds^2(\text{S}^2))\bigg]+\frac{\sqrt{c}}{8 \sqrt{z}}\sqrt{-\frac{\alpha''}{\alpha}}dy^2\nn.
\eeq
If we had $-\frac{\alpha}{\alpha''}=1$, this would be the behaviour one expects of a stack of localised D6 branes on $(\text{AdS}_3,\text{S}^3,y)$, with  NS5 branes inside them of worldvolume  $(\text{AdS}_3,\text{S}^3)$ smeared along $y$. Since  $-\frac{\alpha}{\alpha''}\neq 1$ generically what we actually have is a slight generalisation of this: Rather than the NS5 branes being evenly smeared along $y$ such that the direction is an isometry, they form a $y$ dependent distribution. Finally if $c<0$  we can fix  $c=-b^4$ and the metric is bounded below at $x=b$ where one sees the behaviour of ONS5 fixed planes\footnote{The S-dual of O5-planes.} that are  smeared along $y$. The most attractive of these 3 behaviours is the second\footnote{See our discussion on smeared ONS5s below \eqref{eq:refpoint}.}, so from here we shall assume $c>0$ so that $x\in [0,\infty)$.

In the next subsection we construct the 2d quivers dual to the solutions defined by \eqref{AdS3sliced}-\eqref{AdS3slicedRR}, and show that they can be embedded in the 6d quivers discussed in the previous subsection, dual to the AdS$_7$ solutions. Before we do that we state the value of the holographic central charge computed using the Brown-Henneaux formula \cite{Brown:1986nw} for later comparison with the field theory result,
\begin{equation}
\label{holographic-c-defect}
c_{hol}=\frac{2^6}{3^7\pi^4}\int dx dy\, x^3\, (-\alpha \alpha'').
\end{equation}

\subsection{Surface defect CFTs}\label{surfaceCFT}

In this subsection we construct the 2d quivers that flow in the IR to the CFTs dual to the solutions defined by \eqref{AdS3sliced}-\eqref{AdS3slicedRR}. We show that in a certain limit these quivers can be embedded in the 6d quivers constructed from the D6-NS5-D8 sector of the brane intersection.

We start analysing the brane charges associated to the D2-D4-D6-NS5-D8 brane set-up underlying the solutions. One can see from the expressions for $F_0$ and $F_2$ in \eqref{AdS3slicedRR} that the D8 and D6 quantised charges of the AdS$_3$ solutions coincide with those of the AdS$_7$ backgrounds, given by equations \eqref{QD8charge} and \eqref{QD6charge}. In turn, for finite $x$ there are  NS5-branes located at fixed values in $y$ and also in $x$. Since we are interested in embedding the 2d CFT in the 6d CFT associated to the D6-NS5-D8 subsystem, we will take $x$ large enough such that we can neglect the $(H_3)_{x\text{S}^2}$ component of the NS-NS 3-form flux and take the NS5-branes located at fixed positions in $y$, as in the D6-NS5-D8 subsystem. The fluxes associated to the $\text{AdS}_3$ solutions are then compatible with the brane intersection depicted in Table \ref{D2D4D6D8NS5_1}, that we repeat in Table \ref{D2D4D6D8NS5} below in a generic system of coordinates for a better reading.
 \begin{table}[ht]
	\begin{center}
		\begin{tabular}{| l | c | c | c | c| c | c| c | c| c | c |}
			\hline		    
			& $x^0$ & $x^1$  & $x^2$ & $x^3$ & $x^4$ & $x^5$ & $x^6$ & $x^7$ & $x^8$ & $x^9$\\ \hline
			D2 & x & x & x & & & & & & & \\ \hline
			D4 & x & x & & x & x & x & & & & \\ \hline
			D6 & x & x & x &  &  &  &x  &x   &x   &x   \\ \hline
			D8 & x & x &  &x  & x &  x &x  &x  & x & x  \\ \hline
			NS5 & x & x &  &  &  &   & x  & x  & x & x  \\ \hline
		\end{tabular} 
	\end{center}
	\caption{$\frac18$-BPS brane intersection underlying the AdS$_3$ solutions \eqref{AdS3sliced}-\eqref{AdS3slicedRR}. $(x^0,x^1)$ are the directions where the 2d dual CFT lives. $x^2$ is the field theory direction, that we identify with $y$, where the NS5-branes are located (for $x$ sufficiently large). The D2 and D6-branes are stretched in this direction.  $(x^3, x^4, x^5)$ are the directions associated to the isometries of the S$^2$ while $(x^6,x^7,x^8,x^9)$ are those associated to the S$^3$.}
	\label{D2D4D6D8NS5}	
\end{table} 
Note that the R-symmetry of the 2d field theory living in the brane set-up is the SO(3)$_R\subset $ SO(4) symmetry group of the S$^3$, while for the 6d field theory living in the D6-D8-NS5 brane intersection it is identified with the SO(3) symmetry group of the S$^2$. This is exactly what happens for 2d $(4,4)$ field theories arising upon compactification from 6d $(1,0)$ CFTs, where the SO(3) R-symmetry of the 6d theory becomes the R-symmetry of the Coulomb branch of the 2d theory, and the SO(3)$_L\times$SO(3)$_R$ R-symmetry of the Higgs branch of the 2d theory arises in the dimensional reduction \cite{Diaconescu:1997gu,Witten:1997yu,Aharony:1999dw}. In our $(0,4)$ theories there is just a Higgs branch, since the Coulomb branch contains no scalars, and the R-symmetry is just the  SO(3)$_R$ arising in the dimensional reduction. 

\begin{figure}
\centering
\includegraphics[scale=0.75]{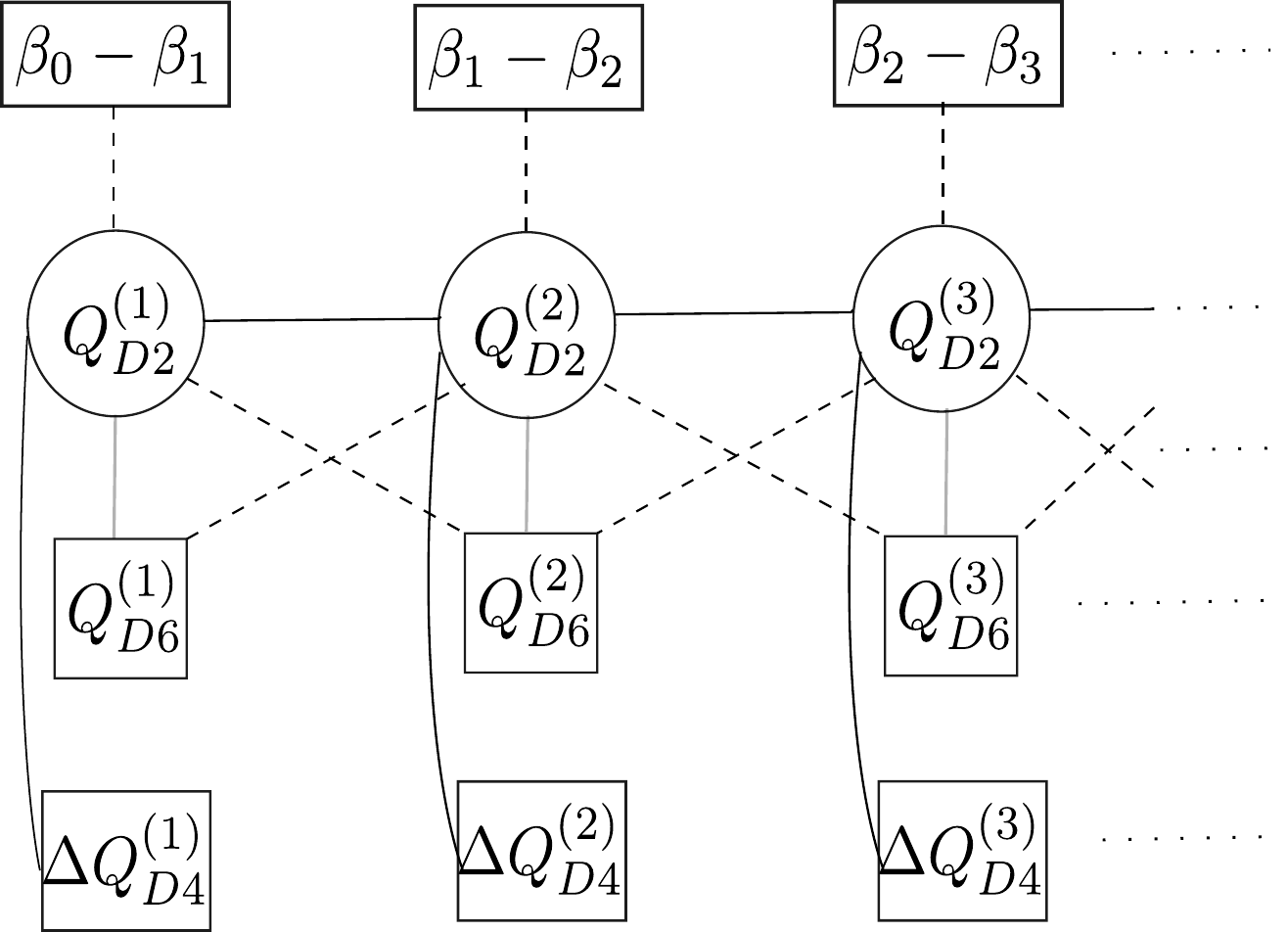}
\caption{2d quivers associated to the AdS$_3$ solutions \eqref{AdS3sliced}-\eqref{AdS3slicedRR}. Circles denote $(4,4)$ vector multiplets, black lines $(4,4)$ bifundamental hypermultiplets, grey lines $(0,4)$ bifundamental twisted hypermultiplets and dashed lines $(0,2)$ bifundamental Fermi multiplets.}
\label{2dquiverdefect}
\end{figure}

The Hanany-Witten brane set-up associated to the brane intersection in Table \ref{D2D4D6D8NS5} is depicted in Figure \ref{brane-setup-defect}.  
In this set-up the D2-branes play the role of colour branes. They are stretched in the $y$-direction, which is divided into intervals of length $1$ in our units, where NS5-branes are located. The D6-branes are also stretched in this direction, however they lie as well along the $x$ direction, which is non-compact, therefore they become flavour branes. The D4-branes lie as well along the $x$ direction, so they are also flavour branes, and so are the D8-branes. In order to construct the quiver that lives in this set-up one needs to look at the quantisation of the open strings stretched between the different branes. This has been studied in detail in various references (see for instance \cite{Couzens:2021veb}), to which we refer the reader for more details. 
There are three types of massless modes to consider:
\begin{figure}
\centering
\includegraphics[scale=0.75]{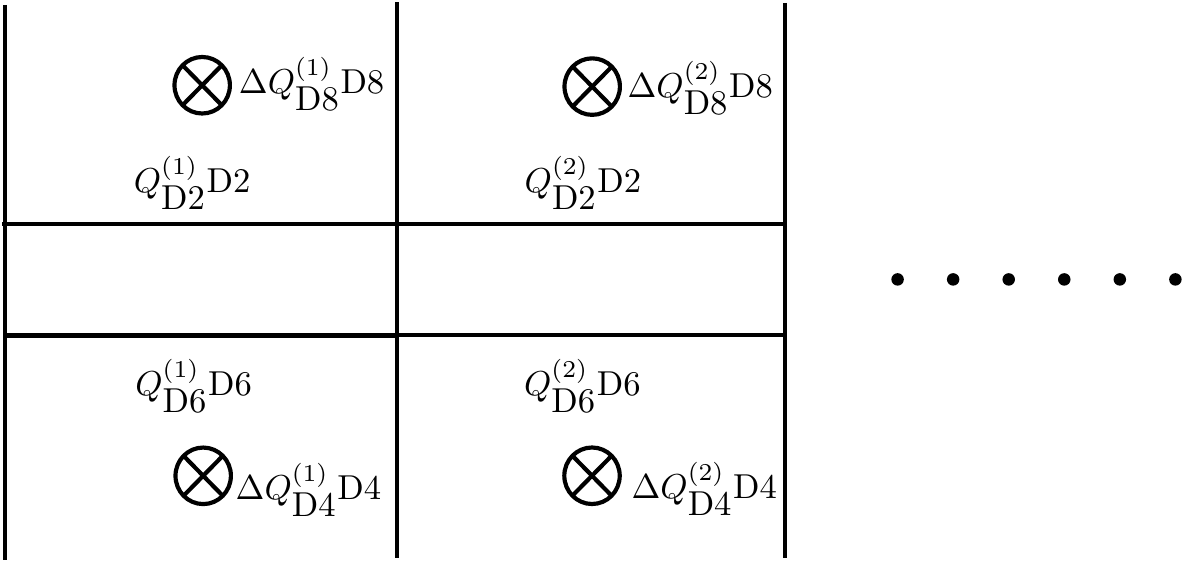}
\caption{Hanany-Witten brane set-up associated to the AdS$_3$ solutions \eqref{AdS3sliced}-\eqref{AdS3slicedRR}.}
\label{brane-setup-defect}
\end{figure} 
\begin{itemize}
\item D2-D2 strings: There are two cases to consider, depending on whether the two end-points of the string lie on the same stack of D2-branes or on two different stacks, separated by an NS5-brane. Let us consider first the case in which the two end-points lie on the same stack. For D2-branes stretched between NS5-branes there is a $\mathcal{N}=(0,4)$ vector multiplet and a $\mathcal{N}=(0,4)$ adjoint twisted hypermultiplet, coming from the motion of the D2-branes along the $(x^6,x^7,x^8,x^9)$ directions. Since these scalars are charged under the R-symmetry of the solution they combine into a twisted hypermultiplet. The $\mathcal{N}=(0,4)$ vector multiplet and the $\mathcal{N}=(0,4)$ adjoint twisted hypermultiplet then build up a $\mathcal{N}=(4,4)$ vector multiplet. 


Let us consider now the case in which the end-points of the string lie on two different stacks of D2-branes, separated by an NS5-brane. The massless modes arise from the intersection of the two stacks of D2-branes and the NS5-brane. This fixes the degrees of freedom moving along the $(x^6,x^7,x^8,x^9)$ directions, leaving behind the scalars associated to the $(x^3,x^4,x^5)$ directions, together with the $A_2$ component of the gauge field. These give rise to a $\mathcal{N}=(4,4)$ hypermultiplet in the bifundamental representation, since the scalars are uncharged under the R-symmetry of the solution. 

\item D2-D4 strings: Strings with one end on D2-branes  and the other end on orthogonal D4-branes in the same interval between NS5-branes contribute with fundamental $(4,4)$ hypermultiplets, associated to the motion of the strings along the $(x^3,x^4,x^5)$ directions plus the $A_2$ component of the gauge field.

\item D2-D6 strings: Strings with one end on D2-branes and the other end on D6-branes in the same interval between NS5-branes contribute with fundamental $(0,4)$ twisted hypermultiplets, associated to the motion of the string along the $(x^6,x^7,x^8,x^9)$ directions, which are charged under the R-symmetry of the solution. Strings with one end on D2-branes and the other end on D6-branes in adjacent intervals between NS5-branes contribute with fundamental $(0,2)$ Fermi multiplets. 

\item D2-D8 strings: Strings with one end on D2-branes and the other end on orthogonal D8-branes in the same interval
contribute with fundamental $(0,2)$ Fermi multiplets.

\end{itemize}
The previous fields give rise to the quivers depicted in Figure \ref{2dquiverdefect}. 
In these quivers the D6 and D8-brane charges are the ones given by equations \eqref{QD6charge} and \eqref{QD8charge}, while the D2 and D4 brane charges at each interval are given by
\begin{equation}
Q_{D2}^{(k)}=\frac{1}{(2\pi)^5}\int_{I_x,\text{S}^2,\text{S}^3}\hat{F}_6=\frac{4}{3^4\pi^2}\int_{I_x} dx\, \frac{2x^3}{\sqrt{c+x^4}}\,\alpha_k 
\end{equation}
and
\begin{equation}
\Delta Q_{D4}^{(k)}=\frac{1}{(2\pi)^3}\int_{I_y,\text{S}^3}\hat{F}_4=\frac{4}{3^4\pi^2}\sqrt{c+x^4}\int_{k}^{k+1}dy\, \alpha_k''.
\end{equation}
As the $x$-direction is semi-infinite the D2-brane charges diverge, as expected from their defect interpretation.  Note that the cancellation of gauge anomalies for the gauge groups associated to them is still given by 
\begin{equation}\label{anomalies}
2 Q_{D6}^{(k)}=Q_{D6}^{(k-1)}+Q_{D6}^{(k+1)}+\Delta Q_{D8}^{(k)},
\end{equation}
as for the 6d quivers depicted in Figure \ref{6dquiver}. Here we have taken into account that $(0,4)$ fundamental multiplets contribute 1 to the gauge anomaly, $(0,2)$ fundamental Fermi multiplets contribute -1/2 and the remaining vector and matter fields do not contribute since they are $(4,4)$ (the reader is referred to \cite{Lozano:2019zvg,Couzens:2021veb} for more details).

Next we turn to the computation of the central charge. We show that, as expected, this quantity diverges, as $x$ is not bounded from above.

\vspace{0.5cm}

\begin{figure}
\centering
\includegraphics[scale=0.75]{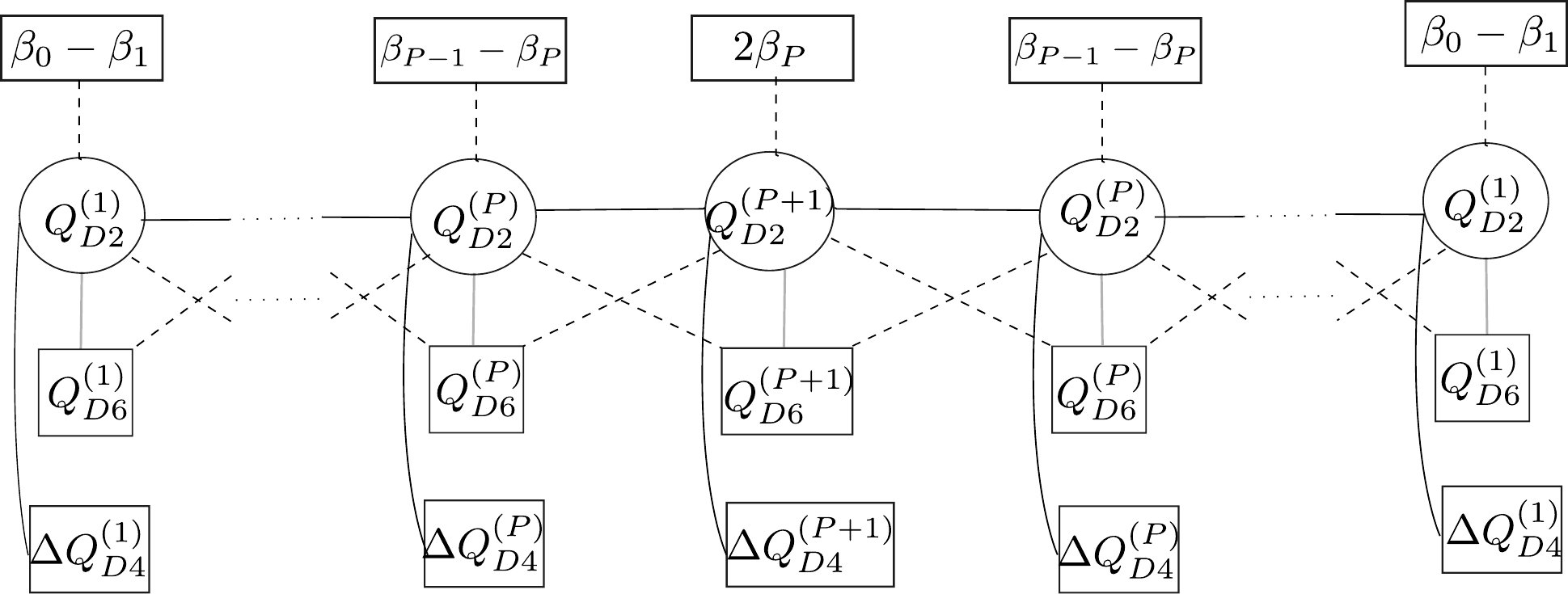}
\caption{2d quivers completed in a symmetric way.}
\label{2dquiverdefectsym}
\end{figure}

\noindent {\bf Central charge:}\\
~~\\
The central charge of a 2d $(0,4)$ CFT can  be computed away from criticality, since it equals the anomaly in the two-point function of the R-symmetry current. In our normalisation this expression is given by \cite{Witten:1997yu}
\begin{equation}
\label{centralcharge(0,4)}
c_R=3{\rm Tr}[\gamma^3 Q_R^2],
\end{equation}
with $Q_R$ the R-charge under the U$(1)_R\subset$ SU$(2)_R$, $\gamma^3$ is the chirality matrix in 2d, and the trace is taken over all fermions in the theory. In order to compute the R-symmetry anomaly we recall the following well-known facts:
\begin{itemize}
\item $(0,4)$ vector multiplets contain two left-moving fermions with R-charge 1. 
\item $(0,4)$ twisted hypermultiplets contain two right-moving fermions with R-charge 0. 
\item $(0,4)$ hypermultiplets contain two right-moving fermions with R-charge -1.  
\item $(0,2)$ Fermi multiplets contain one left-moving fermion with R-charge 0. 
\item $(4,4)$ vector multiplets consist on a $(0,4)$ vector multiplet and a $(0,4)$ adjoint twisted hypermultiplet. Therefore they contribute with 2 to the R-symmetry anomaly.
\item $(4,4)$ hypermultiplets consist on a $(0,4)$ hypermultiplet plus a $(0,4)$ Fermi multiplet. Therefore they contribute with 2 to the R-symmetry anomaly. 
\end{itemize}
This gives the well-known expression for the central charge \cite{Witten:1997yu}
\begin{equation} \label{field-theory-c}
c_R=6(n_{hyp}-n_{vec}),
\end{equation}
where $n_{hyp}$ stands for the number of $(0,4)$ (untwisted) hypermultiplets and $n_{vec}$ for the number of $(0,4)$ vector multiplets. In order to compute these numbers we first need to choose the precise way in which we would like to close the $y$ interval. Our choice is to take $\alpha=\alpha'=\alpha''$ to vanish at both ends of the interval, and to glue the quiver to itself at a given value $y=P+1$, in a continuous way. The resulting quivers are the ones depicted in Figure \ref{2dquiverdefectsym}, where the notation is the same used in Figure \ref{2dquiverdefect}.
This is of course just a possible way to globally define the $y$-direction, and one could consider many others. For the quivers depicted in Figure \ref{2dquiverdefectsym} we have
\begin{equation}
n_{hyp}=2 \sum_{k=1}^{P} Q_{D2}^{(k)}Q_{D4}^{(k)} + Q_{D2}^{(P+1)}Q_{D4}^{(P+1)}+2\sum_{k=1}^{P} Q_{D2}^{(k)}Q_{D2}^{(k+1)},
\end{equation}
and
\begin{equation}
n_{vec}=2\sum_{k=1}^P (Q_{D2}^{(k)})^2+(Q_{D2}^{(P+1)})^2,
\end{equation}
which lead to
\begin{equation}\label{cRcomplete}
c_R=6 \Bigl[\Bigl(2\sum_{k=1}^{P} Q_{D2}^{(k)}Q_{D4}^{(k)}+Q_{D2}^{(P+1)}Q_{D4}^{(P+1)}\Bigr)+ \Bigl(2\sum_{k=1}^{P} Q_{D2}^{(k)}(Q_{D2}^{(k+1)}-Q_{D2}^{(k)})-(Q_{D2}^{(P+1)})^2\Bigr)\Bigr].
\end{equation}
Given that the D2-brane charge is infinite we need a prescription to regularise it. We will evaluate all charges at a given value of $x$ and finally sum over all of them. Doing this one can check that the contribution of the second big bracket to \eqref{cRcomplete} is subleading in $x$ compared to that of the first big bracket. We then find an expression that diverges in $x$ in exactly the same way as the holographic central charge computed in \eqref{holographic-c-defect}, and agrees with it to leading order in $P$ (that is, for long quivers). Explicitly, the leading order in $P$ of \eqref{cRcomplete} gives
\begin{equation}\label{cRcompare}
c_R=\frac{2^7}{3^7 \pi^4}\int_{I_x} dx\, x^3 \sum_{k=1}^P \mu_k\gamma_k.
\end{equation}
In order to show the matching with the holographic central charge we should recall that the holographic central charge is to be identified with \cite{Kraus:2005zm}
\begin{equation}
c_{hol}=\frac{c_L+c_R}{2}.
\end{equation}
Therefore we need to compute first $c_L$. In order to do this we can use that
\begin{equation}
c_L-c_R={\rm Tr}\gamma^3,
\end{equation}
which leads to  \cite{Couzens:2021veb}
\begin{equation}\label{cL}
c_L-c_R=2 n_H^{(0,4)}-n_F^{(0,2)},
\end{equation}
where $n_H^{(0,4)}$ refers to the number of isolated $(0,4)$ hypermultiplets and $n_F^{(0,2)}$ to the number of isolated $(0,2)$ multiplets. It can be checked that $c_L=c_R$ identically for our quivers due to the condition of anomaly cancellation. Therefore $c_{hol}=c_R$ and both quantities can readily be compared. Indeed, we 
find, to leading order in $P$,
\begin{equation}
c_{hol}=\frac{2^6}{3^7 \pi^4}\int dx\, x^3 \Bigl[2\sum_{k=0}^P \int_k^{k+1}dy (-\alpha\alpha'')\Bigr]=
\frac{2^7}{3^7 \pi^4}\int_{I_x} dx\, x^3 \sum_{k=1}^P \mu_k\gamma_k+\dots
\end{equation}
which exactly agrees with \eqref{cRcompare}, to leading order.

As expected, these quantities diverge in $x$ due to its non-compact character. This shows that the 2d quiver CFTs associated to the \eqref{AdS3sliced}-\eqref{AdS3slicedRR} solutions  are ill-defined per se, and only find a meaning in the UV when the deconstructed extra dimensions where the 6d CFTs live emerge. Still, our analysis in this section shows that, for $x$ suitably large, we can nicely embed the D2 and D4 defect branes within the 6d quiver theories associated to the D6-NS5-D8 {\it mother} branes to produce non-anomalous, albeit infinitely charged, 2d quivers.

Note that the quivers discussed in this subsection differ from the quivers constructed in \cite{Faedo:2020nol} for D2-D4-NS5-D6 intersections. The main difference is that in that reference it was wrongly stated that the D2-D6 branes were accounting for bifundamental hypermultiplets and the D2-D4 for bifundamental twisted hypermultiplets, while the careful quantisation of open strings carried out in this section shows that these hypermultiplets are in fact interchanged. This explains why in reference \cite{Faedo:2020nol} it was not possible to match the behaviour in $x$ of the field theory and holographic central charges. 


\section{$\mathcal{N}=(4,4)$ AdS$_3$ from D2-D4-NS5 branes}\label{D2D4NS5}

In this section we consider the particular limiting case in which the coordinates $(z_1,z_2,z_3)$ of the solutions given by \eqref{eq:massiveclassmetric} span a 3-torus $\mathbb{T}^3$ that the warp factors are independent of. We show that the brane intersection reduces to $(4,4)$ theories close to the D2-D4-NS5 Hanany-Witten brane set-ups discussed long ago in \cite{Brodie:1997wn,Alishahiha:1997cm}. These brane set-ups are the two dimensional realisations of the D3-NS5-D5 brane intersections constructed by Hanany and Witten \cite{Hanany:1996ie} and later extended to other dimensions. These D$p$-NS5-D$(p+2)$ brane intersections realise $p$ dimensional field theories with 8 supercharges that flow to CFTs in the IR (for $p<4$), in the UV (for $p>4$), or are conformal per se (for $p=4$). AdS$_{p+1}$ geometries with 16 supercharges dual to these CFTs have been constructed in the literature for $p=6,5,4,3$ (see \cite{Apruzzi:2013yva,Apruzzi:2015wna,Gaiotto:2014lca,Cremonesi:2015bld,Brandhuber:1999np,Bergman:2012kr,DHoker:2016ujz,Lozano:2018pcp,Gaiotto:2009gz,ReidEdwards:2010qs,Aharony:2012tz,Assel:2011xz})\footnote{Also partially for $p=1$ (see \cite{Dibitetto:2019nyz}).}, but the 
$p=2$ case remained an open problem\footnote{In \cite{Chen:2006ps} a probe brane analysis revealed an
AdS$_3\times $S$^3$ geometry as gravity dual of a $(4,4)$ CFT.}.

n this section we make progress toward filling this gap, and construct explicit  AdS$_3\times $S$^3$ duals to $(4,4)$ D2-D4-NS5 brane set-ups, albeit with additional O-planes. In subsection \ref{solutions} we state the main properties of the solutions, consisting of AdS$_3\times $S$^3\times \mathbb{T}^3$ geometries foliated over an interval. In subsection \ref{fieldtheory} we construct the 2d quivers that describe the field theory living in the brane set-ups, and show the agreement between their central charge and the one computed from the supergravity solutions. In subsection \ref{mirror} we discuss the M-theory realisation of these solutions. This allows us to relate them to the AdS$_3\times $S$^2\times \mathbb{T}^4\times I$ solutions of massless Type IIA supergravity constructed in \cite{Lozano:2019emq}. The common M-theory origin of both classes of solutions implies that they flow to the same 2d dual CFT in the IR, that we interpret as a manifestation of mirror symmetry, as discussed in \cite{Brodie:1997wn,Alishahiha:1997cm}. Finally 
in subsection \ref{typeIIB} we construct new $\mathcal{N}=(0,4)$ solutions of Type IIB supergravity related by T-dualities to the previous ones. One such class is holographically dual to D3-brane boxes constructions  \cite{Hanany:2018hlz} with small $\mathcal{N}=(0,4)$ supersymmetry.

\subsection{AdS$_3\times $S$^3\times \mathbb{T}^3$ solutions with $(4,4)$ supersymmetries}\label{solutions}

Imposing the condition that the coordinates $(z_1,z_2,z_3)$ of the solutions given by \eqref{eq:massiveclassmetric} span a 3-torus $\mathbb{T}^3$ that the warp factors are independent of one finds the subclass of solutions 
\begin{align}\label{eq:massiveclassmetricYolanda}
ds^2&= \frac{q}{\sqrt{h}}\bigg[ds^2(\text{AdS}_3)+ds^2(\text{S}^3)\bigg]+ \frac{g}{\sqrt{h}}\, d\rho^2+ g \sqrt{h}\,ds^2(\mathbb{T}^3),\\[2mm]
F_0&= \frac{\partial_{\rho}h}{g}\,,~~~~e^{-\Phi}=\frac{h^{\frac{3}{4}}}{\sqrt{g}},\\
F_4&= 2 q \, \text{vol}(\text{AdS}_3)\wedge d\rho+2 q \, \text{vol}(\text{S}^3)\wedge d\rho,\label{otroF4}\\[2mm]
H_3&=\partial_{\rho}(h g)\, \text{vol}(\mathbb{T}^3), \label{otroH3}\\[2mm]
F_6&= 2q\, g h\,  \text{vol}(\mathbb{T}^3)\wedge (\text{vol}(\text{S}^3) + \text{vol}(\text{AdS}_3)), \label{F6Yolanda}
\end{align}
where $g,h$ are functions of $\rho$ satisfying the Bianchi identities
\begin{equation}\label{bianchisT3}
\partial_\rho(\frac{\partial_\rho h}{g})=0, \quad \partial_\rho^2(gh)=0,\quad F_0\partial_\rho(gh)=0.
\end{equation}
 The smearing of the functions $g$ and $h$ over the $\mathbb{T}^3$ imply that the underlying brane intersection simplifies. In this section we will focus on the massless limit $F_0=0$, to later analyse the non-vanishing Romans' mass case in section \ref{TypeI'}. When $F_0=0$ we have
\beq
h= h_0=\text{constant},
\eeq
and the Bianchi identities imply that
\beq
g''=0.
\eeq
These assumptions imply the exclusion of D8 and D6 branes from the set-up of Table \ref{D2D4D6D8NS5_1}. Moreover, there is a supersymmetry enhancement to $\ma N=(4,4)$, as discussed in subsection \ref{supersymmetry}. We thus obtain a class of $\ma N=(4,4)$ AdS$_3 \times \text{S}^3 \times \mathbb{T}^3$ backgrounds fibered over an interval whose underlying brane intersection is the one depicted in Table \ref{D2-D4-NS5}.
 \begin{table}[ht]
	\begin{center}
		\begin{tabular}{| l | c | c | c | c| c | c| c | c| c | c |}
			\hline		    
			&$x^0$ & $x^1$  & $z_1$ & $z_2$ & $z_3$ & $\rho$ & $\zeta$ & $\theta^1$ & $\theta^2$ & $\theta^3$ \\ \hline
			D2 & x & x & &  &  & x &  &   &   &   \\ \hline
			D4 & x & x & x &x  & x &   &  &  &  &   \\ \hline
			NS5 & x & x &  &  &  &   & x  & x  & x & x  \\ \hline
		\end{tabular} 
	\end{center}
	\caption{$\frac14$-BPS brane intersection underlying the geometry \eqref{eq:massiveclassmetricYolanda} with $F_0=0$. $(x^0,x^1)$ are the directions where the 2d dual CFT lives, $(z_1, z_2, z_3)$ span the $\mathbb{T}^3$ on which the D4-branes are wrapped, $\zeta$ and $\theta^i$ parameterise respectively the radial coordinate of AdS$_3$ and the S$^3$, and  $\rho$ is the field theory direction.
	}
	\label{D2-D4-NS5}	
\end{table} 
The quantised charges of the D2-D4-NS5 branes are computed from the $F_4$, $H_3$ and $F_6$ magnetic fluxes, given by \eqref{otroF4}-\eqref{F6Yolanda}. In order to define the Page fluxes one notes however that it is not possible to define a $B_2$ globally, and that the flux that gives rise to quantised D2-brane charges is rather
\beq \label{newf6}
\hat{f}_6=f_6-C_3\wedge H_3 =2 \,q\, h_0 (g-\rho \,g')\,\text{vol}(\mathbb{T}^3)\wedge \text{vol}(\text{S}^3), 
\eeq
where $\hat{f}_p$ stands for the magnetic component of $F_p$. We will use this definition of the 6-form RR magnetic flux to compute the charge associated to the D2 branes. We will take $h_0=1$ without loss of generality\footnote{This constant can be absorbed through a rescaling of $\rho$ and the radius of AdS$_3$.}.
The definition given by \eqref{newf6} implies that the Page charge associated to D2-branes is sensitive to gauge transformations of the $C_3$ RR potential. In order to carefully account for these we will take as representative of  $C_3$ the one satisfying\footnote{We choose units with $\alpha'=g_s=1$.}
\begin{equation}
\frac{1}{(2\pi)^3}\int_{\text{S}^3}C_3 \in [0,1].
\end{equation}
This is inspired by the more familiar condition that the NS-NS 2-form potential lie in the fundamental region. In order to accomplish this we need to take
\begin{equation}
C_3=-2\,q\,\Bigl(\rho-\frac{2\pi}{q}k\Bigr)\,\text{vol}(\text{S}^3),
\end{equation}
for $\rho\in [\frac{2\pi}{q} k, \frac{2\pi}{q} (k+1)]$.
Given that the D4-brane charge is obtained via computing
\begin{equation}
Q_{D4}^{(k)}=\frac{1}{(2\pi)^3}\int_{I_{\rho},\text{S}^3} \hat{F}_4,
\end{equation}
this gives $Q_{D4}^{(k)}=1$
for $I=[\frac{2\pi}{q}\, k,\, \frac{2\pi}{q}\, (k+1)]$. Therefore there is a single D4 brane in each such interval. This clarifies the role played by the large gauge transformations performed between intervals: a D4-brane is localised on the boundaries of the intervals, generating a strong coupling realisation of the Hanany-Witten brane creation effect\footnote{In the usual Hanany-Witten effect NS5-branes are created. Uplifting this phenomenon to M-theory and reducing along a worldvolume direction of the M5-branes one finds the same effect happening for D4-branes.}.  Taking the whole interval spanned by $\rho$ to be $[0,\frac{2\pi}{q}(P+1)]$, with $P$ as defined below, we then find a total number of $(P+1)$ D4-branes.

We proceed by solving the Bianchi identity $g''=0$. The function $g$ must be continuous, but it can have discontinuities in its first derivative at the locations of the D4-branes, at $\rho=\frac{2\pi}{q} k$. The most general solution is then
\begin{equation}
g_k=\alpha_k+\frac{\beta_k}{2\pi}\Bigl(\rho-\frac{2\pi}{q}k\Bigr), \qquad {\rm for} \qquad  \rho\in [\frac{2\pi}{q}\,k,\, \frac{2\pi}{q}\, (k+1)].
\end{equation}
Imposing that the space begins and ends at $\rho=0, \frac{2\pi}{q} (P+1)$, where $g$ vanishes, we find
  \begin{equation} \label{profileg}
g(\rho) = \left\{ \begin{array}{ccrcl}
                       \frac{\beta_0 }{2\pi}
                       \rho\, , & 0\leq \rho\leq \frac{2\pi}{q} \\
                                      \alpha_k + \frac{\beta_k}{2\pi}(\rho-\frac{2\pi}{q} k )\, , &~~ \frac{2\pi}{q} k\leq \rho \leq \frac{2\pi}{q}(k+1),\;\;\;\; k=1,....,P-1\\
                      \alpha_P+  \frac{\beta_P}{2\pi}(\rho-\frac{2\pi}{q} P)\, , & \frac{2\pi}{q} P\leq \rho \leq \frac{2\pi}{q}(P+1).
                                             \end{array}
\right.
\end{equation}
The condition $g\,(\frac{2\pi}{q}(P+1))=0$ implies $\beta_P=-q\,\alpha_P$, while continuity across the different intervals implies the conditions
\begin{equation}\label{eq:refpoint}
\alpha_k=\frac{1}{q}\sum_{j=0}^{k-1}\beta_j, \qquad k=1,\dots, P.
\end{equation} 
The behaviour close to the zeros of $g$, which bound the solution, is that of an ONS5 plane (the S-dual of an O5 plane) that is smeared over the $\mathbb{T}^3$. Of course for an O-plane in string theory such a smearing is not really physically allowed as the plane should lie  at the fixed point of the orientifold involution. Our solutions here are in supergravity, but as we approach the ONS5 the curvature becomes large and that description should be supplemented by $\alpha'$ corrections. One can hope that such higher order effects conspire to localise the ONS5 behaviour in string theory - indeed \cite{Baines:2020dmu} argues that smeared O-planes can be a good approximation to localised ones in some instances. However if one takes the conservative view and insists on fully localised O-planes in supergravity, all is not lost: The compatibility of a class of solutions with smeared O-planes often suggests that it is also compatible with localised planes. Such solutions are harder to construct, but one can view the solution here as a positive first step in that direction. As this subtly involves the boundaries of the space we expect such generalisations to exhibit qualitatively similar physical behaviour.    

The quantised charges in the different $[\frac{2\pi}{q} k, \frac{2\pi}{q} (k+1)]$ intervals are thus given by
\begin{eqnarray}
&&Q_{D2}^{(k)}=\frac{1}{(2\pi)^5}\int_{\mathbb{T}^3, \text{S}^3}\hat{F}_6=q \Bigl(g-g' (\rho-\frac{2\pi}{q}k)\Bigr)=q\,\alpha_k=\sum_{j=0}^{k-1} \beta_j, \label{chargeD2}\\
&&Q_{NS5}^{(k)}=\frac{1}{(2\pi)^2}\int_{\mathbb{T}^3} H_3=\beta_k,\label{chargeNS5}\\
&&Q_{D4}^{(k)}=\frac{1}{(2\pi)^3}\int_{I_{\rho},\text{S}^3}\hat{F}_4=1\label{chargeD4}.
\end{eqnarray}
This implies that the constants $\beta_k$ must be integer numbers, as they are directly related to the number of branes in the brane set-up. This confirms that the suggested brane configuration is the one given in Table \ref{D2-D4-NS5}.  Substituting our expression for $g$ into the Bianchi identities we find
\begin{eqnarray}
&&dH_3= \frac{h_0}{2\pi} \sum_{k=1}^P (\beta_k-\beta_{k-1}) \delta\left(\rho-\frac{2\pi}{q}k\right) \,d\rho \wedge\text{vol}(\mathbb{T}^3)\,,\\
&&d\hat{f}_6= -\frac{q\,h_0}{\pi} \sum_{k=1}^P (\beta_k-\beta_{k-1})\left(\rho-\frac{2\pi}{q}k\right) \delta\left(\rho-\frac{2\pi}{q}k\right) \,\,d\rho \wedge\text{vol}(\mathbb{T}^3)\wedge \text{vol}(\text{S}^3)=0,\nonumber
\end{eqnarray}
where $\hat{f}_6$ denotes the magnetic component of the 6-form Page flux.
They are thus satisfied up to source terms, which indicate the presence of $(\beta_{k-1}-\beta_k)$ NS5-branes at $\rho=\frac{2\pi}{q}k$, where the slope of $g$ changes. These branes are wrapped on the $\text{AdS}_3\times $S$^3$ subspace of the geometry and smeared over the $\mathbb{T}^3$.

Finally, the central charge computed with the Brown-Henneaux formula gives, for this class of solutions\footnote{Note that this expression is also valid when $F_0\neq 0$.}
\begin{equation}
\label{holographic-c}
c_{hol}=\frac{3}{\pi}\,q^2 \int d\rho\, h\,g.
\end{equation}
This will be later compared to the corresponding field theory expression.


\subsection{2d dual CFTs}\label{fieldtheory}

In order to extract the quiver QFTs associated to the previous solutions we need to account for the ordering of the NS5-branes along the 
$\rho$ direction, together with the net number of D2-branes ending on each of them and the D4-branes orthogonal to both types of branes. The massless modes that give rise to the quiver QFT are then coming from  
the strings stretching between the D-branes in the same interval between NS5-branes, or between adjacent intervals. There are three types of massless modes to consider:
\begin{itemize}
\item D2-D2 strings: There are two cases to consider. Open strings with both end points lying on the same stack of D2-branes give rise to $\mathcal{N}=(4,4)$ vector multiplets, while those with end points on two different stacks separated by an NS5-brane give rise to $\mathcal{N}=(4,\,4)$ hypermultiplets in the bifundamental representation. 

\item D4-D4 strings: Depending on the size of the $\mathbb{T}^3$, on which the D4-branes are wrapped, these strings do not contribute massless modes. Given that D4-D4 strings are T-dual to D2-D2 strings, the open strings would contribute a $(4,\,4)$ vector multiplet for a stringy size $\mathbb{T}^3$.

\item D2-D4 strings: Strings with one end on D2-branes  and the other end on orthogonal D4-branes in the same interval between NS5-branes contribute with fundamental $(4,4)$ hypermultiplets, associated to the motion of the strings along the $(z_1,z_2,z_3)$ directions plus the $A_5$ component of the gauge field.

\end{itemize}

The relevant data to construct the quivers associated to these massless modes are the linking numbers of the D4-branes and the NS5-branes. To define these we use that the brane set-up depicted in Table \ref{D2-D4-NS5} is T-dual to the Type IIB construction studied in \cite{Hanany:1996ie}, and use the definitions
\begin{eqnarray}
&&l_i=n_i+L_i^{NS5}, \qquad \text{for the D4-branes} \\
&&\hat{l}_j=-\hat{n}_j+R_j^{D4}, \qquad \text{for the NS5-branes,}
\end{eqnarray}
where $n_i$ is the number of D2-branes ending on the $i$th D4-brane from the right minus the number of D2-branes ending on it from the left, $\hat{n}_j$ is the same quantity for the $j$th NS5-brane, $L_i^{NS5}$ is the number of NS5-branes lying on the left of the $i$th D4-brane, and $R_j^{D4}$ is the number of D4-branes lying on the right of the $j$th NS5-brane\footnote{Our conventions are related through T and S dualities to the conventions in \cite{Gaiotto:2008ak}.}. Following  \cite{Gaiotto:2008ak} it is then possible to read the data of the QFT living in the brane set-up from the linking numbers, namely, the gauge group $G=U(N_1)\times \dots \times U(N_k)$, the bifundamental fields transforming in the $(N_i,\bar{N}_{i+1})$ representations, and the fundamental matter, transforming under $U(M_i)$ for each group.

The way to proceed is as follows. The linking numbers of both the D4 and NS5 branes define an integer number $N$, as
$N=\sum_{i=1}^p l_i=\sum_{j=1}^{\hat{p}}\hat{l}_j$, where $p$ and $\hat{p}$ are the numbers of D4-branes and NS5-branes, respectively. 
This is the number of D2-branes that end on the left on a collection of D4 branes and on the right on a collection of NS5-branes. Any brane configuration can be pictured in this way after suitable Hanany-Witten moves. Now, in order to read the quiver, we consider the partition $N=\sum_{j=1}^{\hat{p}}\hat{l}_j$, where the NS5-branes have to be ordered such that $\hat{l}_1\ge \hat{l}_2\ge \dots \ge \hat{l}_{\hat{p}}$, and a second partition defined from a list of positive integer numbers satisfying $q_1\ge q_2\ge \dots \ge q_r$, $N=\sum_{s=1}^r M_s q_s$, with the numbers $M_s$ indicating how many times the different integers $q_s$ appear in the partition. The set of integers $q_s$ is defined such that the number of terms in the decomposition that are equal or bigger than a given integer $j$, that we denote as $m_j$, satisfy that 
\begin{equation} \label{keycondition}
\sum_{j=1}^i m_j \ge \sum_{j=1}^i \hat{l}_j, \quad \forall i=1,\dots, \hat{p}.
\end{equation}
From these two partitions the 
ranks of the different $U(N_i)$ gauge groups of the quiver are then computed as
\begin{equation}
N_i=\sum_{j=1}^i (m_j-\hat{l}_j).
\end{equation}
In turn, the numbers $M_s$ appearing in the $N=\sum_{s=1}^r M_s q_s$ decomposition give the ranks of the fundamental matter groups that couple to each of the gauge groups. A detailed account of this construction can be found in \cite{Assel:2011xz}. It will become clearer after we illustrate it with the particular brane set-up that is the subject of our analysis.

\begin{figure}
\centering
\includegraphics[scale=0.75]{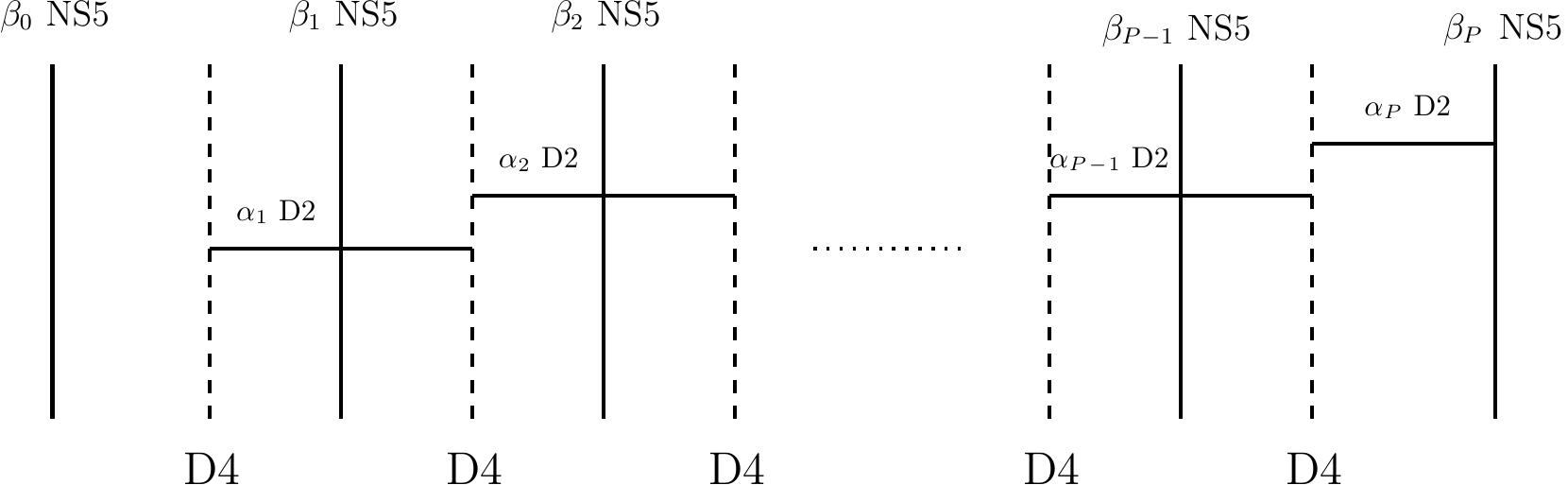}
\caption{Brane set-up associated to the quantised charges \eqref{chargeD2}-\eqref{chargeD4}, in units of $q$.}
\label{brane-setup-T3}
\end{figure}

Let us now apply these rules to the construction of the field theory associated to our solutions, defined by $g(\rho)$ as in \eqref{profileg}. The brane set-up is read from the numbers of branes at each $\rho\in [\frac{2\pi}{q} k, \frac{2\pi}{q} (k+1)]$ interval, determined by equations
 \eqref{chargeD2}-\eqref{chargeD4}. Moreover, as discussed below equation \eqref{profileg}, $\beta_P$ anti-NS5-branes must end the space at $\rho=\frac{2\pi}{q}(P+1)$. The resulting brane set-up is then the one depicted in Figure \ref{brane-setup-T3}.
 From this brane configuration we can read the linking numbers for the D4-branes
\begin{equation}
l_i=\sum_{r=0}^{i-2}\beta_{r}+2\beta_{i-1},\qquad i=1,\dots, P
\end{equation}
and for the NS5-branes
\begin{eqnarray}
&&\hat{l}_1=\hat{l}_2=\dots =\hat{l}_{\beta_0}=P,\nonumber\\
&&\hat{l}_{\beta_0+1}=\hat{l}_{\beta_0+2}=\dots =\hat{l}_{\beta_0+\beta_1}=P-1,\nonumber\\
&& \hspace{4cm}\vdots\nonumber\\
&&\hat{l}_{\beta_0+\beta_1+\dots +\beta_{P-3}+1}=\hat{l}_{\beta_0+\beta_1+\dots +\beta_{P-3}+2}=\dots =
\hat{l}_{\beta_0+\beta_1+\dots +\beta_{P-2}}=2,\nonumber\\
&&\hat{l}_{\beta_0+\beta_1+\dots +\beta_{P-2}+1}=\dots =\hat{l}_{\beta_0+\beta_1+\dots +\beta_{P-1}}=1,\nonumber\\
&&\hat{l}_{\beta_0+\beta_1+\dots +\beta_{P-1}+1}=\dots =\hat{l}_{\beta_0+\beta_1+\dots +\beta_{P-1}+\beta_P}=1.
\end{eqnarray}
From the linking numbers we construct the total number of D2-branes ending on D4-branes on the left and NS5-branes on the right. This is given by
\begin{equation}
N=\sum_{i=1}^P l_i=\sum_{j=1}^{\beta_0+\dots +\beta_P}\hat{l}_j=\sum_{k=0}^{P-1} (P-k+1)\beta_{k}.
\end{equation}
Now, from $N$ we define the two partitions that will allow us to read the quiver CFT. The NS5-branes in our brane set-up are ordered such that $\hat{l}_1\ge \hat{l}_2\ge \dots \ge \hat{l}_{\hat{\beta_0+\dots +\beta_P}}$. These linking numbers define then one of the two partitions, $N=\sum_{j=1}^{\beta_0+\dots +\beta_P}\hat{l}_j$.
In turn, for the D4-branes we take 
\begin{equation} \label{partition}
N=\underbrace{\beta_0}+\underbrace{\beta_0+\beta_1}+\underbrace{\beta_0+\beta_1+\beta_2}+\dots +\underbrace{\beta_0+\beta_1+\dots +\beta_{P-2}}+2\underbrace{(\beta_0+\beta_1+\dots +\beta_{P-1})}
\end{equation}
from where
\begin{eqnarray}
&&m_1=m_2=\dots =m_{\beta_0}=P+1, \nonumber\\
&&m_{\beta_0+1}=\dots=m_{\beta_0+\beta_1}=P, \nonumber\\
&& \hspace{4cm}\vdots\nonumber\\
&&m_{\beta_0+\beta_1+\dots +\beta_{P-3}+1}=\dots =m_{\beta_0+\beta_1+\dots +\beta_{P-2}}=3, \nonumber\\
&&m_{\beta_0+\beta_1+\dots +\beta_{P-2}+1}=\dots =m_{\beta_0+\beta_1+\dots +\beta_{P-1}}=2.
\end{eqnarray}
These numbers satisfy the condition \eqref{keycondition}  $\forall i=1,\dots, (\beta_0+\dots +\beta_P)$.
We then find for the ranks of the gauge groups
\begin{eqnarray}
&&N_1=m_1-\hat{l}_1=P+1-P=1, \quad N_2=N_1+m_2-\hat{l}_2=2, \quad \dots \quad N_{\beta_0}=\beta_0, \nonumber\\
&&N_{\beta_0+1}=\beta_0+1, \quad \dots \quad 
N_{\beta_0+\beta_1+\dots + \beta_{P-1}}= \beta_0+\beta_1+\dots + \beta_{P-1}, 
\end{eqnarray}
to then start decreasing
\begin{equation}
N_{\beta_0+\beta_1+\dots \beta_{P-1}+1}= \beta_0+\beta_1+\dots + \beta_{P-1}-1, \quad \dots \quad 
N_{\beta_0+\beta_1+\dots \beta_{P-1}+\beta_P-1}=1.
\end{equation}
That is, the ranks of the gauge groups increase in units of 1 till the value $\beta_0+\beta_1+\dots +\beta_{P-1}$ is reached, to then start decreasing, again in units of one, till the gauge group of rank 1 is reached, corresponding to the D2-branes stretched between the last pair of NS5-branes. 

\begin{figure}
\centering
\includegraphics[scale=0.65]{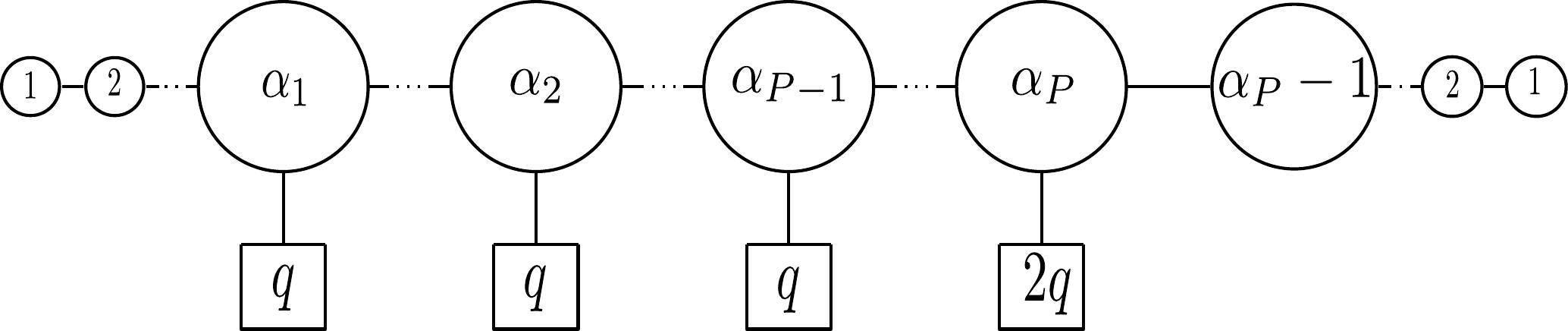}
\caption{2d quiver associated to the brane set-up depicted in Figure \ref{brane-setup-T3}. Circles denote (4,4) vector multiplets and black lines (4,4) bifundamental hypermultiplets. The gauge groups with ranks $\alpha_k$, with $k=1,\dots, P-1$ to the left of the gauge group with rank $\alpha_P$ have U(1) flavour symmetries. The gauge group with rank $\alpha_P$ has U(2) flavour symmetry. The rest of gauge groups do not have attached any flavours.}
\label{quiverT3_1}
\end{figure}

Finally, from the partition \eqref{partition} we have that
\begin{equation}\label{flavourgroups}
M_{\beta_0}=M_{\beta_0+\beta_1}=\dots = M_{\beta_0+\beta_1+\dots +\beta_{P-2}}=1,\quad M_{\beta_0+\beta_1+\dots +\beta_{P-1}}=2.
\end{equation}
This implies that the gauge groups with ranks $\beta_0=q \,\alpha_1$, $\beta_0+\beta_1=q\, \alpha_2$, till $\beta_0+\dots + \beta_{P-2}=q\,\alpha_{P-1}$ have U(1) flavour groups, while the gauge group with rank $\beta_0+\beta_1\dots +\beta_{P-1}=q\,\alpha_P$ has flavour group U(2). The rest of gauge groups have no flavour groups attached. The resulting quiver is depicted in Figure \ref{quiverT3_1}.  
One can check that the number of gauge nodes equals the total number of NS5-branes minus 1, as it should be. In this quiver circles denote $(4,4)$ vector multiplets and black lines $(4,4)$ bifundamental hypermultiplets. Note that we have rescaled it such that the intervals have length $[0,2\pi]$, as it is more standard in the literature, and therefore
\begin{equation}\label{nuevasalphas}
\alpha_k=\sum_{j=0}^{k-1}\beta_j, \qquad \text{for} \qquad k=1,\dots, P.
\end{equation}

Our proposal is that the QFTs defined by these quivers flow in the IR to the 2d CFTs dual to the class of solutions defined by \eqref{eq:massiveclassmetricYolanda}-\eqref{F6Yolanda}, with $h=$ constant and $g$ given by \eqref{profileg}. Next we will provide a non-trivial check of this proposal, consisting on the matching between the field theory and holographic central charges. However, before we do that we should recall that the Higgs and Coulomb branches of 2d $(4,4)$ theories are described by different CFTs, with the different branches having different R-symmetries and usually different central charges \cite{Diaconescu:1997gu,Witten:1997yu}. Thus, the question arises as to which 
of these branches of the theory is described holographically by our class of solutions. 
The basis of the argument in  \cite{Witten:1997yu} is that the scalars should be singlets under the SO(4) R-symmetry of the 2d CFT. In our case this symmetry is associated to the isometry group of the 3-sphere in the internal space. Since the scalars in the Higgs branch are singlets under this group the Higgs branch flows to a CFT with R-symmetry coming from this SO(4). In turn, the scalars in the Coulomb branch transform in the $(\textbf{2},\textbf{2})$ representation of SO(4), so the Coulomb branch must flow to a 2d CFT with R-symmetry coming from the SU(2) associated to the S$^2$ living in the $\mathbb{T}^3$ (this is locally $\mathbb{R}^3$), which should be enhanced to SO(4) at strong coupling (see below). Based on this argument our solutions must be holographically dual to the Higgs branch 2d CFT. Accordingly, the holographic central charge must match the central charge of the Higgs branch.

 Given that our theories are $(4,4)$ supersymmetric, we can use the expression that gives the central charge of the left or right-moving SU(2) group of R-symmetries to compute the central charge of the Higgs branch, given by equation \eqref{field-theory-c}, $c=6(n_{hyp}-n_{vec})$,
where $n_{hyp}$ stands for the number of $(0,4)$ hypermultiplets and $n_{vec}$ for the number of $(0,4)$ vector multiplets. Note that they can also stand, respectively, for the number of $(4,4)$ hypermultiplets and $(4,4)$ vector multiplets, more useful for our quiver constructions, since their respective $(0,4)$ Fermi multiplets and $(0,4)$ adjoint twisted hypermultiplets do not contribute to the R-symmetry anomaly. For the quivers depicted in Figure \ref{quiverT3_1} we have
\begin{equation}
n_{hyp}=2  \sum_{k=1}^{\alpha_P-1}k(k+1)+q\,\sum_{k=1}^{P-1}\alpha_k+2q\, \alpha_P \qquad \text{and} \qquad n_{vec}=2\sum_{k=1}^{\alpha_P-1}k^2+ \alpha_P^2.
\end{equation}
This gives
\begin{equation}
\label{result-c}
c=6\,q\,\sum_{k=1}^{P}\alpha_k.
\end{equation}

The holographic central charge was computed in the previous section. It is given by expression \eqref{holographic-c}. Taking $h=1$ and $g$ as defined by \eqref{profileg}, \eqref{nuevasalphas} it reduces to
\begin{equation}\label{holoc}
c_{hol}=6\,q\,\sum_{k=1}^{P}\alpha_k.
\end{equation}
We thus find exact agreement with the field theory calculation.

\begin{figure}
\centering
\includegraphics[scale=0.45]{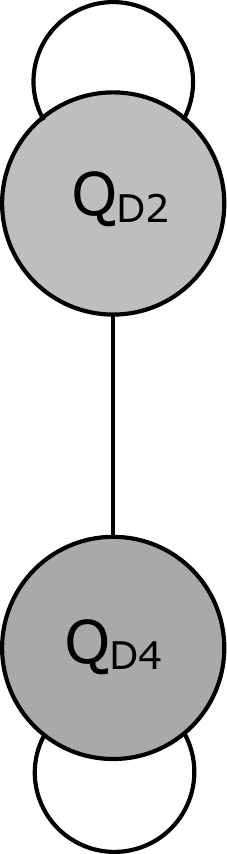}
\caption{Quiver associated to the solution with $g=\text{constant}$, corresponding to the T-dual of the D1-D5 system.}
\label{quiverTD}
\end{figure}

A particular example in our class of solutions is when the interval is periodically identified, in which case the function $g$ has to be constant. This gives the quantised charges
$Q_{\text{D}2}=q\, g$, $Q_{\text{D}4}=1$, for $\rho\in [0,\frac{2\pi}{q}]$. For  $\rho\in [0,2\pi]$ we have $Q_{\text{D}2}=g$, $Q_{\text{D}4}=q$.
This solution describes the T-dual of the D1-D5 system when the $\text{CY}_2$ is a $\mathbb{T}^4$, and the T-duality takes place along one of the directions of the $\mathbb{T}^4$. The D5-branes become D4-branes smeared on the T-duality direction and the quiver collapses to the one describing the D1-D5 system, depicted in Figure \ref{quiverTD} (for $\rho\in [0,2\pi]$). Equation \eqref{field-theory-c} gives the well-known result $c=6\, Q_{\text{D}2}Q_{\text{D}4}$ for the central charge, in agreement with the holographic result. 

\subsection{Realisation in M-theory}\label{mirror}

In this subsection we look into the M-theory regime of the brane intersection depicted in Figure \ref{D2-D4-NS5}. At strong coupling the D4-branes become M5-branes wrapped on the 11th direction, while the NS5-branes become M5'-branes transverse to it. Thus, the Hanany-Witten configuration consists on M2-branes stretched between M5'-branes with M5-branes orthogonal to them. In M-theory the M5 and the M5' branes are however equally non-perturbative, so one could alternatively consider the configuration in which the M2-branes are stretched between the M5-branes with the M5'-branes orthogonal to them. 

\begin{figure}
\centering
\includegraphics[scale=0.7]{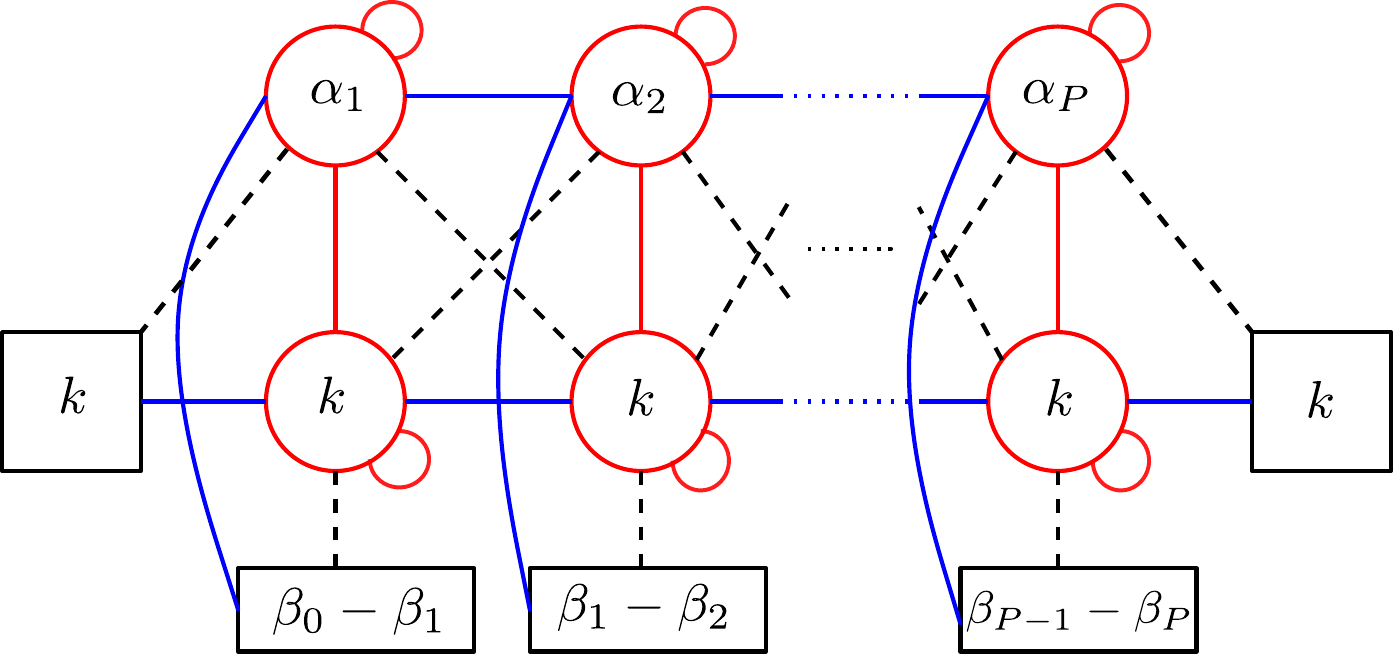}
\caption{2d quiver associated to the $\text{AdS}_3\times \text{S}^2\times \mathbb{T}^4\times I$ solutions with $\alpha_k$ D2-branes and $k$ D6-branes wrapped on the $\mathbb{T}^4$. Circles denote $(0,4)$ vector multiplets, blue lines $(4,4)$ twisted hypermultiplets, red lines $(0,4)$ hypermultiplets and dashed lines $(0,2)$ Fermi multiplets.}
\label{quiverT3_2}
\end{figure}

In order to read off the field content associated to this  configuration in weakly coupled string theory we need to reduce to ten dimensions in a direction in which the M5-branes become NS5-branes. In our set-up this can be achieved reducing along the Hopf-fibre direction of the S$^3$, which is transverse to the M5-branes. This halves the number of supersymmetries and creates a D6-brane. Moreover, in the reduction the $\mathbb{T}^3$ combines with the S$^1$ (that played before the role of eleventh direction, that we denote by $\psi$) to produce a $\mathbb{T}^4$. The resulting brane set-up is the one depicted in Table 
\ref{D2-D4-NS5-D6}, which is the one underlying the AdS$_3\times $S$^2\times \mathbb{T}^4\times I$ solutions constructed in \cite{Lozano:2019emq}, restricted to the massless case.
 \begin{table}[ht]
	\begin{center}
		\begin{tabular}{| l | c | c | c | c| c | c| c | c| c | c |}
			\hline		    
			&$x^0$ & $x^1$  & $z_1$ & $z_2$ & $z_3$ & $\psi$ & $\rho$ & $\zeta$ & $\theta^1$ & $\theta^2$ \\ \hline
			D2 & x & x & &  &  & & x &   &   &   \\ \hline
			NS5 & x & x & x &x  & x & x  &  &  &  &   \\ \hline
			D4 & x & x &  &  &  &   &   & x  & x & x  \\ \hline
			D6 & x & x & x & x & x & x & x & & &   \\ \hline 
		\end{tabular} 
	\end{center}
	\caption{$\frac18$-BPS brane intersection associated to the solutions in \cite{Lozano:2019emq}. $(x^0,x^1)$ are the directions where the 2d dual CFT lives. $(z_1, z_2, z_3, \psi)$ span the $\mathbb{T}^4$, on which the NS5 and D6 branes are wrapped. The coordinates $(\zeta,\theta^1,\theta)$ are the transverse directions realising the SO(3)-symmetry associated with the isometries of the S$^2$.}
	\label{D2-D4-NS5-D6}	
\end{table} 
In the particular brane intersection associated to our solutions there are $\alpha_j$ D2-branes\footnote{With the $\alpha_j$ defined as in \eqref{nuevasalphas}.} and a D6-brane wrapped on the $\mathbb{T}^4$ stretched between NS5-branes, that play the role of colour branes. 
Note however that in order to have a consistent IIA supergravity background the number of D6-branes should be large,  which implies that prior to the reduction the S$^3$ has to be modded by $\mathbb{Z}_k$, such that $k$ D6-branes are obtained upon reduction. 
Additional $(\beta_{j-1}-\beta_j)$ orthogonal D4-branes at each interval play the role of flavour branes. The holographic central charge can be obtained from the result in \cite{Lozano:2019zvg}, where this quantity was computed for the general class of solutions in  \cite{Lozano:2019emq}. One can check that for our configuration it agrees with the holographic central charge  computed in \eqref{holoc}, multiplied by $k$ due to the $\mathbb{Z}_k$ orbifolding, that mods out the S$^3$ by $\mathbb{Z}_k$. The field theory living in the brane intersection can also be determined from the general study  in \cite{Lozano:2019zvg}\footnote{See also \cite{Couzens:2021veb}, where some corrections to the analysis in \cite{Lozano:2019zvg} were pointed out.}. The result is the quiver gauge theory depicted in  Figure \ref{quiverT3_2}. 
In this figure circles denote $(0,4)$ vector multiplets, blue lines $(4,4)$ twisted hypermultiplets, red lines $(0,4)$ hypermultiplets and dashed lines $(0,2)$ Fermi multiplets. 2d $(0,4)$ theories do not have a Coulomb branch, since $(0,4)$ vector multiplets contain no scalars. In turn, for the Higgs branch one can use expression \eqref{field-theory-c}, which gives rise to\footnote{Here the factor of $q$ arises because the quiver has to be rescaled by $q$ in order to account for the rescaling of the quiver of Figure \ref{quiverT3_1}.}
\begin{equation}
c_R=6\,q\,k\, \sum_{j=1}^P \alpha_j.
\end{equation}
In this case this is the right-moving central charge, since the theory is $(0,4)$ supersymmetric. However, using expression \eqref{cL} one can see that $c_L=c_R$, due to the condition of anomaly cancellation. One can now see that this expression agrees with the central charge of (the Higgs branch of) the quiver depicted in  Figure \ref{quiverT3_1}, given by expression 
\eqref{result-c}\footnote{Multiplied by $k$ due to the orbifolding by $\mathbb{Z}_k$.}.  This result shows that the different light multiplets appearing in the quivers depicted in Figures \ref{quiverT3_1}  and \ref{quiverT3_2}, both of which are  precise deductions of perturbative string theory, lead to the same central charge. Of course the reason for this agreement is the common origin in M-theory of both classes of solutions. Field theoretically what we find is a realisation of the mirror symmetry of the dual CFT, in the precise sense discussed below.

\subsection{Realisation in Type IIB}\label{typeIIB}

The common M-theory origin of both classes of solutions implies that they are related by S-duality once they are T-dualised onto Type IIB string theory. This is another reason why they should flow to the same CFT in the IR.
Which deformation would be more convenient to use away from the critical point depends as usual on the concrete value of the gauge coupling. At the level of the solutions, once they have been T-dualised to Type IIB supergravity  both classes become $(0,4)$ supersymmetric, because the T-duality on the $\text{AdS}_3\times\text{S}^3\times\mathbb{T}^3\times I$ solutions, $(4,4)$ supersymmetric in Type IIA, takes place along the Hopf-fibre of the $\text S^3$, and this halves the supersymmetries to $(0,4)$\footnote{Still, the 2d dual CFT does not change,  independently on the number of supersymmetries that are manifest in the UV.}. These Type IIB solutions are interesting on their own, since they provide explicit holographic duals to D3-brane boxes constructions \cite{Hanany:2018hlz}, realising in this case small $\mathcal{N}=(0,4)$ supersymmetry\footnote{Recall that, instead, the D3-brane boxes constructed in \cite{Hanany:2018hlz} have SO$(4)_R$ symmetry, and should therefore be dual to $\text{AdS}_3$ solutions with large $\mathcal{N}=(0,4)$ supersymmetry.}. In the next subsection we construct these Type IIB backgrounds, and show that they are related by an SL(2,$\mathbb{R}$) transformation to the T-duals (along a circle on the $\mathbb{T}^4$) of the AdS$_3\times $S$^2\times \mathbb{T}^4\times I$ solutions of massless Type IIA constructed in \cite{Lozano:2019emq}. 

The brane set-up associated to the T-dual of the AdS$_3\times $S$^3\times \mathbb{T}^3\times I$ solutions studied in section \ref{D2D4NS5} is depicted in Table \ref{D3-D5-NS5-NS5'}, while that associated to the T-dual of the AdS$_3\times $S$^2\times \mathbb{T}^4\times I$ solutions constructed in \cite{Lozano:2019emq} is depicted in Table \ref{D3-NS5-D5-D5'}. One can check that these brane set-ups are S-dual to each other.
 \begin{table}[ht]
	\begin{center}
		\begin{tabular}{| l | c | c | c | c| c | c| c | c| c | c |}
			\hline		    
			&$x^0$ & $x^1$  & $z_1$ & $z_2$ & $z_3$ & $\rho$ & $\zeta$ & $\psi$ & $\theta^1$ & $\theta^2$\\ \hline
			D3 & x & x & &  &  &x &  & x  &   &   \\ \hline
			D5 & x & x & x &x  & x &  &  & x &  &   \\ \hline
			NS5 & x & x &  &  &  &   & x  & x  & x & x  \\ \hline
			NS5' & x & x & x & x & x & x & & & &   \\ \hline 
		\end{tabular} 
	\end{center}
	\caption{$\frac18$-BPS brane intersection T-dual to the brane intersection depicted in Table \ref{D2-D4-NS5}, realising a D3-brane box model. $(x^0,x^1)$ are the directions where the field theory lives, $(z_1, z_2, z_3)$ span the $\mathbb{T}^3$, $\rho$ is the direction where the NS5-branes are located, $\zeta$ and $\theta^i$ are respectively the radial coordinate of AdS$_3$ and the angles that parameterise the S$^2$, and $\psi$ parameterises the S$^1$ generated upon the dualisation, where the NS5'-branes are located. $(\rho,\psi)$ are thus the two directions of the brane box.}   
	\label{D3-D5-NS5-NS5'}	
\end{table} 
 \begin{table}[http!]
	\begin{center}
		\begin{tabular}{| l | c | c | c | c| c | c| c | c| c | c |}\hline
			&$x^0$ & $x^1$  & $z_1$ & $z_2$ & $z_3$ & $\psi$ & $\rho$ & $\zeta$ & $\theta^1$ & $\theta^2$  \\ \hline
			D3 & x & x & &  &  &x & x &   &   &   \\ \hline
			NS5 & x & x & x &x  & x &x  &  &  &  &   \\ \hline
			D5 & x & x &  &  &  & x  &  & x  & x & x  \\ \hline
			D5' & x & x & x & x & x &  &x & & &   \\ \hline 
		\end{tabular} 
	\end{center}
	\caption{$\frac18$-BPS brane intersection T-dual to the brane intersection depicted in Table \ref{D2-D4-NS5-D6}. $(x^0,x^1)$ are the directions where the field theory lives, $(z_1, z_2, z_3)$ span a $\mathbb{T}^3$, $\psi$ is the T-duality circle and $\rho$ is the field theory direction. This configuration is S-dual to the configuration in Table \ref{D3-D5-NS5-NS5'}.}   
	\label{D3-NS5-D5-D5'}	
\end{table} 
Furthermore, one can see that the S-duality of Type IIB string theory interchanges the $(0,4)$ hypermultiplets and $(0,4)$ twisted hypermultiplets associated to the massless string modes living in the respective Type IIB configurations.
This is the 2d manifestation of the mirror symmetry present in 3d gauge theories  \cite{Intriligator:1996ex,Hanany:1996ie}, which besides inverting the coupling constant, interchanges the scalars in the hypermultiplets and vector multiplets, and therefore the Higgs and Coulomb branches of the 3d theory. Given that 2d (0,4) field theories do not have a Coulomb branch, since (0,4) vector multiplets contain no scalars, 2d mirror symmetry cannot be realised as the interchange between the Higgs and Coulomb branches. Remarkably, mirror symmetry is realised in this set-up as the interchange between the scalars transforming under the SU$(2)_R$ symmetry, i.e those belonging to the twisted hypermultiplets, with those that are singlets under the SU(2)$_R$, i.e the ones belonging to the untwisted hypermultiplets. This extends very naturally the mirror symmetry present in 3d gauge theories to these 2d theories, and parallels the interchange between chiral and twisted chiral superfields inherent to mirror symmetry in supersymmetric sigma models.

\subsubsection{Solutions of Type IIB supergravity}

In this subsection we complement the above holographic discussion with the explicit construction of the Type IIB supergravity solutions. 
    
We start presenting the T-dual of the solutions studied in section \ref{D2D4NS5}. T-dualising along the Hopf fiber of the S$^3$ of the $\text{AdS}_3\times\text{S}^3\times\mathbb{T}^3$ solutions given by \eqref{eq:massiveclassmetricYolanda}-\eqref{F6Yolanda} we obtain the Type IIB backgrounds
  \begin{align}\label{eq:solD3D5NS5}
ds^2&= q\,h^{-1/2}\left[ds^2(\text{AdS}_3)+4^{-1}\,ds^2(\text{S}^2)\right]+q^{-1}\,h^{1/2}\,d\psi^2+ g\,\left[h^{-1/2} d\rho^2+ h^{1/2}\,ds^2(\mathbb{T}^3)\right]\,,\nonumber\\[2mm]
e^{-\Phi}&=(q\,h)^{1/2}\, g^{-1/2}\,,\qquad \qquad H_3=\partial_{\rho}\,(hg)\, \text{vol}(\mathbb{T}^3)-2^{-1}\,\text{vol}(\text{S}^2)\wedge d\psi,\nonumber \\[2mm]
F_1&=g^{-1}\,\partial_\rho\,h\,d\psi\,,\nonumber\\[2mm]
F_3&=-2^{-1} q \, \text{vol}(\text{S}^2)\wedge d\rho,\nonumber\\[2mm]
F_5&= 2\, q \, \text{vol}(\text{AdS}_3)\wedge d\rho\wedge d\psi+2^{-1}\,q\,g \,h\,\text{vol}(\mathbb{T}^3)\wedge \text{vol}(\text{S}^2)\,, 
\end{align}
where $\psi$ parameterises the T-duality circle\footnote{This solution is an example  contained in  the class of \cite{Macpherson:2022sbs} section 3.1: one should identify $(h,g)$ and $(P,G)$ there, restrict $u'=0$ and impose that $\partial_{z_i}$ are all isometries.}.
In order to provide the local representation of the brane set-ups of Tables \ref{D3-D5-NS5-NS5'} and \ref{D3-NS5-D5-D5'} we need to focus on the particular situation
\begin{equation}
h=\text{constant}, \qquad g''=0,
\end{equation}
corresponding to the massless solutions in Type IIA. In this case the metric exhibits the characteristic behaviour of NS5-branes wrapped on an AdS$_3\times$S$^2\times$S$^1$ geometry. Indeed, it 
can be verified that these solutions arise in the near horizon limit of a D3-D5-NS5-NS5' brane solution representing the bound state of Table \ref{D3-D5-NS5-NS5'}, where the D3-D5-NS5' branes have been fully localised within the worldvolume of the NS5 branes (as it was done for the $H_{\text D 2}(\zeta)$ and $H_{\text D 4}(\zeta)$ harmonic functions in section \ref{branepicture}). We will restrict to this subclass of solutions, characterised by a vanishing axion, in the remainder of this section.

Let us now perform an $\mrm{SL}(2,\mathbb R)$ rotation parameterised by an angle $\xi \in[0,\frac{\pi}{2}]$,\begin{equation}
R=
\left(\begin{array}{cc}
\cos \xi & - \sin \xi \\
\sin\xi &   \cos\xi \\
\end{array}\right)\,,
 \end{equation}
 in this subclass of solutions.
Starting with a ``seed" background described by fluxes, dilaton, metric and axio-dilaton $F_{(n),s}$, $\Phi_{s}$, $ds^2_{10,s}$ and $\tau_s=C_{0,s}+ie^{-\Phi_s}$, $R$ acts as usual,
\begin{equation}
 \begin{split}\label{Srotation_fluxes}
&\left(\begin{array}{c}
\tilde F_{3} \\
H_{3}\\
\end{array}\right)=\left(\begin{array}{cc}
\cos \xi & - \sin \xi \\
\sin\xi &   \cos\xi \\
\end{array}\right)\left(\begin{array}{c}
F_{3,s} \\
H_{3,s}\\
\end{array}\right)\,,\\
 & \tau=\frac{\cos\xi\,\tau_s-\sin\xi}{\sin\xi\,\tau_s+\cos\xi}\,, \qquad F_{5}=F_{5,s}\,.\\
 \end{split}
\end{equation}
Note that even if our seed solutions are characterized by a vanishing axion, this transformation generates a non-trivial profile for $C_{0}$. This implies that the 3-form flux associated to the rotated solution is given by $F_{3}=\tilde F_{3}-C_{0}H_{3}$.
Finally, the metric in the string frame transforms as $ds^2_{10}=|\cos\xi+\sin\xi\,\tau|\,ds^2_{10,s}$. Applying these rules to the backgrounds given by \eqref{eq:solD3D5NS5} the following one-parameter family of solutions is obtained,
  \begin{align}\label{eq:solD3D5NS5_SLrotated}
ds^2&=\Delta^{1/2}\left[q\,h^{-1/2}\left[ds^2(\text{AdS}_3)+4^{-1}\,ds^2(\text{S}^2)\right]+q^{-1}\,h^{1/2}\,d\psi^2+ g\,\left[h^{-1/2} d\rho^2+ h^{1/2}\,ds^2(\mathbb{T}^3)\right]\right]\,,\nonumber\\[2mm]
\Delta&=c^2+q\,h\,g^{-1}\,s^2\,,\nonumber\\[2mm]
e^{-\Phi}&=\Delta^{-1}(h\,q)^{1/2}\,g^{-1/2}\,,\qquad \qquad C_0=sc\,\Delta^{-1}\left(h\,q\,g^{-1}-1  \right)\,,\nonumber\\[2mm]
H_3&=c\,h\,\partial_{\rho}\,g\, \text{vol}(\mathbb{T}^3)-2^{-1}\,c\text{vol}(\text{S}^2)\wedge d\psi-2^{-1}\,s\,q \, \text{vol}(\text{S}^2)\wedge d\rho, \nonumber\\[2mm]
F_3&= -2^{-1}\,q \,c\,\Delta^{-1} \text{vol}(\text{S}^2)\wedge d\rho-s\,q\,h^2g^{-1}\Delta^{-1}\partial_{\rho}\,g\, \text{vol}(\mathbb{T}^3)+2^{-1}s\,q\,h\,g^{-1}\,\Delta^{-1}\text{vol}(\text{S}^2)\wedge d\psi,\nonumber\\[2mm]
F_5&= 2\, q \text{vol}(\text{AdS}_3)\wedge d\rho\wedge d\psi+2^{-1}\,q\,g \,h\,\text{vol}(\mathbb{T}^3)\wedge \text{vol}(\text{S}^2)\,,
\end{align}
where $s=\sin\xi$ and $c=\cos\xi$\footnote{This generalised solution is an example contained in  the class of \cite{Macpherson:2022sbs} section 3.2: again one should identify $(h,g)$ with $(P,G)$ there, restrict $u'=0$ and impose that $\partial_{z_i}$ are all isometries.}. In particular, the family of S-dual solutions is obtained by setting $\xi=\frac{\pi}{2}$ in the above class, giving rise to
 \begin{align}\label{eq:solD3D5NS5_Sdual}
ds^2&= q^{3/2}g^{-1/2}\left(ds^2(\text{AdS}_3)+4^{-1}ds^2(\text{S}^2)\right)+q^{-1/2}h\,g^{-1/2}d\psi^2\nonumber\\
&+q^{1/2} g^{1/2} d\rho^2+ q^{1/2} g^{1/2}\,h\,ds^2(\mathbb{T}^3),\nonumber\\[2mm]
e^{-\Phi}&=(q\,h)^{-1/2}g^{1/2}\,,\qquad \qquad H_3=-2^{-1}q \, \text{vol}(\text{S}^2)\wedge d\rho\,\nonumber\\[2mm]
F_3&=-h\, \partial_{\rho}\,g\, \text{vol}(\mathbb{T}^3)+2^{-1}\text{vol}(\text{S}^2)\wedge d\psi,\nonumber\\[2mm]
F_5&= 2\, q  \text{vol}(\text{AdS}_3)\wedge d\rho\wedge d\psi+2^{-1}\,q\,g \,h\,\text{vol}(\mathbb{T}^3)\wedge \text{vol}(\text{S}^2)\,.
\end{align}
One can observe that, as expected, the 5-branes exchange their roles, with the metric now exhibiting the characteristic behaviour of  D5-branes wrapped on a AdS$_3\times$S$^2\times$S$^1$ geometry, originated by a D3-D5'-NS5 fully-backreacted intersection. One can also verify that these solutions arise in the near horizon limit of the brane 
intersection depicted in Table \ref{D3-NS5-D5-D5'}, where the D3-D5'-NS5 branes are fully localised within the worldvolume of the D5-branes.

The class of solutions presented in this section can be related to the Type IIB $\ma N=(0,4)$ AdS$_3$ solutions constructed in \cite{Faedo:2020lyw}, from slightly more general, D3-D5-NS5-D5'-NS5', brane set-ups. The easiest way to show this is by relating the solution with $\xi=\frac{\pi}{2}$ given in \eqref{eq:solD3D5NS5_Sdual} with equation (2.7) in \cite{Faedo:2020lyw}. One needs to impose that $H_{\text{NS5}'}=1$, rename $H_{\text{D5}'}=g$ and smear the solution in \cite{Faedo:2020lyw} in such a way that $H_{\text{D5}'}=g$ is delocalised over the internal $\mathbb R^3$, such that it can be replaced by a $\mathbb{T}^3$.


\section{AdS$_3\times$S$^3\times \mathbb{T}^3$ in Type I'}\label{TypeI'}

In this section we return to the solutions constructed in section \ref{D2D4NS5} but we now focus on the massive case  
$F_0\neq 0$. Recall that we had the AdS$_3\times$S$^3\times \mathbb{T}^3$ geometries fibered over an interval given by \eqref{eq:massiveclassmetricYolanda}-\eqref{F6Yolanda}, with defining functions satisfying the Bianchi identities \eqref{bianchisT3}.
In the massive case we choose to write $(g,h)$ in terms of a function $u$ as
\beq \label{massiveT3}
h=\sqrt{u} ,~~~~g=\frac{c}{\sqrt{u}},
\eeq
such that the Bianchi identities are satisfied with $c$ constant and $u$ a linear function.
The solutions then take the form
\begin{align}\label{eq:massiveclassmetricmassive}
&ds^2= \frac{q}{u^{\frac{1}{4}}}\bigg[ds^2(\text{AdS}_3)+ds^2(\text{S}^3)\bigg]+ \frac{c}{u^{\frac{1}{4}}}\bigg[ds^2(\mathbb{T}^3)+\frac{1}{\sqrt{u}}d\rho^2\bigg],~~~~e^{-\Phi}=\frac{u^{\frac{5}{8}}}{\sqrt{c}},\\
&F_0=\frac{u'}{2c}, \qquad F_4=2\,q\Bigl(\text{vol}(\text{AdS}_3)+\text{vol}(\text{S}^3)\Bigr)\wedge d\rho, \label{F0F4}\\
&F_6= 2\,q\,c\text{vol}(\mathbb{T}^3)\wedge (\text{vol}(\text{S}^3) +  \text{vol}(\text{AdS}_3))  \label{massiveclasslast}
\end{align}
The underlying brane set-up is the one depicted in Table \ref{D2D4D8}.
\begin{table}[http!]
	\begin{center}
		\begin{tabular}{| l | c | c | c | c| c | c| c | c| c | c |}\hline
			&$x^0$ & $x^1$  & $z_1$ & $z_2$ & $z_3$ & $\rho$ & $\zeta$ & $\theta^1$ & $\theta^2$ & $\theta^3$ \\ \hline
			D2 & x & x & &  &  &x &  &   &   &   \\ \hline
			D4 & x & x &x  &x  &x  &   &   &   &  &   \\ \hline
			D8 & x & x & x & x & x &  &x &x &x &x   \\ \hline 
		\end{tabular} 
	\end{center}
	\caption{$\frac18$-BPS brane intersection underlying the geometry \eqref{eq:massiveclassmetricmassive}-\eqref{massiveclasslast}. $(x^0,x^1)$ are the directions where the 2d dual CFT lives, $(z_1, z_2, z_3)$ span the $\mathbb{T}^3$, where the D4's and the D8's are wrapped, $\rho$ is the field theory direction, where the D2 branes are stretched, and $\theta^i$ parameterise the S$^3$.}
	\label{D2D4D8} 
	\end{table} 
As mentioned above, $u$ has to be a linear function in order to satisfy the Bianchi identities. We will take it to be piece-wise linear such that D8-branes can be introduced at the different jumps of its derivative, according to the expression for $F_0$ in \eqref{F0F4}.
We take the space to begin at $\rho=0$ and end at $\rho_P$, where $u$ vanishes. At the zeros of $u$ the solutions behave as 
\begin{align}\label{eq:massiveclassmetricmassiveII}
ds^2&= \frac{q}{\sqrt{x}}\bigg[ds^2(\text{AdS}_3)+ds^2(\text{S}^3)+c \,ds^2(\mathbb{T}^3) \bigg]+4c \sqrt{x}dx^2,~~~~e^{-\Phi}=\frac{x^{\frac{5}{4}}}{\sqrt{c}},
\end{align}
where $\rho=x^2$, which is the behaviour of a localised D8/O8 system on AdS$_3\times $S$^3\times \mathbb{T}^3$. 
We will then define the solutions globally by embedding them into Type I' string theory, that is, introducing O8 orientifold fixed points at both ends of the space and 16 D8-branes (together with their mirrors under $\mathbb{Z}_2$) at arbitrary positions in $\rho$. Taking $\rho_P=\rho_{17}=\pi$ and the 16 D8-branes located at arbitrary points $\rho_1,\dots, \rho_{16}$ between $\rho=0$ and $\rho_{17}=\pi$, we have that $u(\rho)$ is given by
  \begin{equation} \label{profileu}
u(\rho) = \left\{ \begin{array}{ccrcl}
                       - \frac{16c}{2\pi}\rho , &0\leq \rho_1 \\
                       \alpha_1- \frac{14c}{2\pi}(\rho-\rho_1), &\rho_1\leq \rho \leq \rho_2 \\
                       \vdots&\\
                       \alpha_k+ \frac{2c(k-8)}{2\pi}(\rho-\rho_k), &\rho_k\leq \rho \leq \rho_{k+1} \\
                       \vdots&\\
                       \alpha_{15}+\frac{14c}{2\pi}(\rho-\rho_{15}), &\rho_{15}\leq \rho\leq \rho_{16}\\
                      \alpha_{16}+ \frac{16c}{2\pi}(\rho-\pi), &\rho_{16}\leq \rho\leq \pi ,\\
                      \end{array}
\right.
\end{equation}
where, for continuity the $\alpha_k$ must satisfy 
\begin{equation}
\alpha_k=\alpha_{k-1}-\frac{2c}{2\pi}(9-k)(\rho_k-\rho_{k-1}), \qquad \text{for} \qquad k=1,\dots, 16.
\end{equation}
In turn, in order to satisfy the condition $u(\pi)=0$ the positions of the D8-branes must be such that
\begin{equation}
\sum_{k=1}^{17}(9-k)(\rho_k-\rho_{k-1})=0.
\end{equation}
Note that this is trivially satisfied when $\rho_{17-k}=\pi-\rho_k$, with $k=1,\dots, 8$, i.e. when the D8-branes are symmetrically distributed along the interval, and also when the D8-branes are equally spaced, such that $\rho_k-\rho_{k-1}=\pi/16$ for all $k$.

Besides the D8-brane charge jumping by +1 at the position of each D8-brane, we have the quantised charges
\begin{eqnarray}
&&Q_{D2}^{(k)}=\frac{1}{(2\pi)^5}\int_{\mathbb{T}^3,\text{S}^3}f_6= c\, q \label{QD2}\\
&&Q_{D4}^{(k)}=\frac{1}{(2\pi)^3}\int_{I_\rho,\text{S}^3}F_4=\frac{q}{2\pi}(\rho_{k+1}-\rho_k). \label{QD4}
\end{eqnarray}
The number of D2-branes must thus be the same in all intervals, with $c=Q_{D2}/q$, while the jump in the D4-brane charge must be given by \eqref{QD4}. 

\begin{figure}
\centering
\includegraphics[scale=0.45]{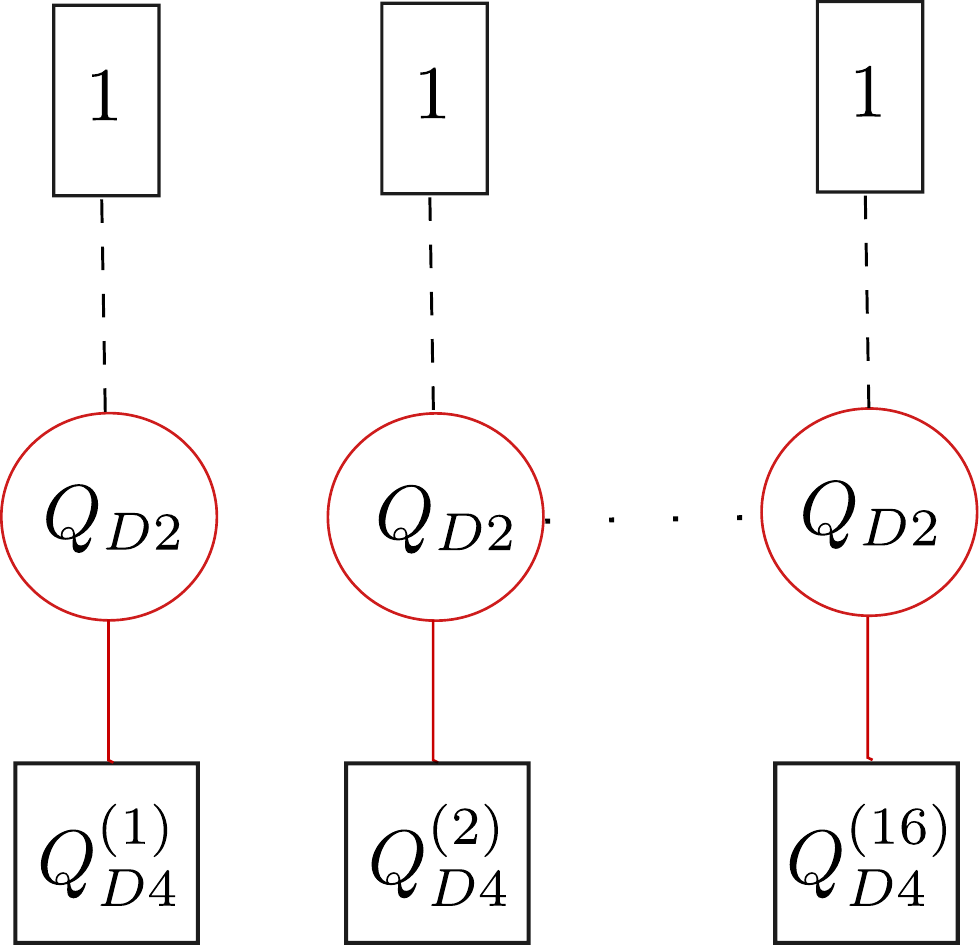}
\caption{Quiver associated to the AdS$_3\times $S$^3\times \mathbb{T}^4$ solutions in Type I'.}
\label{quiverTypeI'}
\end{figure}

With these ingredients we can proceed to construct the quiver gauge theories that flow in the IR to the CFTs dual to our solutions. In order to account for the different massless fields that build the quivers we look at the quantisation of the open strings stretched between the different branes in the brane set-up depicted in Table \ref{D2D4D8}. Following \cite{Douglas:1996uz}\footnote{In this reference the projection induced by the orientifold fixed points was carefully analysed for the Type I D1-D5 system, T-dual to our D2-D4-D8 brane set-up.} we find:
\begin{itemize}
\item D2-D2 strings: Open strings with both ends on the same stack of D2-branes give rise to $(0,4)$ SO($Q_{D2}$) vector multiplets and $(0,4)$ hypermultiplets in the symmetric representation of SO($Q_{D2}$).
\item D2-D4 strings: Open strings stretched between D2 and D4 branes give rise to $\mathcal{N}=(0,4)$ hypermultiplets in the bifundamental representation of SO($Q_{D2}) \times$ Sp($2Q_{D4}$).
\item D2-D8 strings: Open strings stretched between D2 and D8 branes give rise to $(0,2)$ Fermi multiplets in the bifundamental representation of SO($Q_{D2}) \times$ SO($Q_{D8}$).
\end{itemize}
These massless modes give rise to the $(0,4)$ disconnected quivers depicted in Figure \ref{quiverTypeI'}. 
In these quivers anomaly cancellation imposes that 
\begin{equation}
2Q_{D4}^{(k)}=\Delta Q_{D8}^{(k)}=1,
\end{equation}
as explained below equation \eqref{anomalies}. Given that D4-branes in Type I' carry $1/2$ units of charge \cite{Witten:1995gx}, in order to obtain a consistent CFT in the IR the D4-branes must be located in exactly the same positions in $\rho$ as the D8-branes. This fixes the total number of D4-branes to 16\footnote{Note that it is possible to consider the situation in which some of the $\rho_k$ coincide, such that a group of D8-D4 branes is added at that position.}. 
This condition needs to be imposed on the supergravity solution in order to describe a proper Type I' background with a well-defined 2d dual CFT.  It is likely that this condition arises as a consistency condition of the supergravity solution itself, however we leave confirmation of this for future work.

Finally, substituting \eqref{massiveT3} in \eqref{holographic-c} it is straightforward to see that the holographic central charge for this class of solutions is given by
\begin{equation}
c_{hol}= 48\, Q_{D2},
\end{equation}
and that this matches exactly the field theory result, obtained from \eqref{field-theory-c}, which gives in this case
\begin{equation}
c_R=c_L=6\sum_{k=1}^{16} Q_{D2}Q_{D4}^{(k)}= 48\, Q_{D2}.
\end{equation}

\section{Conclusions}\label{conclusions}

In this paper we have constructed a new class of AdS$_3\times $S$^3\times $M$_4$ solutions of massive Type IIA supergravity with $\mathcal{N}=(0,4)$ supersymmetries and SU(3) structure. We have then analysed separately two interesting subclasses of solutions. The first one is when M$_4=~$S$^2\times \Sigma_2$, with $\Sigma_2$ a 2d Riemann surface, and the geometry is foliated over the $\Sigma_2$. We have shown that the AdS$_3\times $S$^3\times $S$^2\times \Sigma_2$ geometries flow in the UV, asymptotically locally, to the AdS$_7\times $S$^2\times I$ geometries constructed in \cite{Apruzzi:2013yva}. This points at a possible interpretation of the solutions as describing surface defect CFTs within the 6d $(1,0)$ CFTs dual to the AdS$_7$ solutions. We have checked that this interpretation is correct by explicitly embedding the 2d $(0,4)$ quivers associated to the AdS$_3$ solutions into the 6d quivers that describe the 6d $(1,0)$ CFTs dual to the AdS$_7$ spaces. Our analysis extends\footnote{And corrects, as explained in section \ref{surfaceCFT}.}  the results in \cite{Faedo:2020nol}, where AdS$_3$ solutions dual to surface defect CFTs embedded in the 6d $(1,0)$ CFT dual to the AdS$_7$ solution to massless Type IIA supergravity \cite{Cvetic:2000cj} were constructed, allowing now for $F_0\neq0$. In our analysis we have been able to show the exact agreement between the field theory and holographic central charges, even if both quantities are divergent due to the existence of the non-compact direction inherent to the defect. Indeed, the whole point of the defect interpretation is that the presence of the non-compact direction allows to build up the AdS$_7$ geometry asymptotically and therefore to complete the non-compact AdS$_3$ solutions in the UV.

The second case that we have addressed in detail is when M$_4=\mathbb{T}^3\times I$ and the AdS$_3\times $S$^3\times \mathbb{T}^3$ geometry is foliated over the interval. We have studied separately the massless and massive cases, starting with the former. We have shown that in this case there is a supersymmetry enhancement to $(4,4)$, and that the solutions are holographically dual to 2d CFTs with 8 supercharges living in D2-D4-NS5 Hanany-Witten brane set-ups. These are the trivial extension to 2d of the 3d Hanany-Witten brane set-ups constructed in \cite{Hanany:1996ie}, and even if they were studied long ago \cite{Brodie:1997wn,Alishahiha:1997cm} the holographic duals were still missing in the literature. In this paper we have taken the first step towards filling this gap. A point of concern is that the global completion that we have found for our AdS$_3$ constructions is in terms of smeared ONS5 orientifold fixed planes. ONS5 orientifold fixed planes are perfectly well-defined objects in string theory (one way of defining them is as S-duals of O5-planes), but in our construction they are smeared  on the $\mathbb{T}^3$. As we discuss below \eqref{eq:refpoint}, it is possible that the smearing of the ONS5s is an artifact of the supergravity approximation and is resolved in string theory. The existence of our solutions with smeared ONS5s also suggests the existence of similar solutions in supergravity with localised ONS5s, as is often the case in constructions involving O-planes.  These solutions would then be holographically dual to D2-D4-NS5 Hanany-Witten brane set-ups with extra ONS5s. Such solutions are generically far harder to construct, and lie outside the scope of this work, but one can view our solutions as an important first step in this direction. 
We have shown that the embedding of our class of solutions within M-theory and Type IIB supergravity sheds some light onto some of their properties. 
The first realisation relates them to the AdS$_3\times $S$^2\times \mathbb{T}^4\times I$ solutions of massless Type IIA supergravity constructed in \cite{Lozano:2019emq} \footnote{The class of solutions in \cite{Lozano:2019emq}  is more general, since they allow for a non-vanishing Romans' mass. Here we come across the massless subclass due to the connection via M-theory.}. This realisation allows one to interpret the quiver CFTs dual to the solutions studied in this paper, which exhibit $(4,4)$ supersymmetry, and the quiver CFTs associated to the massless solutions in \cite{Lozano:2019emq}, $(0,4)$ supersymmetric, as deformations of a unique 2d $(4,4)$ CFT, which exhibits different supersymmetries depending on how it is deformed  in the UV. We have completed our analysis with a study in Type IIB string theory, where both AdS$_3$/CFT$_2$ pairs are related by S-duality. The realisation in Type IIB shows that mirror symmetry in 2d interchanges the scalars in the hypermultiplets and twisted hypermultiplets, instead of the scalars in the vector multiplets and hypermultiplets (and therefore the Coulomb and Higgs branches) as in 3d  \cite{Intriligator:1996ex,Hanany:1996ie}. That mirror symmetry can still be realised in this way in theories without a Coulomb branch is a remarkable output of our analysis. These AdS$_3$ solutions in Type IIB provide concrete examples within the broad classification of AdS$_3\times$S$^2\times$M$_5$ vacua with M$_5$ supporting an identity-structure derived in \cite{Macpherson:2022sbs}.

Finally, we have extended our study of the AdS$_3\times $S$^3\times \mathbb{T}^3\times I$ solutions by turning on a Romans' mass. We find solutions with local non-compact parts glued together with localised D8-branes, bounded between D8/O8s.  The solutions so constructed can be globally embedded within Type I' string theory allowing us to propose a dual AdS/CFT pair: We provide evidence for our proposal by comparing the central charges of the two theories, finding exact agreement.  In this case the condition for anomaly cancellation of the 2d quivers required that we impose an additional constraint on the dual supergravity background by hand - it would be interesting to reproduce this condition with a gravity computation. The solutions constructed in this section are the small ${\cal N}=(0,4)$ analogues of a similar class of geometries on AdS$_3\times $S$^3\times $S$^3\times I$ constructed in \cite{Macpherson:2018mif}. It would be interesting to explore what the CFT dual of these solutions is also, and to what extent it is similar to our proposal here.

\section*{Acknowledgements}

YL and CR are partially supported by the AEI through the Spanish grant PGC2018-096894-B-100 and by the FICYT through the Asturian grant SV-PA-21-AYUD/2021/52177. CR is supported by a Severo Ochoa Fellowship by the Principality of Asturias (Spain). NM  is also supported by the AEI through the project PID2020-114157GB-I00 and the Unidad de Excelencia Mar\'\i a de Maetzu MDM-2016-0692, and by Xunta de Galicia-Conseller\'\i a de Educaci\'on (Centro singular de investigaci\'on de Galicia accreditation 2019-2022, and project ED431C-2021/14), and the European Union FEDER. NP is supported by the Israel Science Foundation (grant No. 741/20) and by the German Research Foundation through a German-Israeli Project Cooperation (DIP) grant ''Holography and the Swampland".

\appendix

\section{Defect interpretation as a 7d domain wall}\label{appendix}

In this appendix we complement the analysis of section \ref{defectAnsatz} by showing that the coordinates in which the AdS$_7$ geometry appears asymptotically emerge naturally in a 7d domain wall solution to 7d $\mathcal{N}=1$ minimal supergravity \cite{Passias:2015gya}. 

We start considering $\ma N=1$ minimal gauged supergravity in seven dimensions \cite{Passias:2015gya}. The minimal field content (excluding the presence of vectors) is given by the gravitational field, a real scalar $X$ and a 3-form gauge potential $\ma B_{3}$. The 7d background in which we are interested was introduced in~\cite{Dibitetto:2017tve} and further studied in~\cite{Dibitetto:2017klx}. It has the following form
\begin{equation}
\begin{split}\label{7dAdS3}
& ds^2_7=e^{2U(\mu)}\left(ds^2(\text{AdS}_3)+ds^2(\text S^3) \right)+e^{2V(\mu)}d\mu^2\,,\\
&\ma B_{3}=b(\mu)\,\left(\text{vol}(\text{AdS}_3)+\text{vol}(\text S^3)\right)\,,\\
&X=X(\mu)\,.
\end{split}
\end{equation}
The BPS equations were worked out in \cite{Dibitetto:2017tve}. They are given by
\begin{equation}
 \begin{split}
  U^\prime= \frac{2}{5}\,e^{V}\,f\,,\qquad X^\prime=-\frac{2}{5}\,e^{V}\,X^2\,D_Xf\,,\qquad b^\prime=- \frac{2\,e^{2U+V}}{X^2}\,,
  \label{chargedDW7d}
 \end{split}
\end{equation}
where the BPS superpotential has the form
\begin{equation}
\label{7dsuperpotential}
f(h,\mathfrak{g},X) = \frac12 \left(h \, X^{-4} + \sqrt{2} \, \mathfrak{g} \, X \right) \,.
\end{equation}
The flow \eqref{chargedDW7d} preserves 8 real supercharges (it is BPS/2 in 7d) and in order to be consistent with the field equations has to be endowed by an odd-dimensional self-duality condition, which takes the form
\begin{equation}  \label{chargedDW7d1}
b=-\frac{e^{2U}\,X^2}{h}\,.
\end{equation}
The truncation from massive IIA performed in \cite{Passias:2015gya} requires that the two gauging parameters $\mathfrak{g}$ and $h$ respect the relation $h=\frac{\mathfrak{g}}{2\sqrt 2}$.
We anticipate that in order to derive the change of coordinates linking the aforementioned 7d geometry to the near-horizon \eqref{brane_metric_D2D4NS5D6D8_nh} one does not need the explicit solution of the 7d BPS equations \eqref{chargedDW7d}. Nevertheless we present here the solution that can be obtained by imposing the following gauge,
\begin{equation}
 e^{-V}=-\frac25\,X^2\,D_Xf\,.
\end{equation}
In this situation the BPS equations can be easily integrated out to give~\cite{Dibitetto:2017tve}
\begin{equation}
 \begin{split}
  e^{2U}= &\ 2^{-1/4}\mathfrak{g}^{-1/2}\,\left(\frac{\mu}{1-\mu^5}\right)^{1/2}\ , \qquad e^{2V}=\frac{25}{2\,\mathfrak{g}^2}\, 
  \frac{\mu^6}{\left(1- \mu^5\right)^2}\ ,\\
   b=&\ -2^{1/4}\,\mathfrak{g}^{-3/2}\,\frac{\mu^{5/2}}{(1-\mu^5)^{1/2}}\ ,\qquad \ X=\mu\ ,
   \label{chargedDWsol7d}
 \end{split}
\end{equation}
with $\mu$ running between 0 and 1. The behaviour at the boundaries is such that when  $\mu \rightarrow 1$ the domain wall \eqref{7dAdS3} is locally $\mathrm{AdS}_7$, since in this limit we have
\begin{equation}
 \begin{split}
  \ma {R}_{7}= -\frac{21}{4}\,\mathfrak{g}^2+\ma O (1-\mu)^{2}\,,\qquad X=&\ 1+\ma O (1-\mu)\ ,
  \label{UVchargedDW7d}
 \end{split}
\end{equation}
where $\mathcal{R}_{7}$ is the 7d scalar curvature. In turn, when $\mu \rightarrow 0$ the 7d spacetime exhibits a singular behaviour. 

We show now that the $\mrm{AdS}_3$ solutions given by \eqref{brane_metric_D2D4NS5D6D8_nh} can be related to the 7d domain wall geometries defined by the BPS equations \eqref{chargedDW7d}. We first consider the embedding of the 7d geometry within massive IIA. The consistent truncation has been derived in \cite{Passias:2015gya} and in what follows we will use the notation of \cite{Apruzzi:2013yva,Cremonesi:2015bld}.
The uplift of the 7d domain wall \eqref{7dAdS3} to massive IIA reads
\begin{equation}
\begin{split}
ds^2 &= \frac{16\pi}{\mathfrak{g}} \Bigl(-\frac{\alpha}{\ddot{\alpha}}\Bigr)^{1/2} X^{-1/2} \biggl[e^{2U(\mu)}\left(ds^2(\text{AdS}_3)+ds^2(\text S^3) \right)+e^{2V(\mu)}d\mu^2 \biggr] +\\
&\frac{16\pi}{\mathfrak{g}^3} X^{5/2} \biggl[ \Bigl(-\frac{\ddot{\alpha}}{\alpha}\Bigr)^{1/2} dy^2 +\Bigl(-\frac{\alpha}{\ddot{\alpha}}\Bigr)^{1/2} \frac{(-\alpha \ddot{\alpha})}{\dot{\alpha}^2-2\alpha\ddot{\alpha} X^5} ds^2(\text S^2) \bigg] \,, \\
e^{\Phi} &= \frac{3^4 2^3 \pi^{5/2}}{\mathfrak{g}^{3/2}} \frac{X^{5/4}}{(\dot{\alpha}^2-2\alpha\ddot{\alpha} X^5)^{1/2}} \Bigl(-\frac{\alpha}{\ddot{\alpha}}\Bigr)^{3/4} \,,\\
B_{2} &= \frac{2^3 \sqrt{2} \pi}{\mathfrak{g}^3} \biggl( -y + \frac{\alpha\dot{\alpha}}{\dot{\alpha}^2-2\alpha\ddot{\alpha} X^5} \biggr) \, \text{vol}(\text S^2)\,, \\
F_{2} &= -\biggl( \frac{\ddot{\alpha}}{3^4 2 \pi^2} + \frac{2^3 \sqrt{2} \pi}{\mathfrak{g}^3} \, F_{0} \, \frac{\alpha\dot{\alpha}}{\dot{\alpha}^2-2\alpha\ddot{\alpha} X^5} \biggr) \, \text{vol}(\text S^2) \\
F_{4} &= \frac{2^3}{3^4 \pi} \bigl(\ddot{\alpha} k dy+\dot{\alpha} k^\prime d\mu\bigr)\wedge \bigl(\text{vol}(\text{AdS}_3)+\text{vol}(\text S^3)  \bigr) \,, \\
\label{AdS7X}
\end{split}
\end{equation}
where $ds_7^2$, $X_7$ and $\mathcal{B}_{3}$ are the 7d fields defined in (\ref{7dAdS3}), satisfying the BPS equations \eqref{chargedDW7d}.
The function $\alpha(y)$ defines the internal geometry associated to the AdS$_7$ vacuum and was already introduced in section \ref{defectAnsatz}.
We can now relate this domain wall solution to the AdS$_3\times  $S$^3$ geometry \eqref{brane_metric_D2D4NS5D6D8_nh}. First, one shows that the near horizon geometry~\eqref{brane_metric_D2D4NS5D6D8_nh} takes the form given by~\eqref{AdS7X} if one redefines the $(\rho,r)$ coordinates in terms of the domain wall coordinates $(\mu,y)$ as
\begin{equation}
\label{changeofcoord}
\rho= \frac{8\sqrt2}{3^4 \pi \mathfrak{g} \, q} \, \dot{\alpha}\, X^{2} e^{2U}  \,,  \qquad  r = \frac{2^7}{3^4q^2\,\mathfrak{g}^2} \, \alpha \, X^{-1} e^{4U}\, ,
\end{equation}
and fixes
\begin{equation}
\label{S&K}
h = -\frac{\mathfrak{g}^2q^2e^{-4U} X }{2^{8} \pi^2} \left(\frac{\ddot{\alpha}}{\alpha}\right)\,,  \qquad  g = \frac{3^8\,2^{-6} \pi^2  q^3 X^4 e^{-6U}}{\dot{\alpha}^2 - 2\alpha \ddot{\alpha} X^5} \,.
\end{equation}
In this calculation the 7d BPS equations~\eqref{chargedDW7d} and the self-duality condition~\eqref{chargedDW7d1} need to be used, together with $h=\frac{\mathfrak{g}}{2\sqrt{2}}$. Moreover, one needs to set $\mathfrak{g}^3=2^{7/2}$ in order to match the 2-form fluxes. This is exactly the value needed to reproduce locally the AdS$_7$ solutions of \cite{Apruzzi:2013yva} in the limit $\mu\rightarrow 1$.


\begin{thebibliography}{99}

\bibitem{Tong:2014yna}
D.~Tong,
``The holographic dual of $AdS_{3} \times  S^{3} \times S^{3} \times S^{1}$,''
JHEP \textbf{04} (2014), 193
doi:10.1007/JHEP04(2014)193
[arXiv:1402.5135 [hep-th]].

\bibitem{Lozano:2015bra}
Y.~Lozano, N.~T.~Macpherson, J.~Montero and E.~\'O.~Colg\'ain,
``New $AdS_3 \times S^2$ T-duals with $ \mathcal{N}=\left(0,4\right) $ supersymmetry,''
JHEP \textbf{08} (2015), 121
doi:10.1007/JHEP08(2015)121
[arXiv:1507.02659 [hep-th]].

\bibitem{Kelekci:2016uqv}
\"O.~Kelekci, Y.~Lozano, J.~Montero, E.~\'O.~Colg\'ain and M.~Park,
``Large superconformal near-horizons from M-theory,''
Phys. Rev. D \textbf{93} (2016) no.8, 086010
doi:10.1103/PhysRevD.93.086010
[arXiv:1602.02802 [hep-th]].

\bibitem{Couzens:2017way}
C.~Couzens, C.~Lawrie, D.~Martelli, S.~Schafer-Nameki and J.~M.~Wong,
``F-theory and AdS$_{3}$/CFT$_{2}$,''
JHEP \textbf{08} (2017), 043
doi:10.1007/JHEP08(2017)043
[arXiv:1705.04679 [hep-th]].

\bibitem{Eberhardt:2017pty}
L.~Eberhardt, M.~R.~Gaberdiel and W.~Li,
``A holographic dual for string theory on AdS$_{3}\times$S$^{3} \times$S$^{3} \times$S$^{1}$,''
JHEP \textbf{08} (2017), 111
[arXiv:1707.02705 [hep-th]].

\bibitem{Dibitetto:2017tve}
G.~Dibitetto and N.~Petri,
``BPS objects in D = 7 supergravity and their M-theory origin,''
JHEP \textbf{12} (2017), 041
[arXiv:1707.06152 [hep-th]].

\bibitem{Dibitetto:2017klx}
G.~Dibitetto and N.~Petri,
``6d surface defects from massive type IIA,''
JHEP \textbf{01} (2018), 039
[arXiv:1707.06154 [hep-th]].

\bibitem{Dibitetto:2018iar}
G.~Dibitetto and N.~Petri,
``Surface defects in the D4 $-$ D8 brane system,''
JHEP \textbf{01} (2019), 193
doi:10.1007/JHEP01(2019)193
[arXiv:1807.07768 [hep-th]].

\bibitem{Datta:2017ert}
S.~Datta, L.~Eberhardt and M.~R.~Gaberdiel,
``Stringy $\mathcal{N}=(2,2)$ holography for AdS${_3}$,''
JHEP \textbf{01} (2018), 146
doi:10.1007/JHEP01(2018)146
[arXiv:1709.06393 [hep-th]].

\bibitem{Corbino:2017tfl}
D.~Corbino, E.~D'Hoker and C.~F.~Uhlemann,
``AdS$_{2} \times $S$^{6}$ versus AdS$_{6}\times$S$^{2}$ in Type IIB supergravity,''
JHEP \textbf{03} (2018), 120
doi:10.1007/JHEP03(2018)120
[arXiv:1712.04463 [hep-th]].

\bibitem{Couzens:2017nnr}
C.~Couzens, D.~Martelli and S.~Schafer-Nameki,
``F-theory and AdS$_{3}$/CFT$_{2}$ (2, 0),''
JHEP \textbf{06} (2018), 008
doi:10.1007/JHEP06(2018)008
[arXiv:1712.07631 [hep-th]].

\bibitem{Gaberdiel:2018rqv}
M.~R.~Gaberdiel and R.~Gopakumar,
``Tensionless string spectra on AdS$_{3}$,''
JHEP \textbf{05} (2018), 085
doi:10.1007/JHEP05(2018)085
[arXiv:1803.04423 [hep-th]].

\bibitem{Eberhardt:2018sce}
L.~Eberhardt and I.~G.~Zadeh,
``$\mathcal{N}=(3,3)$ holography on ${\rm AdS}_3 \times ({\rm S}^3 \times {\rm S}^3 \times {\rm S}^1)/\mathbb Z_2$,''
JHEP \textbf{07} (2018), 143
doi:10.1007/JHEP07(2018)143
[arXiv:1805.09832 [hep-th]].

\bibitem{Dibitetto:2018gbk}
G.~Dibitetto and A.~Passias,
``AdS$_{2}\times$ S$^{7}$ solutions from D0-F1-D8 intersections,''
JHEP \textbf{10} (2018), 190
doi:10.1007/JHEP10(2018)190
[arXiv:1807.00555 [hep-th]].

\bibitem{Dibitetto:2018ftj}
G.~Dibitetto, G.~Lo Monaco, A.~Passias, N.~Petri and A.~Tomasiello,
``AdS$_3$ Solutions with Exceptional Supersymmetry,''
Fortsch. Phys. \textbf{66} (2018) no.10, 1800060
doi:10.1002/prop.201800060
[arXiv:1807.06602 [hep-th]].

\bibitem{Dibitetto:2018gtk}
G.~Dibitetto and N.~Petri,
``AdS$_{2}$ solutions and their massive IIA origin,''
JHEP \textbf{05} (2019), 107
doi:10.1007/JHEP05(2019)107
[arXiv:1811.11572 [hep-th]].

\bibitem{Eberhardt:2018ouy}
L.~Eberhardt, M.~R.~Gaberdiel and R.~Gopakumar,
``The Worldsheet Dual of the Symmetric Product CFT,''
JHEP \textbf{04} (2019), 103
doi:10.1007/JHEP04(2019)103
[arXiv:1812.01007 [hep-th]].

\bibitem{Corbino:2018fwb}
D.~Corbino, E.~D'Hoker, J.~Kaidi and C.~F.~Uhlemann,
``Global half-BPS $AdS_2\times S^6$ solutions in Type IIB,''
JHEP \textbf{03} (2019), 039
doi:10.1007/JHEP03(2019)039
[arXiv:1812.10206 [hep-th]].

\bibitem{Macpherson:2018mif}
N.~T.~Macpherson,
``Type II solutions on AdS$_{3} \times$ S$^{3} \times$ S$^{3}$ with large superconformal symmetry,''
JHEP \textbf{05} (2019), 089
doi:10.1007/JHEP05(2019)089
[arXiv:1812.10172 [hep-th]].

\bibitem{Hong:2019wyi}
J.~Hong, N.~T.~Macpherson and L.~A.~Pando Zayas,
``Aspects of AdS$_{2}$ classification in M-theory: solutions with mesonic and baryonic charges,''
JHEP \textbf{11} (2019), 127
doi:10.1007/JHEP11(2019)127
[arXiv:1908.08518 [hep-th]].

\bibitem{Lozano:2019emq}
Y.~Lozano, N.~T.~Macpherson, C.~Nunez and A.~Ramirez,
``AdS$_3$ solutions in Massive IIA with small $\mathcal{N}=(4,0)$ supersymmetry,''
JHEP \textbf{01} (2020), 129
doi:10.1007/JHEP01(2020)129
[arXiv:1908.09851 [hep-th]].

\bibitem{Lozano:2019jza}
Y.~Lozano, N.~T.~Macpherson, C.~Nunez and A.~Ramirez,
``1/4 BPS solutions and the AdS$_3$/CFT$_2$ correspondence,''
Phys. Rev. D \textbf{101} (2020) no.2, 026014
doi:10.1103/PhysRevD.101.026014
[arXiv:1909.09636 [hep-th]].

\bibitem{Lozano:2019zvg}
Y.~Lozano, N.~T.~Macpherson, C.~Nunez and A.~Ramirez,
``Two dimensional ${\cal N}=(0,4)$ quivers dual to AdS$_3$ solutions in massive IIA,''
JHEP \textbf{01} (2020), 140
doi:10.1007/JHEP01(2020)140
[arXiv:1909.10510 [hep-th]].

\bibitem{Lozano:2019ywa}
Y.~Lozano, N.~T.~Macpherson, C.~Nunez and A.~Ramirez,
``AdS$_3$ solutions in massive IIA, defect CFTs and T-duality,''
JHEP \textbf{12} (2019), 013
doi:10.1007/JHEP12(2019)013
[arXiv:1909.11669 [hep-th]].

\bibitem{Passias:2019rga}
A.~Passias and D.~Prins,
``On AdS$_3$ solutions of Type IIB,''
JHEP \textbf{05} (2020), 048
doi:10.1007/JHEP05(2020)048
[arXiv:1910.06326 [hep-th]].

\bibitem{Eberhardt:2019ywk}
L.~Eberhardt, M.~R.~Gaberdiel and R.~Gopakumar,
``Deriving the AdS$_{3}$/CFT$_{2}$ correspondence,''
JHEP \textbf{02} (2020), 136
doi:10.1007/JHEP02(2020)136
[arXiv:1911.00378 [hep-th]].

\bibitem{Couzens:2019iog}
C.~Couzens,
``$ \mathcal{N} $ = (0, 2) AdS$_{3}$ solutions of type IIB and F-theory with generic fluxes,''
JHEP \textbf{04} (2021), 038
doi:10.1007/JHEP04(2021)038
[arXiv:1911.04439 [hep-th]].

\bibitem{Couzens:2019mkh}
C.~Couzens, H.~het Lam and K.~Mayer,
``Twisted $ \mathcal{N} $ = 1 SCFTs and their AdS$_{3}$ duals,''
JHEP \textbf{03} (2020), 032
doi:10.1007/JHEP03(2020)032
[arXiv:1912.07605 [hep-th]].

\bibitem{Dibitetto:2019nyz}
G.~Dibitetto, Y.~Lozano, N.~Petri and A.~Ramirez,
``Holographic description of M-branes via AdS$_{2}$,''
JHEP \textbf{04} (2020), 037
doi:10.1007/JHEP04(2020)037
[arXiv:1912.09932 [hep-th]].

\bibitem{Legramandi:2019xqd}
A.~Legramandi and N.~T.~Macpherson,
``AdS$_3$ solutions with from $\mathcal{N}=(3,0)$ from S$^3\times$S$^3$  fibrations,''
Fortsch. Phys. \textbf{68} (2020) no.3-4, 2000014
doi:10.1002/prop.202000014
[arXiv:1912.10509 [hep-th]].

\bibitem{Lust:2020npd}
D.~L\"ust and D.~Tsimpis,
``AdS$_{2}$ type-IIA solutions and scale separation,''
JHEP \textbf{07} (2020), 060
doi:10.1007/JHEP07(2020)060
[arXiv:2004.07582 [hep-th]].

\bibitem{Corbino:2020lzq}
D.~Corbino,
``Warped AdS$_{2}$ and SU(1, 1|4) symmetry in Type IIB,''
JHEP \textbf{03} (2021), 060
doi:10.1007/JHEP03(2021)060
[arXiv:2004.12613 [hep-th]].

\bibitem{Chen:2020mtv}
K.~Chen, M.~Gutperle and M.~Vicino,
``Holographic Line Defects in $D=4$, $N=2$ Gauged Supergravity,''
Phys. Rev. D \textbf{102} (2020) no.2, 026025
doi:10.1103/PhysRevD.102.026025
[arXiv:2005.03046 [hep-th]].

\bibitem{Lozano:2020bxo}
Y.~Lozano, C.~Nunez, A.~Ramirez and S.~Speziali,
``$M$-strings and AdS$_3$ solutions to M-theory with small $\mathcal{N}=(0,4)$ supersymmetry,''
JHEP \textbf{08} (2020), 118
doi:10.1007/JHEP08(2020)118
[arXiv:2005.06561 [hep-th]].

\bibitem{Faedo:2020nol}
F.~Faedo, Y.~Lozano and N.~Petri,
``Searching for surface defect CFTs within AdS$_3$,''
JHEP \textbf{11} (2020), 052
doi:10.1007/JHEP11(2020)052
[arXiv:2007.16167 [hep-th]].

\bibitem{Dibitetto:2020bsh}
G.~Dibitetto and N.~Petri,
``AdS$_{3}$ from M-branes at conical singularities,''
JHEP \textbf{01} (2021), 129
doi:10.1007/JHEP01(2021)129
[arXiv:2010.12323 [hep-th]].

\bibitem{Lozano:2020txg}
Y.~Lozano, C.~Nunez, A.~Ramirez and S.~Speziali,
``New AdS$_{2}$ backgrounds and $ \mathcal{N} $ = 4 conformal quantum mechanics,''
JHEP \textbf{03} (2021), 277
doi:10.1007/JHEP03(2021)277
[arXiv:2011.00005 [hep-th]].

\bibitem{Lozano:2020sae}
Y.~Lozano, C.~Nunez, A.~Ramirez and S.~Speziali,
``AdS$_{2}$ duals to ADHM quivers with Wilson lines,''
JHEP \textbf{03} (2021), 145
doi:10.1007/JHEP03(2021)145
[arXiv:2011.13932 [hep-th]].

\bibitem{Passias:2020ubv}
A.~Passias and D.~Prins,
``On supersymmetric AdS$_{3}$ solutions of Type II,''
JHEP \textbf{08} (2021), 168
doi:10.1007/JHEP08(2021)168
[arXiv:2011.00008 [hep-th]].

\bibitem{Faedo:2020lyw}
F.~Faedo, Y.~Lozano and N.~Petri,
``New $\mathcal{N}=(0,4)$ AdS$_3$ near-horizons in Type IIB,''
JHEP \textbf{04} (2021), 028
doi:10.1007/JHEP04(2021)028
[arXiv:2012.07148 [hep-th]].

\bibitem{Legramandi:2020txf}
A.~Legramandi, G.~Lo Monaco and N.~T.~Macpherson,
``All $\mathcal{N}=(8,0)$ AdS$_3$ solutions in 10 and 11 dimensions,''
JHEP \textbf{05} (2021), 263
doi:10.1007/JHEP05(2021)263f
[arXiv:2012.10507 [hep-th]].

\bibitem{Lozano:2021rmk}
Y.~Lozano, C.~Nunez and A.~Ramirez,
``$\text{AdS}_2\times \text{S}^2\times \text{CY}_2$ solutions in Type IIB with 8 supersymmetries,''
JHEP \textbf{04} (2021), 110
doi:10.1007/JHEP04(2021)110
[arXiv:2101.04682 [hep-th]].

\bibitem{Ramirez:2021tkd}
A.~Ramirez,
``AdS$_{2}$ geometries and non-Abelian T-duality in non-compact spaces,''
JHEP \textbf{10} (2021), 020
doi:10.1007/JHEP10(2021)020
[arXiv:2106.09735 [hep-th]].

\bibitem{Lozano:2021fkk}
Y.~Lozano, N.~Petri and C.~Risco,
``New AdS$_{2}$ supergravity duals of 4d SCFTs with defects,''
JHEP \textbf{10} (2021), 217
doi:10.1007/JHEP10(2021)217
[arXiv:2107.12277 [hep-th]].

\bibitem{Couzens:2021tnv}
C.~Couzens, N.~T.~Macpherson and A.~Passias,
``$ \mathcal{N} $ = (2, 2) AdS$_{3}$ from D3-branes wrapped on Riemann surfaces,''
JHEP \textbf{02} (2022), 189
doi:10.1007/JHEP02(2022)189
[arXiv:2107.13562 [hep-th]].

\bibitem{Couzens:2021veb}
C.~Couzens, Y.~Lozano, N.~Petri and S.~Vandoren,
``N=(0,4) black string chains,''
Phys. Rev. D \textbf{105} (2022) no.8, 086015
doi:10.1103/PhysRevD.105.086015
[arXiv:2109.10413 [hep-th]].

\bibitem{Macpherson:2021lbr}
N.~T.~Macpherson and A.~Tomasiello,
``$ \mathcal{N} $ = (1, 1) supersymmetric AdS$_{3}$ in 10 dimensions,''
JHEP \textbf{03} (2022), 112
doi:10.1007/JHEP03(2022)112
[arXiv:2110.01627 [hep-th]].

\bibitem{Macpherson:2022sbs}
N.~T.~Macpherson and A.~Ramirez,
``AdS$_{3}\times$S$^{2}$ in IIB with small $ \mathcal{N} $ = (4, 0) supersymmetry,''
JHEP \textbf{04} (2022), 143
doi:10.1007/JHEP04(2022)143
[arXiv:2202.00352 [hep-th]].

\bibitem{Couzens:2022agr}
C.~Couzens, N.~T.~Macpherson and A.~Passias,
``On Type IIA AdS$_3$ solutions and massive GK geometries,''
[arXiv:2203.09532 [hep-th]].

\bibitem{Strominger:1996sh}
A.~Strominger and C.~Vafa,
``Microscopic origin of the Bekenstein-Hawking entropy,''
Phys. Lett. B \textbf{379} (1996), 99-104
doi:10.1016/0370-2693(96)00345-0
[arXiv:hep-th/9601029 [hep-th]].

\bibitem{Minasian:1999qn}
R.~Minasian, G.~W.~Moore and D.~Tsimpis,
``Calabi-Yau black holes and (0,4) sigma models,''
Commun. Math. Phys. \textbf{209} (2000), 325-352
[arXiv:hep-th/9904217 [hep-th]].

\bibitem{Karch:2000gx}
A.~Karch and L.~Randall,
``Open and closed string interpretation of SUSY CFT's on branes with boundaries,''
JHEP \textbf{06} (2001), 063
doi:10.1088/1126-6708/2001/06/063
[arXiv:hep-th/0105132 [hep-th]].

\bibitem{DeWolfe:2001pq}
O.~DeWolfe, D.~Z.~Freedman and H.~Ooguri,
``Holography and defect conformal field theories,''
Phys. Rev. D \textbf{66} (2002), 025009
doi:10.1103/PhysRevD.66.025009
[arXiv:hep-th/0111135 [hep-th]].

\bibitem{Aharony:2003qf}
O.~Aharony, O.~DeWolfe, D.~Z.~Freedman and A.~Karch,
``Defect conformal field theory and locally localized gravity,''
JHEP \textbf{07} (2003), 030
doi:10.1088/1126-6708/2003/07/030
[arXiv:hep-th/0303249 [hep-th]].

\bibitem{DHoker:2006vfr}
E.~D'Hoker, J.~Estes and M.~Gutperle,
``Ten-dimensional supersymmetric Janus solutions,''
Nucl. Phys. B \textbf{757} (2006), 79-116
doi:10.1016/j.nuclphysb.2006.08.017
[arXiv:hep-th/0603012 [hep-th]].

\bibitem{Lunin:2007ab}
O.~Lunin,
``1/2-BPS states in M theory and defects in the dual CFTs,''
JHEP \textbf{10} (2007), 014
doi:10.1088/1126-6708/2007/10/014
[arXiv:0704.3442 [hep-th]].

\bibitem{DHoker:2007mci}
E.~D'Hoker, J.~Estes and M.~Gutperle,
``Gravity duals of half-BPS Wilson loops,''
JHEP \textbf{06} (2007), 063
doi:10.1088/1126-6708/2007/06/063
[arXiv:0705.1004 [hep-th]].

\bibitem{Chiodaroli:2009yw}
M.~Chiodaroli, M.~Gutperle and D.~Krym,
``Half-BPS Solutions locally asymptotic to AdS(3) x S**3 and interface conformal field theories,''
JHEP \textbf{02} (2010), 066
doi:10.1007/JHEP02(2010)066
[arXiv:0910.0466 [hep-th]].

\bibitem{Chiodaroli:2009xh}
M.~Chiodaroli, E.~D'Hoker and M.~Gutperle,
``Open Worldsheets for Holographic Interfaces,''
JHEP \textbf{03} (2010), 060
doi:10.1007/JHEP03(2010)060
[arXiv:0912.4679 [hep-th]].

\bibitem{Maldacena:1998uz}
J.~M.~Maldacena, J.~Michelson and A.~Strominger,
``Anti-de Sitter fragmentation,''
JHEP \textbf{02} (1999), 011
doi:10.1088/1126-6708/1999/02/011
[arXiv:hep-th/9812073 [hep-th]].

\bibitem{Strominger:1998yg}
A.~Strominger,
``AdS(2) quantum gravity and string theory,''
JHEP \textbf{01} (1999), 007
doi:10.1088/1126-6708/1999/01/007
[arXiv:hep-th/9809027 [hep-th]].

\bibitem{Balasubramanian:2003kq}
V.~Balasubramanian, A.~Naqvi and J.~Simon,
``A Multiboundary AdS orbifold and DLCQ holography: A Universal holographic description of extremal black hole horizons,''
JHEP \textbf{08} (2004), 023
doi:10.1088/1126-6708/2004/08/023
[arXiv:hep-th/0311237 [hep-th]].

\bibitem{Hartman:2008dq}
T.~Hartman and A.~Strominger,
``Central Charge for AdS(2) Quantum Gravity,''
JHEP \textbf{04} (2009), 026
doi:10.1088/1126-6708/2009/04/026
[arXiv:0803.3621 [hep-th]].

\bibitem{Balasubramanian:2009bg}
V.~Balasubramanian, J.~de Boer, M.~M.~Sheikh-Jabbari and J.~Simon,
``What is a chiral 2d CFT? And what does it have to do with extremal black holes?,''
JHEP \textbf{02} (2010), 017
doi:10.1007/JHEP02(2010)017
[arXiv:0906.3272 [hep-th]].

\bibitem{Almheiri:2014cka}
A.~Almheiri and J.~Polchinski,
``Models of AdS$_{2}$ backreaction and holography,''
JHEP \textbf{11} (2015), 014
doi:10.1007/JHEP11(2015)014
[arXiv:1402.6334 [hep-th]].

\bibitem{Maldacena:2016hyu}
J.~Maldacena and D.~Stanford,
``Remarks on the Sachdev-Ye-Kitaev model,''
Phys. Rev. D \textbf{94} (2016) no.10, 106002
doi:10.1103/PhysRevD.94.106002
[arXiv:1604.07818 [hep-th]].

\bibitem{Maldacena:2016upp}
J.~Maldacena, D.~Stanford and Z.~Yang,
``Conformal symmetry and its breaking in two dimensional Nearly Anti-de-Sitter space,''
PTEP \textbf{2016} (2016) no.12, 12C104
doi:10.1093/ptep/ptw124
[arXiv:1606.01857 [hep-th]].

\bibitem{Harlow:2018tqv}
D.~Harlow and D.~Jafferis,
``The Factorization Problem in Jackiw-Teitelboim Gravity,''
JHEP \textbf{02} (2020), 177
doi:10.1007/JHEP02(2020)177
[arXiv:1804.01081 [hep-th]].

\bibitem{Maldacena:1997re}
J.~M.~Maldacena,
``The Large N limit of superconformal field theories and supergravity,''
Adv. Theor. Math. Phys. \textbf{2} (1998), 231-252
doi:10.1023/A:1026654312961
[arXiv:hep-th/9711200 [hep-th]].

\bibitem{Lin:2004nb}
H.~Lin, O.~Lunin and J.~M.~Maldacena,
``Bubbling AdS space and 1/2 BPS geometries,''
JHEP \textbf{10} (2004), 025
doi:10.1088/1126-6708/2004/10/025
[arXiv:hep-th/0409174 [hep-th]].

\bibitem{Legramandi:2019ulq}
A.~Legramandi and A.~Tomasiello,
``Breaking supersymmetry with pure spinors,''
JHEP \textbf{11} (2020), 098
doi:10.1007/JHEP11(2020)098
[arXiv:1912.00001 [hep-th]].

\bibitem{Imamura:2001cr}
Y.~Imamura,
``1/4 BPS solutions in massive IIA supergravity,''
Prog. Theor. Phys. \textbf{106} (2001), 653-670
doi:10.1143/PTP.106.653
[arXiv:hep-th/0105263 [hep-th]].

\bibitem{Bena:2015bea}
I.~Bena, S.~Giusto, R.~Russo, M.~Shigemori and N.~P.~Warner,
``Habemus Superstratum! A constructive proof of the existence of superstrata,''
JHEP \textbf{05} (2015), 110
[arXiv:1503.01463 [hep-th]].

\bibitem{Apruzzi:2013yva}
F.~Apruzzi, M.~Fazzi, D.~Rosa and A.~Tomasiello,
``All AdS$_7$ solutions of type II supergravity,''
JHEP \textbf{04} (2014), 064
doi:10.1007/JHEP04(2014)064
[arXiv:1309.2949 [hep-th]].

\bibitem{Gaiotto:2014lca}
D.~Gaiotto and A.~Tomasiello,
``Holography for (1,0) theories in six dimensions,''
JHEP \textbf{12} (2014), 003
doi:10.1007/JHEP12(2014)003
[arXiv:1404.0711 [hep-th]].

\bibitem{Cremonesi:2015bld}
S.~Cremonesi and A.~Tomasiello,
``6d holographic anomaly match as a continuum limit,''
JHEP \textbf{05} (2016), 031
doi:10.1007/JHEP05(2016)031
[arXiv:1512.02225 [hep-th]].

\bibitem{Brodie:1997wn}
J.~H.~Brodie,
``Two-dimensional mirror symmetry from M theory,''
Nucl. Phys. B \textbf{517} (1998), 36-52
doi:10.1016/S0550-3213(97)00755-4
[arXiv:hep-th/9709228 [hep-th]].

\bibitem{Alishahiha:1997cm}
M.~Alishahiha,
``N=(4,4) 2-D supersymmetric gauge theory and brane configuration,''
Phys. Lett. B \textbf{420} (1998), 51-54
[arXiv:hep-th/9710020 [hep-th]];
M.~Alishahiha,
``On the brane configuration of N=(4,4) 2-D supersymmetric gauge theories,''
Nucl. Phys. B \textbf{528} (1998), 171-184
[arXiv:hep-th/9802151 [hep-th]].

\bibitem{us}
Y.~Lozano, N.T.~Macpherson, N.~Petri, C.~Risco, ``New AdS$_2$ solutions to massive IIA with $\mathcal{N}=4$ supersymmetries.''

\bibitem{Macpherson:2016xwk}
N.~T.~Macpherson and A.~Tomasiello,
``Minimal flux Minkowski classification,''
JHEP \textbf{09} (2017), 126
doi:10.1007/JHEP09(2017)126
[arXiv:1612.06885 [hep-th]].

\bibitem{Bobev:2016phc}
N.~Bobev, G.~Dibitetto, F.~F.~Gautason and B.~Truijen,
``Holography, Brane Intersections and Six-dimensional SCFTs,''
JHEP \textbf{02} (2017), 116
doi:10.1007/JHEP02(2017)116
[arXiv:1612.06324 [hep-th]].

\bibitem{Hristov:2014eba}
K.~Hristov and A.~Rota,
``6d-5d-4d reduction of BPS attractors in flat gauged supergravities,''
Nucl. Phys. B \textbf{897} (2015), 213-228
doi:10.1016/j.nuclphysb.2015.05.023
[arXiv:1410.5386 [hep-th]].

\bibitem{Gutowski:2003rg}
J.~B.~Gutowski, D.~Martelli and H.~S.~Reall,
``All Supersymmetric solutions of minimal supergravity in six- dimensions,''
Class. Quant. Grav. \textbf{20} (2003), 5049-5078
doi:10.1088/0264-9381/20/23/008
[arXiv:hep-th/0306235 [hep-th]].

\bibitem{Passias:2015gya}
A.~Passias, A.~Rota and A.~Tomasiello,
``Universal consistent truncation for 6d/7d gauge/gravity duals,''
JHEP \textbf{10} (2015), 187
doi:10.1007/JHEP10(2015)187
[arXiv:1506.05462 [hep-th]].

\bibitem{Nunez:2018ags}
C.~N\'u\~nez, J.~M.~Pen\'\i{}n, D.~Roychowdhury and J.~Van Gorsel,
``The non-Integrability of Strings in Massive Type IIA and their Holographic duals,''
JHEP \textbf{06} (2018), 078
doi:10.1007/JHEP06(2018)078
[arXiv:1802.04269 [hep-th]].

\bibitem{Brunner:1997gk}
I.~Brunner and A.~Karch,
``Branes and six-dimensional fixed points,''
Phys. Lett. B \textbf{409} (1997), 109-116
doi:10.1016/S0370-2693(97)00935-0
[arXiv:hep-th/9705022 [hep-th]].

\bibitem{Hanany:1997sa}
A.~Hanany and A.~Zaffaroni,
``Chiral symmetry from type IIA branes,''
Nucl. Phys. B \textbf{509} (1998), 145-168
doi:10.1016/S0550-3213(97)00595-6
[arXiv:hep-th/9706047 [hep-th]].

\bibitem{Brown:1986nw}
J.~D.~Brown and M.~Henneaux,
``Central Charges in the Canonical Realization of Asymptotic Symmetries: An Example from Three-Dimensional Gravity,''
Commun. Math. Phys. \textbf{104} (1986), 207-226
doi:10.1007/BF01211590

\bibitem{Diaconescu:1997gu}
D.~E.~Diaconescu and N.~Seiberg,
``The Coulomb branch of (4,4) supersymmetric field theories in two-dimensions,''
JHEP \textbf{07} (1997), 001
doi:10.1088/1126-6708/1997/07/001
[arXiv:hep-th/9707158 [hep-th]].

\bibitem{Witten:1997yu}
E.~Witten,
``On the conformal field theory of the Higgs branch,''
JHEP \textbf{07} (1997), 003
doi:10.1088/1126-6708/1997/07/003
[arXiv:hep-th/9707093 [hep-th]].

\bibitem{Aharony:1999dw}
O.~Aharony and M.~Berkooz,
``IR dynamics of D = 2, N=(4,4) gauge theories and DLCQ of 'little string theories',''
JHEP \textbf{10} (1999), 030
doi:10.1088/1126-6708/1999/10/030
[arXiv:hep-th/9909101 [hep-th]].

\bibitem{Kraus:2005zm}
P.~Kraus and F.~Larsen,
``Holographic gravitational anomalies,''
JHEP \textbf{01} (2006), 022
doi:10.1088/1126-6708/2006/01/022
[arXiv:hep-th/0508218 [hep-th]].

\bibitem{Hanany:1996ie}
A.~Hanany and E.~Witten,
``Type IIB superstrings, BPS monopoles, and three-dimensional gauge dynamics,''
Nucl. Phys. B \textbf{492} (1997), 152-190
doi:10.1016/S0550-3213(97)00157-0
[arXiv:hep-th/9611230 [hep-th]].

\bibitem{Apruzzi:2015wna}
F.~Apruzzi, M.~Fazzi, A.~Passias, A.~Rota and A.~Tomasiello,
``Six-Dimensional Superconformal Theories and their Compactifications from Type IIA Supergravity,''
Phys. Rev. Lett. \textbf{115} (2015) no.6, 061601
doi:10.1103/PhysRevLett.115.061601
[arXiv:1502.06616 [hep-th]].

\bibitem{Brandhuber:1999np}
A.~Brandhuber and Y.~Oz,
``The D-4 - D-8 brane system and five-dimensional fixed points,''
Phys. Lett. B \textbf{460} (1999), 307-312
doi:10.1016/S0370-2693(99)00763-7
[arXiv:hep-th/9905148 [hep-th]].

\bibitem{Bergman:2012kr}
O.~Bergman and D.~Rodriguez-Gomez,
``5d quivers and their AdS(6) duals,''
JHEP \textbf{07} (2012), 171
doi:10.1007/JHEP07(2012)171
[arXiv:1206.3503 [hep-th]].

\bibitem{DHoker:2016ujz} 
  E.~D'Hoker, M.~Gutperle, A.~Karch and C.~F.~Uhlemann,
  ``Warped $AdS_6\times S^2$ in Type IIB supergravity I: Local solutions,''
  JHEP {\bf 1608}, 046 (2016)
  [arXiv:1606.01254 [hep-th]].
 E.~D'Hoker, M.~Gutperle and C.~F.~Uhlemann,
  ``Holographic duals for five-dimensional superconformal quantum field theories,''
  Phys.\ Rev.\ Lett.\  {\bf 118}, no. 10, 101601 (2017)
  [arXiv:1611.09411 [hep-th]].
  E.~D'Hoker, M.~Gutperle and C.~F.~Uhlemann,
  ``Warped $AdS_6\times S^2$ in Type IIB supergravity II: Global solutions and five-brane webs,''
  JHEP {\bf 1705}, 131 (2017)
  [arXiv:1703.08186 [hep-th]].
  M.~Gutperle, C.~Marasinou, A.~Trivella and C.~F.~Uhlemann,
``Entanglement entropy vs. free energy in IIB supergravity duals for 5d SCFTs,''
JHEP \textbf{09} (2017), 125
[arXiv:1705.01561 [hep-th]].
  E.~D'Hoker, M.~Gutperle and C.~F.~Uhlemann,
``Warped $AdS_6\times S^2$ in Type IIB supergravity III: Global solutions with seven-branes,''
JHEP \textbf{11} (2017), 200
[arXiv:1706.00433 [hep-th]].


\bibitem{Lozano:2018pcp}
Y.~Lozano, N.~T.~Macpherson and J.~Montero,
``AdS$_{6}$ T-duals and type IIB AdS$_{6} \times$ S$^{2}$ geometries with 7-branes,''
JHEP \textbf{01} (2019), 116
doi:10.1007/JHEP01(2019)116
[arXiv:1810.08093 [hep-th]].

\bibitem{Gaiotto:2009gz}
D.~Gaiotto and J.~Maldacena,
``The Gravity duals of N=2 superconformal field theories,''
JHEP \textbf{10} (2012), 189
doi:10.1007/JHEP10(2012)189
[arXiv:0904.4466 [hep-th]].

\bibitem{ReidEdwards:2010qs}
R.~A.~Reid-Edwards and B.~Stefanski, jr.,
``On Type IIA geometries dual to N = 2 SCFTs,''
Nucl. Phys. B \textbf{849} (2011), 549-572
doi:10.1016/j.nuclphysb.2011.04.002
[arXiv:1011.0216 [hep-th]].

\bibitem{Aharony:2012tz}
O.~Aharony, L.~Berdichevsky and M.~Berkooz,
``4d N=2 superconformal linear quivers with type IIA duals,''
JHEP \textbf{08} (2012), 131
doi:10.1007/JHEP08(2012)131
[arXiv:1206.5916 [hep-th]].

\bibitem{Assel:2011xz}
B.~Assel, C.~Bachas, J.~Estes and J.~Gomis,
``Holographic Duals of D=3 N=4 Superconformal Field Theories,''
JHEP \textbf{08} (2011), 087
doi:10.1007/JHEP08(2011)087
[arXiv:1106.4253 [hep-th]].

\bibitem{Chen:2006ps}
H.~Y.~Chen and D.~Tong,
``Instantons and Emergent AdS(3) x S**3 Geometry,''
JHEP \textbf{06} (2006), 017
doi:10.1088/1126-6708/2006/06/017
[arXiv:hep-th/0604090 [hep-th]].

\bibitem{Hanany:2018hlz}
A.~Hanany and T.~Okazaki,
``(0,4) brane box models,''
JHEP \textbf{03} (2019), 027
doi:10.1007/JHEP03(2019)027
[arXiv:1811.09117 [hep-th]].

\bibitem{Baines:2020dmu}
S.~Baines and T.~Van Riet,
``Smearing orientifolds in flux compactifications can be OK,''
Class. Quant. Grav. \textbf{37} (2020) no.19, 195015
doi:10.1088/1361-6382/aba8e0
[arXiv:2005.09501 [hep-th]].

\bibitem{Gaiotto:2008ak}
D.~Gaiotto and E.~Witten,
``S-Duality of Boundary Conditions In N=4 Super Yang-Mills Theory,''
Adv. Theor. Math. Phys. \textbf{13} (2009) no.3, 721-896
doi:10.4310/ATMP.2009.v13.n3.a5
[arXiv:0807.3720 [hep-th]].

\bibitem{Intriligator:1996ex}
K.~A.~Intriligator and N.~Seiberg,
``Mirror symmetry in three-dimensional gauge theories,''
Phys. Lett. B \textbf{387} (1996), 513-519
doi:10.1016/0370-2693(96)01088-X
[arXiv:hep-th/9607207 [hep-th]].

\bibitem{Douglas:1996uz}
M.~R.~Douglas,
``Gauge fields and D-branes,''
J. Geom. Phys. \textbf{28} (1998), 255-262
doi:10.1016/S0393-0440(97)00024-7
[arXiv:hep-th/9604198 [hep-th]].

\bibitem{Witten:1995gx}
E.~Witten,
``Small instantons in string theory,''
Nucl. Phys. B \textbf{460} (1996), 541-559
doi:10.1016/0550-3213(95)00625-7
[arXiv:hep-th/9511030 [hep-th]].

\bibitem{Cvetic:2000cj}
M.~Cvetic, H.~Lu, C.~N.~Pope and J.~F.~Vazquez-Poritz,
``AdS in warped space-times,''
Phys. Rev. D \textbf{62} (2000), 122003
doi:10.1103/PhysRevD.62.122003
[arXiv:hep-th/0005246 [hep-th]].

\end{thebibliography}
\end{document}